\documentclass{aa}

\usepackage{amsmath,graphicx}
\usepackage{multirow}
\usepackage{booktabs}
\usepackage{color}
\usepackage{soul}
\usepackage{txfonts}
\usepackage{natbib}
\bibpunct{(}{)}{;}{a}{}{,}

\definecolor{Green}{RGB}{255,0,0}

\newcommand{\ascension}[4]{#1$^\mathrm{h}$#2$^\mathrm{m}$#3$^\mathrm{s}$.#4}
\newcommand{\declination}[4]{#1$^\circ$#2$^\prime$#3\; #4$^{\prime\prime}$}

\graphicspath{ {./images/} }

\begin{document}


\title{Polytropic spheres modelling dark matter haloes of dwarf galaxies}


\author{Jan Novotn\'{y}\inst{1}, Zden\v{e}k Stuchl\'{i}k\inst{1} and Jan Hlad\'{i}k\inst{1}}
\institute{Research Centre for Theoretical Physics and Astrophysics, Institute of Physics, Silesian University in Opava, Bezru\v{c}ovo n\'{a}m. 13, CZ-746\,01 Opava, Czech Republic}
          
\date{Received August 19, 2020; accepted --}

\abstract{Dwarf galaxies and their dark matter (DM) haloes have velocity curves of a different character than those in large galaxies. These velocity curves are modelled by a simple pseudo-isothermal model containing only two parameters, which do not give us insight into the physics of the DM halo.}{We seek to obtain some insight into the physical conditions in DM haloes of dwarf galaxies by using a simple physically based model of DM haloes.}{To treat the diversity of the dwarf galaxy velocity profiles in a unifying framework, we applied polytropic spheres characterised by the polytropic index $n$ and the relativistic parameter $\sigma$ as a model of dwarf-galaxy DM haloes and matched the velocity of circular geodesics of the polytropes to the velocity curves observed in the dwarf galaxies from the LITTLE THINGS ensemble.}{We introduce three classes of the LITTLE THINGS dwarf galaxies in relation to the polytrope models due to the different character of the velocity profile. The first class corresponds to polytropes that have $n < 1$ with linearly increasing velocity along the whole profile, the second class has $1 < n < 2$ and the velocity profile becomes flat in the external region, the third class has $n > 2,$ and the velocity profile reaches a maximum and demonstrates a decline in the external region. The $\sigma$ parameter has to be strongly non-relativistic ($\sigma < 10^{-8}$) for all dwarf galaxy models; this parameter varies for the models of each class, but these variations have negligible influence on the character of the velocity profile.}{Our results indicate a possibility that at least two different kinds of DM are behind the composition of DM haloes. The matches of the observational velocity curves are of the same quality as those obtained by the pseudo-isothermal, core-like models of dwarf galaxy DM haloes.}

\titlerunning{Polytropic spheres modelling DM haloes of dwarf galaxies}
\authorrunning{J. Novotn{\'y}, Z. Stuchl\'{i}k, J. Hlad\'{i}k}

\maketitle

\section{Introduction}\label{intro}

The general relativistic polytropic spheres represent a physically relevant idealisation of fluid configurations, giving a coherent picture of the physical (especially relativistic) phenomena influencing matter configurations across different distance scales. The polytropic equations of state are used to model neutron (or quark) stars. In the basic approximation, degenerated Fermi gas can be represented by an equation of state with the polytropic index $n = 3/2$ in the non-relativistic limit, and $n = 3$ in the ultra-relativistic limit \citep{Sha-Teu:1983:BHWDNS:}. Polytropic state equations with various values of the polytropic index $n$ are applied for precise approximations of the relativistic equations of state in neutron stars \citep{Oze-Psa:2009:PRD:, Lat-Pra:2001:ASTRJ2:NS}; even several polytropic state equations can be applied to cover the behaviour of the neutron star core \citep{Alv-Bla:2017:PRC}. In the models of neutron stars, the relativistic parameter giving the ratio of the central pressure to the central energy density, $\sigma = p_\mathrm{c}/\rho_\mathrm{c} \geq 0.1$. However, the polytropic spheres could also be relevant as very extended objects representing dark matter (DM) haloes of galaxies, or even galaxy clusters; then the parameter $\sigma$ can take both the large relativistic values $\sigma \geq 0.1$, or small values corresponding to the non-relativistic limit  $\sigma \leq 0.001$ \citep{Stu-Hle-Nov:2016:PHYSR4:,Stu-etal:2017:JCAP:,Stu-Nov-Hla-Hle:2021:preparation:}.

Despite  long-term efforts in theoretical and experimental particle physics, the composition of DM remains unknown. There are a large variety of possible, but not confirmed, candidates of both cold and hot (warm) DM that are still acceptable; this is similar to the possibility of the relevance of self-interacting DM. Since a clear DM candidate predicted by the particle physics is unknown, we are approved to choose any parameters of the polytropic equation of state to test predictions of the general relativistic polytropic spheres related to their extension, mass, and velocity curves, with respect to the observational data from galaxies or their clusters. Details of the physical processes inside the polytropic spheres are not considered; the power law relating the total pressure to the total energy density of matter is assumed. The polytropic approximation seems to be applicable in the DM models that assume weakly interacting particles (see for example \citet{Bor:1993:EarlyUniv:}, \citet{Kol-Tur:1990:EarUni:} and \citet{Cre-Stu:2013:IJMPD:}).

The general relativistic polytropic spheres in spacetimes with the repulsive cosmological constant have been numerically studied in detail \citep{Stu-Hle-Nov:2016:PHYSR4:}, generalising the case of uniform energy density spheres, corresponding to the polytropes with an index $n = 0$ that can be solved in terms of elementary functions~\citep{Stu:2000:ACTPS2:,Boh:2004:GENRG2:,Nil-Ugg:2000:ANNPH1:GRStarPoEqSt:,Boe-Fod:2008:PHYSR4:}. Two important results follow from this study. First, the polytropic spheres cannot exceed the static radius \citep{Stu:1983:BULAI:,Stu-Hle:1999:PHYSR4:,Arra:2014:PHYSR4:,Far-Lap-Pra:2015:JCAP:,Stu-Char-Sche:2018:EPJC:,Stu-etal:2020:Universe}; the radius is  governed by the polytrope mass and value of the cosmological constant, where the gravitational attraction of the polytrope is just balanced by the cosmic repulsion, giving a natural limit on gravitationally bound systems in the expanding Universe \citep{Stu:2005:MODPLA:}. The static radius thus also represents a limit on the extension of the polytropes with critical values of the relativistic parameter $\sigma_\mathrm{crit}(n),$ which are unlimited in spacetimes with a vanishing cosmological constant \citep{Nil-Ugg:2000:ANNPH1:GRStarPoEqSt:,Stu-Hle-Nov:2016:PHYSR4:, Stu-etal:2020:Universe}. The role of the cosmological constant and the static radius is also crucial for the motion of gravitationally bound galaxies, for example in the case of the Milky Way and its small satellite galaxies as shown in \citet{Stu-Sch:2011:JCAP:CCMagOnCloud} and \citet{Stu-Sch:2012:INTJMD:GRvsPsNewtMagClou}. Second, the extension and mass of the polytropic spheres can even be comparable with extension and mass of DM haloes of large galaxies and galaxy clusters \citep{Stu-Hle-Nov:2016:PHYSR4:}. Extremely extended general relativistic polytropic spheres can well serve as models of DM in large galaxies or even galaxy clusters. This is the case both for non-relativistic polytropes with $\sigma \ll 0.1,$ which can represent very diluted haloes made of cold DM; and for the relativistic polytropes with sufficiently large polytropic indices $3.3 < n < 5$ when such polytropes can exist, while $\sigma \sim \sigma_\mathrm{crit}(n) \geq 0.1$, representing haloes constituted by hot DM \citep{Stu-Nov-Hla-Hle:2021:preparation:}. For such large galaxies (or their clusters), the role of the cosmological constant is crucial, as an extension of such galaxies is comparable to the values of the static radius corresponding to their mass. This statement holds even for Milky Way \citep{Stu-Sch:2011:JCAP:CCMagOnCloud}. As expected, the cosmological constant is not significant for dwarf galaxies, as their extension is much smaller than the static radius corresponding to their mass.

Of special interest are highly relativistic trapping polytropes containing a region of trapped null geodesics existing for polytropes with $n > 2.167$ and $\sigma > 0.667$ \citep{Nov-Hla-Stu:2017:PRD:trapping}. The trapping of null geodesics is relevant in the region in which circular null geodesics exist an inner (stable) one at $r_{\mathrm{ph(s)}}$, and outer (unstable) one at $r_{\mathrm{ph(u)}}$. Surprisingly, the local compactness of the trapping configuration does not exceed the global compactness of ultra-compact objects whose surfaces are located under the unstable photon circular geodesic of the external vacuum Schwarzschild (Schwarzschild-de Sitter) spacetime \citep{Stu:2000:ACTPS2:}, namely,
\begin{equation}
  \frac{m(r_\mathrm{ph(s)})}{r_\mathrm{ph(s)}} < \frac{m(r_\mathrm{ph(u)})}{r_\mathrm{ph(u)}} < \frac{1}{3}\, ;\end{equation}
details are given in \citet{Nov-Hla-Stu:2017:PRD:trapping}. An analytic confirmation of this unexpected phenomenon can be found in \citet{Hod:2018:PHYSR4:,Hod:2018:EPJC:}. The trapping zone demonstrates an efficient gravitational instability causing gravitational collapse of its matter content and its conversion to a central black hole \citep{Stu-etal:2017:JCAP:}. For the polytropic parameters properly tuned, the trapping polytrope can be extremely extended, thus representing a DM halo of a large galaxy with mass $M_\mathrm{halo} > 10^{12}\,M_\odot$; the central super-massive black hole created in the trapping zone can have  mass $M_\mathrm{BH} > 10^9\,M_\odot$ \citep{Stu-etal:2017:JCAP:}. The gravitational instability of the ``trapping'' region of extremely extended trapping polytropes has thus been proposed as an alternative model of formation of the super-massive black holes in high-redshift ($z > 6$) large galaxies or their clusters \citep{Stu-etal:2017:JCAP:,Stu-Nov-Hla-Hle:2021:preparation:}. There is another interesting family of relativistic polytropes that contains a region of trapped test particles \citep{Stu-Nov-Hla-Hle:2021:preparation:}, which could be relevant for capturing DM in the interior of neutron stars.

The cold DM haloes are considered the natural explanation of hidden gravitating structures in large galaxies or galaxy clusters \citep{Bin-Tre:1988:GalacDynam:,Nav-Fre-Whi:1997:ASTRJ2:UniDeProHiCl,Bar:etal:2015:MNRAS:GalLensing:}. However, there are indications of presence of warm (hot) DM haloes, especially in the primeval galaxies \citep{Lap-Dan:2015:JCAP:}. Even mixed cold-warm \citep{And-etal:2012:JCAP:} or self-interacting \citep{Bon-Boy-etal:2018:JCAP:} DM halo models are considered. In our recent study, we demonstrated \citep{Stu-Nov-Hla-Hle:2021:preparation:} that extremely extended polytropes can explain the DM haloes (their mass, extension, and velocity curves) of large galaxies and galaxy clusters, both for the polytropes with any value of polytropic index $n \in (0,4\mbox{--}5)$ and a small value of the relativistic parameter $\sigma \ll 0.1$ corresponding to cold matter (or warm matter with a substantially decreased temperature owing to the cosmic evolution), and polytropes of high polytropic index $n > 3.3$ and high relativistic parameter $\sigma \geq 0.1$ corresponding to hot matter.

In the present paper, we focus our attention on the type of galaxies that  complement large galaxies, namely, dwarf galaxies, which demonstrate a behaviour of velocity curves that is not relevant for large galaxies. There is an extensive study of the dwarf galaxies concentrated in the so-called ``LITTLE THINGS'' (Local Irregulars That Trace Luminosity Extremes, The H~I Nearby Galaxy Survey) \citep{Oh-etal:2015:AJ:} where velocity curves are presented along the whole regions corresponding to the dwarf galaxies; we can assume that the polytrope model can be applied along the entire dwarf galaxy region, including its innermost part. This is true at least for almost all of the galaxies of the LITTLE THINGS, where the role of the visible parts of the galaxy on their motion seems to be negligible. However, there are some dwarf galaxies that demonstrate a strong deviation from regular behaviour of velocity curves, for which we have to expect a strong influence of baryonic matter causing the irregularities; see for example the dwarf galaxies DDO~50 and IC~1613, with the observed velocity curves divided into several parts, presented in \citet{Oh-etal:2015:AJ:}. Even in these very special cases, we can still realise an average matching of the observed velocity curves, keeping in mind the limits of this matching that could be modified by considering the possible influence of baryonic matter causing the strong deviations from the smooth lines predicted by the DM halo polytropic model. Extension and mass of dwarf galaxies are  several orders smaller in comparison with those of large galaxies. Therefore, in the polytrope models of DM haloes related to the dwarf galaxies, we can abandon the role of the cosmological constant that can be very important in the case of large galaxies \citep{Stu-Hle-Nov:2016:PHYSR4:}. We use the polytrope models to propose a classification of the dwarf galaxies according to the values of the polytrope index $n$, characterising the models of concrete dwarf galaxies included in the LITTLE THINGS. In these polytrope models, the relativistic parameter $\sigma$ can vary between polytropes modelling various dwarf galaxies of the ensemble, having no influence on the classification of the polytropes. We have to stress that the polytrope models enable a systematic study of dwarf galaxies and search for a unique physical phenomenon hidden behind their DM haloes.

In previous studies of the dwarf galaxies of the LITTLE THINGS, the velocity curves were matched by the standard phenomenological model of DM haloes, namely the pseudo-isothermal model. The pseudo-isothermal model is oversimplified, as it contains only two parameters characterising the core density $\rho_0$ and core radius $R_\mathrm{C}$. This model thus enables us to match the observational data, but we could hardly extract relevant information on the physics behind such a halo model. On the other hand, the polytrope model contains parameters that reflect physical model conditions, especially the polytrope parameter $n$ , which facilitates a representation of physical conditions. Moreover, we are able to demonstrate that the precision of our matches by the polytrope models is comparable to those of the pseudo-isothermal model of dwarf galaxies. However, the fact that in almost half of the dwarf galaxies the polytrope model give better fits, while the phenomenological iso-thermal model gives better fits in the other half of the ensemble, indicates the need to include the role of the baryonic matter in the galaxy structure. Such a complex model will be considered in future studies.

In our paper we summarise first the polytrope structure equations, their solutions, and properties of the polytropic spheres. Then we present an overview of the circular geodesics in the internal spacetime of the polytropic spheres, finally giving the velocity radial profiles of the circular geodesics that should correspond to the velocity curves of stars at the dwarf galaxies. We assume applicability of the geodesic velocity profiles in the whole polytropic sphere to the complete observed velocity curves of dwarf galaxies. Finally, we present results of the matching procedure realised for each of the selected dwarf galaxies of the LITTLE THINGS; the parameters of the polytrope yielding the best match of the observational velocity curve of the galaxy are explicitly established and the corresponding theoretical velocity curve is compared to the observational data for each dwarf galaxy of the LITTLE THINGS. We then present the conclusions resulting from the best matches.


\section{General relativistic polytropes}\label{eost}

The line element of spherically symmetric, static spacetimes, can be expressed in terms of the standard Schwarzschild coordinates in the form
\begin{equation}
    \mathrm{d}s^{2} = - \mathrm{e}^{2\Phi} c^{2} \mathrm{d}t^{2} + \mathrm{e}^{2\Psi} \mathrm{d}r^{2} + r^{2} (\mathrm{d}\theta^{2} + \sin^{2} \theta\, \mathrm{d}\phi^{2})\, .
\end{equation}
The line element contains two unknown functions of the radial coordinate, $\Phi(r)$ and $\Psi(r)$. The static spherically symmetric perfect fluid configurations are in the fluid rest frame represented by the mass-energy density radial profile $\rho = \rho (r)$, and isotropic pressure radial profile $p = p(r)$.

The mass-energy density and pressure are related by the polytropic equation of state
\begin{equation}
    p = K \rho^{1+1/n}\, ,
\end{equation}
where $n$ denotes the constant polytropic index and $K$ denotes the constant governed by the thermal characteristics of the considered polytropic configuration, given by the central density $\rho_\mathrm{c}$ and the central pressure $p_\mathrm{c}$. The parameter $K$ is determined by the total mass and radius of the polytropic configuration, and the relativistic parameter \citep{1964T-GRP}
\begin{equation}
    \sigma  \equiv \frac{p_{\mathrm{c}}}{\rho_{\mathrm{c}} c^{2}}\  = \frac{K}{c^{2}}\rho_{\mathrm{c}}^{1/n}\, .
\end{equation}
The constant $K$ contains the temperature implicitly \citep{1964T-GRP}. The polytropic equation represents a limiting form of the parametric equations of state for the completely degenerate gas at zero temperature; therefore, this equation can be relevant for a basic description of neutron stars. Then, both $n$ and $K$ are universal physical constants \citep{1964T-GRP}. The ultra-relativistic degenerate Fermi gas is determined by the polytropic equation with the adiabatic index $\Gamma = 4/3$ and the polytropic index $n = 3$; for the non-relativistic degenerate Fermi gas, the polytropic equation of state has $\Gamma = 5/3$, and $n = 3/2$ \citep{Sha-Teu:1983:BHWDNS:}.

\subsection{Structure equations}
The structure of the polytropic spheres is governed by the two structure functions; their behaviour is determined by the Einstein field equations and the local energy-momentum conservation law. The first structure function, $\theta(r)$, is related to the mass-energy density radial profile $\rho(r)$ and the central density $\rho_\mathrm{c}$ through the relation \citep{1964T-GRP}
\begin{equation}
    \rho = \rho_{\mathrm{c}} \theta^{n}\, ,
\end{equation}
with the boundary condition $\theta(r=0)=1$. The second structure function is the mass function, which is determined by the relation
\begin{equation}
    m(r) =  \int^{r}_{0} {4 \pi r'^{2} \rho\, \mathrm{d}r'}\, ;
\end{equation}
the integration constant $m(0) = 0$ is chosen to guarantee the smooth spacetime geometry at the origin \citep{1973MTW-G}. At the surface of the configuration, $r=R$, there is $\rho(R)=p(R)=0$. The total mass of the polytropic configuration is denoted as $M=m(R)$, and outside the polytropic configuration, the spacetime is described by the vacuum Schwarzschild metric with mass parameter $M$.

The structure equations of the polytropic spheres that relate the two structure functions, $\theta(r)$ and $m(r)$, and the polytrope parameters $n$, $\sigma$, can be put into the form \citep{1964T-GRP, Stu-Hle-Nov:2016:PHYSR4:}
\begin{align}
    \frac{\sigma(n+1)}{1+\sigma\theta}\,r\,\frac{\mathrm{d}\theta}{\mathrm{d}r} \left(1-\frac{2G m}{c^{2}r}\right) + \frac{G m}{c^{2}r} &= -\frac{G}{c^{2}}\sigma\theta\frac{\mathrm{d}m}{\mathrm{d}r},   \label{grp24}\\
    \frac{\mathrm{d}m}{\mathrm{d}r} &= 4\pi r^{2} \rho_{\mathrm{c}}\theta^{n}.    \label{grp25}
\end{align}

It is useful to introduce two characteristic scales \citep{1964T-GRP} for the polytropic spheres: the characteristic length scale $\mathcal{L}$ by the relation
\begin{equation}
    \mathcal{L} = \left[\frac{(n+1)K\rho_{\mathrm{c}}^{1/n}}{4\pi G\rho_{\mathrm{c}}}\right]^{1/2} = \left[\frac{\sigma(n+1)c^{2}}{4\pi G\rho_{\mathrm{c}}}\right]^{1/2}\, ,
\end{equation}
and the characteristic mass scale $\mathcal{M}$ by the relation
\begin{equation}
    \mathcal{M} = 4\pi \mathcal{L}^3 \rho_\mathrm{c} = \frac{c^2}{G}\sigma (n+1)\mathcal{L}\, .
\end{equation}
The structure equations, Eqs.~(\ref{grp24}) and (\ref{grp25}), can be then transformed into dimensionless form by introducing a dimensionless radial coordinate
\begin{equation}
    \xi = \frac{r}{\mathcal{L}}\, ,
\end{equation}
and dimensionless gravitational mass function
\begin{align}
    v(\xi) &= \frac{m(r)}{4\pi \mathcal{L}^{3}\rho_{\mathrm{c}}}\, .
\end{align}

The dimensionless structure equations take the form (for details see \citet{Stu-Hle-Nov:2016:PHYSR4:} and \citet{1964T-GRP})
\begin{align}
    \xi^{2}\frac{\mathrm{d}\theta}{\mathrm{d}\xi}\frac{1-2\sigma(n+1)\left(v/\xi\right)}{1+\sigma\theta} + v(\xi)&= - \sigma\xi\theta\frac{\mathrm{d}v}{\mathrm{d}\xi}, \label{grp31}\\
    \frac{\mathrm{d}v}{\mathrm{d}\xi} &= \xi^{2}\theta^{n}\, . \label{grp32}
\end{align}
The dimensionless polytropic structure equations are fully determined only by the polytrope parameters $n$ and $\sigma$; the equations are independent of the central energy density $\rho_\mathrm{c}$ that governs the extension and mass scales of the polytrope. This degeneracy in the structure equations can be violated by the presence of the cosmological constant when the parameter determining the ratio of the central energy density and the vacuum energy density related to the cosmological constant enters the structure equations \citep{Stu-Hle-Nov:2016:PHYSR4:}. The solutions of the polytropic structure equations can only be obtained by numerical methods \citep{1964T-GRP}; the exception to this is the $n=0$ polytropes governing the spheres with a uniform distribution of the energy density when the solution can be given in terms of the elementary functions \citep{Stu-Hle-Nov:2016:PHYSR4:,Stu:2000:ACTPS2:}.

For fixed parameters $n$, $\sigma$, the structure equations (\ref{grp31}) and (\ref{grp32}) are solved simultaneously under the boundary conditions
\begin{equation}
    \theta(0) = 1\, , \quad v(0) = 0\,.    \label{grp33}
\end{equation}
Eqs.~(\ref{grp32}) and (\ref{grp33}) imply that $v(\xi) \sim \xi^{3}$ for $\xi \to 0$, according to Eq.\,(\ref{grp31}), we find
\begin{equation}
    \lim_{\xi\to 0_+}\frac{\mathrm{d}\theta}{\mathrm{d}\xi} = 0\, .
\end{equation}
The polytrope surface, $r = R$, is represented by the first zero point of $\theta(\xi)$, denoted as $\xi_{1}$, that is by
\begin{equation}
    \theta(\xi_{1}) = 0\, .
\end{equation}
The solution $\xi_{1}$ determines the surface radius of the polytropic sphere, and the solution $v(\xi_{1})$ determines the gravitational mass of the polytrope in the general dimensionless form. The dimensional form is then obtained using the length and mass scales where the central energy density is also in play.

\subsection{Characteristics of the polytropic spheres}

A polytropic sphere constructed for given parameters $n$, $\sigma$, and $\rho_\mathrm{c}$ is characterised by two solutions of the dimensionless structure equations, $\xi_{1}$ and $v(\xi_{1})$, and by the scale factors $\mathcal{L}$ and $\mathcal{M}$. We note that the mass scale $\mathcal{M}$ is not independent, being determined by the length scale $\mathcal{L}$ and parameters $n$, $\sigma$; both $\mathcal{L}$ and $\mathcal{M}$ are determined by the central density $\rho_c$.

The radius of the polytropic sphere is given by
\begin{equation}
    R = \mathcal{L} \xi_{1}\, ,
\end{equation}
while its gravitational mass is written as
\begin{equation}
    M = \mathcal{M} v(\xi_1) = \frac{c^{2}}{G} \mathcal{L}\sigma(n+1) v(\xi_{1})\, .
\end{equation}
The radial profiles of the energy density, pressure, and mass distribution are given by the relations
\begin{eqnarray}
    \rho(\xi) &=& \rho_{\mathrm{c}}\theta^{n}(\xi)\, ,\\
    p(\xi) &=& \sigma\rho_{\mathrm{c}}\theta^{n+1}(\xi)\, ,\\
    M(\xi) &=& M\frac{v(\xi)}{v(\xi_{1})}\, .
\end{eqnarray}
The temporal metric coefficient is
written as\begin{equation}
    \mathrm{e}^{2\Phi_\mathrm{int}} = (1+\sigma\theta)^{-2(n+1)} \left\{1-2\sigma(n+1) \frac{v(\xi_{1})}{\xi_{1}} \right\}\, ,
\end{equation}
and the radial metric coefficient is written as
\begin{equation}
    \mathrm{e}^{-2\Psi_\mathrm{int}} = 1 - 2\sigma(n+1) \frac{v(\xi)}{\xi}\, .
\end{equation}
Detailed discussion of the properties of the polytropic spheres, including their gravitational binding energy and the internal energy, can be found in \citet{Stu-Hle-Nov:2016:PHYSR4:} and \citet{1964T-GRP}. The effectiveness of the gravitational binding of the polytropic spheres is given by the compactness parameter, which is written as
\begin{equation}
    \mathcal{C} \equiv \frac{GM}{c^{2}R} = \frac{1}{2}\frac{r_{\mathrm{g}}}{R} = \frac{\sigma(n+1)v(\xi_{1})}{\xi_{1}}\, ,
\end{equation}
where the standard gravitational radius of the polytropic sphere is given by
\begin{equation}
    r_{\mathrm{g}} = \frac{2GM}{c^{2}}.
\end{equation}
The compactness $\mathcal{C}$ of the polytropic sphere can be reflected by the gravitational redshift of radiation emitted from the polytrope surface \citep{2011HS-PNR}. The exterior of the polytropic sphere is represented by the vacuum Schwarzschild spacetime with the same gravitational mass parameter $M$ as those characterising the internal spacetime of the polytropic sphere, being given by the metric coefficients
\begin{equation}
    \mathrm{e}^{2\Phi_\mathrm{ext}} = \mathrm{e}^{-2\Psi_\mathrm{ext}} = 1 - \frac{2GM}{c^{2} r}\, .
\end{equation}


\section{Circular geodesics of the internal polytrope spacetime}

The velocity curves of the stars observed in the galaxy plane of dwarf galaxies must be compared to the circular geodesics of the spherically symmetric internal spacetime of the polytropic sphere in the central plane, which can be selected as the equatorial plane, $\theta=\pi/2$. The circular geodesic motion of the test particles (stars) in the internal polytrope spacetime is characterised by two constants of motion --- specific energy $E$ and specific angular momentum $L$, which are defined as the ratio of the covariant energy (axial angular momentum) related to the rest energy of the star $m$ that is also considered as a constant of the motion. In general spherically symmetric static spacetime, the equations of the equatorial geodesic motion can be given in the separated and integrated form as follows:
\begin{eqnarray}
    \frac{\mathrm{d} t}{\mathrm{d} w} & = & \frac{E g_{\phi\phi}}{-g_{tt}g_{\phi\phi}}\, ,\\
    \frac{\mathrm{d} \phi}{\mathrm{d} w}& = & \frac{L}{g_{\phi\phi}}\, ,\\
    g_{rr}\left(\frac{\mathrm{d} r}{\mathrm{d} w}\right)^2 & = & R(r)\, ,
\end{eqnarray}
where $w$ is the proper time of the particle and the function governing the radial motion takes the form
\begin{equation}
    R(r) = -1 + \frac{E^2 g_{\phi\phi} + L^2 g_{tt}}{-g_{tt}g_{\phi\phi}}\, .
\end{equation}

Using the simultaneous conditions of the circular motion, $R=0$, $\mathrm{d} R/\mathrm{d} r=0$, we can express the constants of the motion, $E, L$, and the angular velocity relative to the distant observers, $\Omega$, in terms of the metric coefficients in the form (assuming corotating orbits with $L>0$)
\begin{eqnarray}
    E &=& \frac{-g_{tt}}{\sqrt{-g_{tt}-g_{\phi\phi}\Omega^2}}\, , \\
    L &=& \frac{g_{\phi\phi}\Omega}{\sqrt{-g_{tt}-g_{\phi\phi}\Omega^2}}\, ,\\
    \Omega & = & \frac{\mathrm{d} \phi}{\mathrm{d} t} = \frac{\sqrt{-g_{tt,r}g_{\phi\phi,r}}}{g_{\phi\phi,r}}\, ,
\end{eqnarray}
where $,r$ ($,rr$) denotes the first (second) derivative in the radial direction. The condition of marginal stability of the circular geodesics, $R_{,rr} = 0$, can be expressed in the form
\begin{equation}
    E^2 g_{\phi\phi,rr} + L^2 g_{tt,rr} + (g_{tt}g_{\phi\phi})_{,rr} = 0\, .
\end{equation}
The velocity profile of the circular geodesics is determined by the simple formula
\begin{equation}
    v(r) = \frac{L}{\sqrt{g_{\phi\phi}}}\, .
\end{equation}

In the internal polytrope spacetime, where the metric coefficient $g_{tt} = - \mathrm{e}^{2\Phi}$ is governed by the structure function $\theta(\xi)$ and the polytrope parameters, and $g_{\phi\phi} = \xi^2$, the radial profile of the specific energy of the circular geodesics is written as
\begin{multline}
    E^2(\xi) =  (1 + \sigma\theta)^{-2(n+1)} \left\{1 - 2\sigma(n+1)\frac{\nu(\xi_{1})}{\xi_{1}} \right\} \times \\
       \left[1 + \frac{\sigma\xi(n+1)}{1+\sigma\theta}\frac{\mathrm{d}\theta}{\mathrm{d}\xi}\right]^{-1}\, ,
\end{multline}
while for the specific axial angular momentum we obtain
\begin{equation}
           L^2(\xi) = - (n+1)\sigma \xi^3 \frac{\mathrm{d}\theta}{\mathrm{d}\xi} \left[1 + \sigma\theta + (n+1)\sigma\xi \frac{\mathrm{d}\theta}{\mathrm{d}\xi}\right]^{-1}\, ,
\end{equation}
and the angular velocity relative to static distant observers is written as
\begin{equation}
    \Omega^2 = \frac{- \left\{1 - 2\sigma(n+1) {\nu(\xi_1)}/{\xi_1}\right\}(n+1)\sigma\, {\mathrm{d}\theta}/{\mathrm{d}\xi}}{\xi(1+\sigma\theta)^{2n+3}}\, .
\end{equation}
We note that the relevant structure function satisfies the condition $\mathrm{d}\theta/\mathrm{d}\xi < 0$, which guarantees positiveness of the radial profiles of the circular geodesic motion constants $E^2, L^2$, and the angular velocity $\Omega^2$. The condition for the marginally stable circular geodesics takes the form

\begin{equation}
\begin{split}
    -1 + \frac{4(n + 1)\sigma\xi}{1 + \sigma\theta}\frac{\mathrm{d}\theta}{\mathrm{d}\xi} + \left[1 + \frac{(n + 1)\sigma\xi}{1 + \sigma\theta}\frac{\mathrm{d}\theta}{\mathrm{d}\xi}\right]^{-1} + & \\
    \left\{\frac{(n + 1)\sigma\xi^2}{1 + \sigma\theta} - \frac{(n + 1)^2 \sigma^2\xi^3 \left(\mathrm{d}\theta/\mathrm{d}\xi\right)}{\left(1 + \sigma\theta\right)^2\left[1 + \frac{(n + 1)\sigma\xi}{1 + \sigma\theta}\frac{\mathrm{d}\theta}{\mathrm{d}\xi}\right]}\right\} \times & \\
     \left[\frac{\mathrm{d}^2\theta}{\mathrm{d}\xi^2} - \frac{(2n + 3)\sigma}{1 + \sigma\theta} \left(\frac{\mathrm{d}\theta}{\mathrm{d}\xi}\right)^2\right] & = 0\, .
\end{split}
\end{equation}

The radial profile of the orbital velocity of the circular geodesics that is applied in the matching procedure to the observational data is written as
\begin{equation}
    v(\xi) = \left[\frac{- (n + 1)\sigma \xi \frac{\mathrm{d}\theta}{\mathrm{d}\xi}}{{1 + \sigma\theta} + (n + 1)\sigma\xi\, \frac{\mathrm{d}\theta}{\mathrm{d}\xi}}\right]^{1/2}\,.\label{eq:pol_radial_velocity}
\end{equation}

\begin{figure}[htb!]
        \begin{center}
                \includegraphics[width=\linewidth]{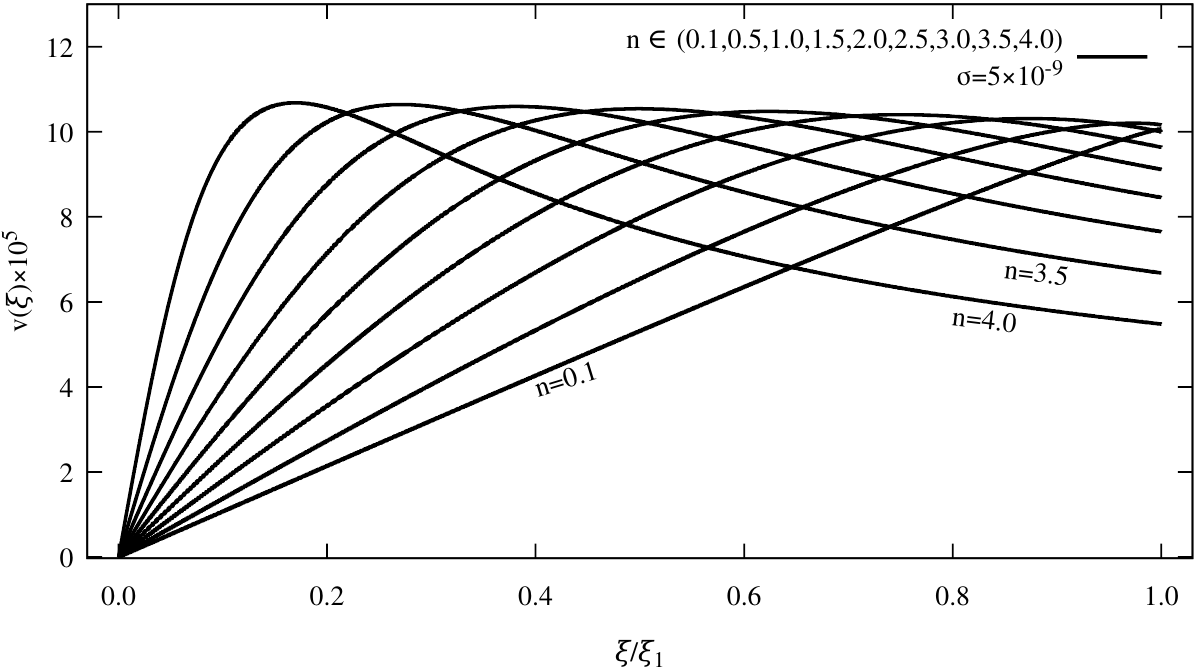}\\
        \includegraphics[width=\linewidth]{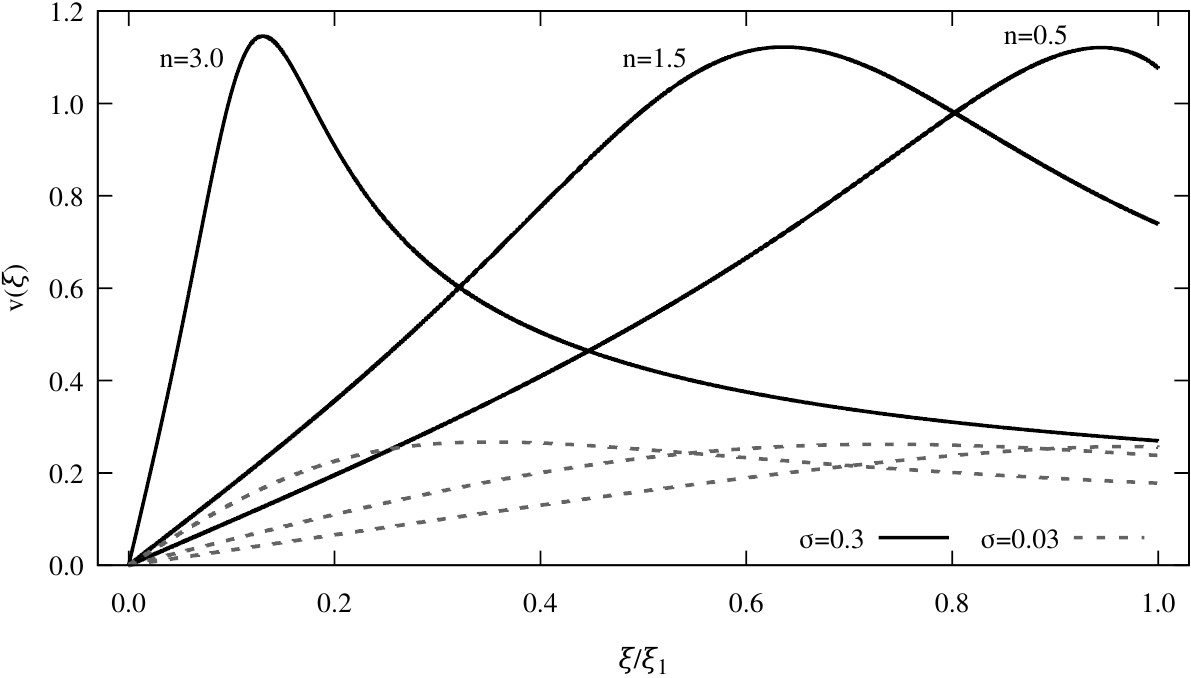}
        \end{center}    
                \caption{\label{fig1}Polytrope velocity profiles. \emph{Top}: Types of velocity radial profiles in the interior of polytropic spheres. The dependence of the radial velocity profiles on the polytropic index $n$ for fixed relativistic parameter $\sigma$ typical for the dwarf-galaxy polytropes are given. \emph{Bottom}: Profiles given for fixed representative values of $n$ and varying relativistic parameter $\sigma$ to demonstrate that in the second case the variations keep the character of the profile but shift their range.}
\end{figure}

\begin{figure}[htb!]
        \begin{center}
                \includegraphics[width=\linewidth]{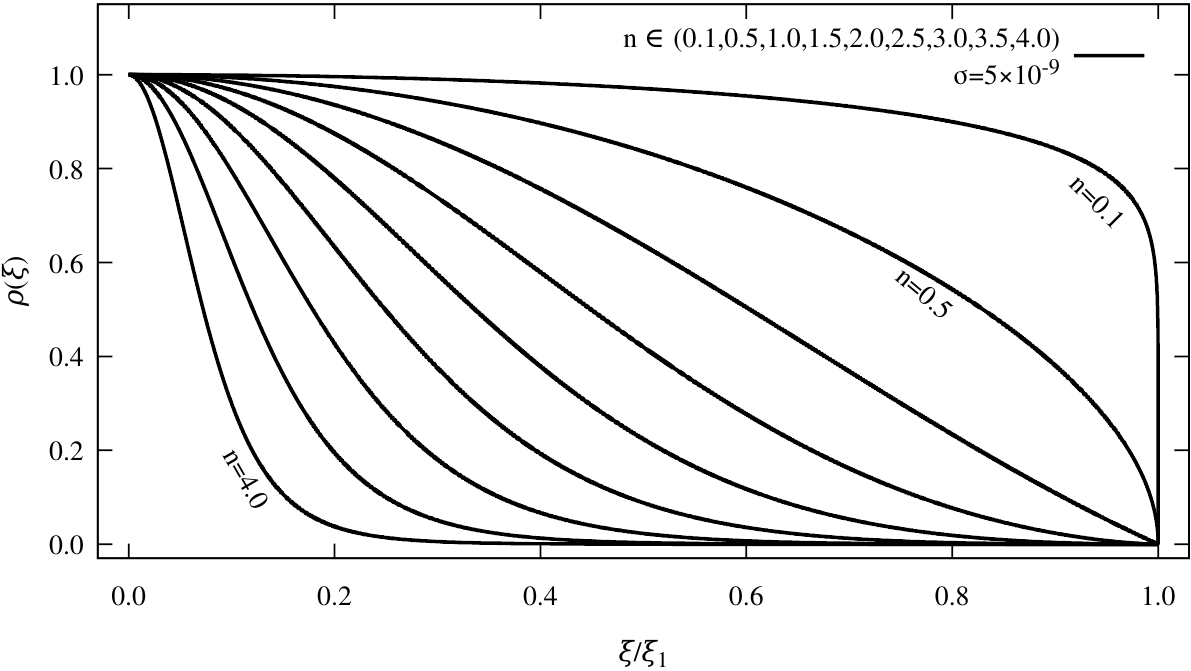}\\
        \includegraphics[width=\linewidth]{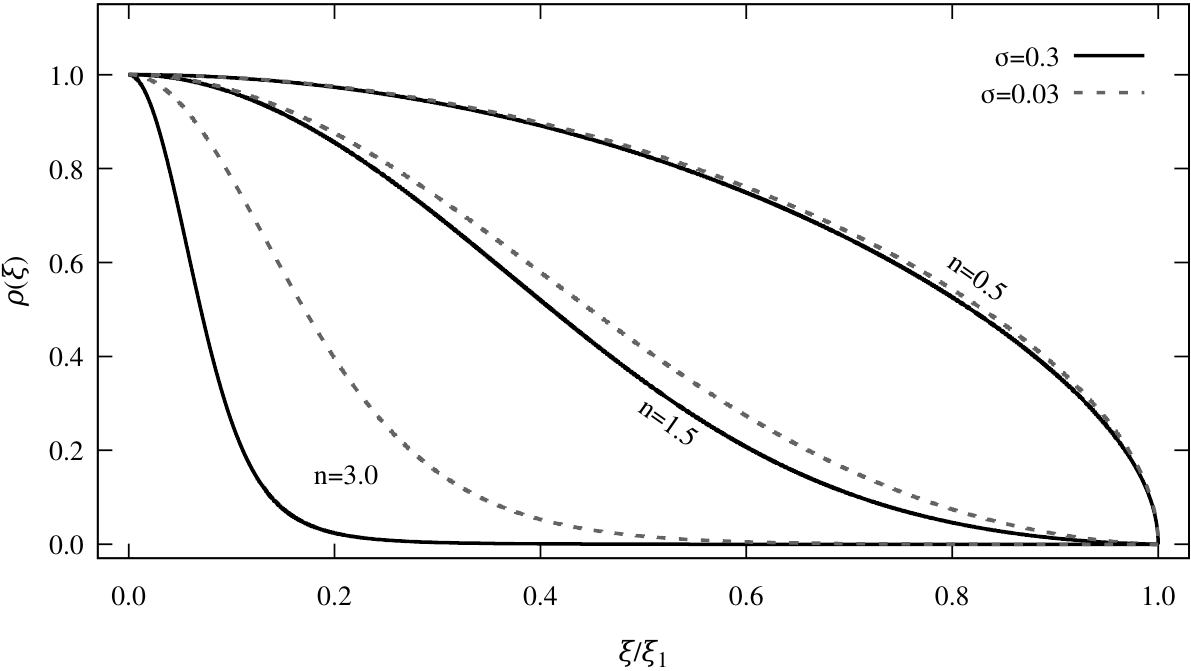}
        \end{center}    
                \caption{\label{fig:density}Dependence of the density radial profiles in the interior of polytropic spheres with polytropic index $n$ and relativistic parameter $\sigma$ as specified in Figure~\ref{fig1}.}
\end{figure}

The functions characterising the circular geodesics are governed by the ``density'' structure function $\theta(\xi)$, but these functions are independent of the ``mass'' structure function $\nu(\xi)$, which is only affected by the parameter $\nu(\xi_1)$ characterising the total mass of the polytropic sphere. A detailed discussion of the properties of the circular geodesics of the internal polytrope spacetimes can be found in \citet{Stu-Nov-Hla-Hle:2021:preparation:}. In this work, we use the expression giving the velocity radial profiles in the interior of the polytropes to match to the velocity curves observed in the individual concrete dwarf galaxies. We give several examples of the behaviour of the theoretical velocity curves representing the fundamental types in Figure~\ref{fig1} and their appropriate density profiles in Figure~\ref{fig:density}. The first type demonstrates almost a linear increase of the velocity, in the second type we observe a flat tail of the velocity profile, while the third type contains a maximum followed by the slowly descending part of the profile, which could correspond to the profiles observed in large galaxies.

The energy profiles clearly demonstrate the tendency to concentrate mass in the central region with increasing polytropic index $n$ and this tendency is strengthened by increasing polytropic relativistic parameter $\sigma$. Nevertheless, the polytropic index plays the crucial role, so the core-like character of the density profiles holds for high $n$, even for strongly non-relativistic polytropes.

\section{Dwarf galaxies}
We tested the applicability of our polytrope model of galactic haloes for the very interesting, and in some sense mysterious, family of dwarf galaxies. Their basic mystery is related to the numerical simulations of the evolution of the mass distribution in the Universe, which predict the number of dwarf galaxies to be higher (by an order) than the observed number \citep{Moore-etal:1999:APJL:}. One of the potential solutions of the so-called missing dwarf galaxy problem is that a small part of these galaxies are able to attract baryonic matter in amounts sufficient to become visible galaxies. Therefore, the whole structure of the dwarf galaxies could be generally DM dominated \citep{Sim-Geh:2007:APJ:}, and whole the profile of the velocity curves observed in such galaxies could be used for direct comparison with the velocity profiles predicted by the models of DM haloes. The other crucial open problem of the dwarf galaxies is the so-called core-cusp problem related to the character of the density profile in the central region of the DM halo \citep{Bul-etal:2001:MNRAS}, as the density profiles of dwarf galaxies are rather ``core-like'' because they are described by the pseudo-isothermal (ISO) models, not the ``cusp-like'' standard CDM Navarro-Frenk-White (NFW) model. One possible solution to this problem could be realised if we put aside the assumption of cold DM and take into consideration warm or self-interacting DM \citep{Lov-etal:2012:MNRAS:WDM,Elb-etal:2015:MNRAS:selfint, Spe-Ste:2000}. Another solution is connected with the overall effect of a repeating DM heating caused by a suddenly removed accreted gas (by stellar winds or supernovae feedback) during the star formation leading to a complete DM core formation \citep{Read-Gil:2005}. Moreover, recent numerical simulations \citep{Tes-etal:2013:MNRAS, Mas-etal:2008,Pon-Gov:2012,Read-Age-Col:2016} show that the gas flows transforming DM cusps to cores conduct to cause a bursty star formation history and the stars are dynamically heated similar to the DM, leading to a stellar velocity dispersion approaching the local rotational velocity of the stars ($v/\sigma{\sim}{}1$). These results are supported by observations of dwarf galaxies \citep{Dohm-etal:1998, Dohm-etal:2002, Young:2007, Leaman:2012,Kau:2014,Spa-etal:2017, Whe-etal:2017}. Moreover, \citet{Read-Wal-Ste:2019:MNRAS} show that the dwarfs fell into two classes related by the central DM density at a common radius 150~pc ($\rho_\mathrm{DM}$), respectively, galaxies with old stars (>6 Gyr old) with central DM densities $\rho_\mathrm{DM}(150~\mathrm{pc})>10^8~\mathrm{M}_\odot{}\mathrm{kpc}^{-3}$ consistent with DM cusps, and galaxies with stars at least 3~Gyr old with $\rho_\mathrm{DM}(150~\mathrm{pc})<10^8~\mathrm{M}_\odot{}\mathrm{kpc}^{-3}$ consistent with DM cores. Other possible solutions of the core-cusp problem include ``wave-like'' DM \citep{Schive-etal:2014}, ``fuzzy'' DM \citep{Hu-Bar-Gru:2000,Hui-etal:2017}, and ``fluid'' DM \citep{Pee:2000}. 

In this paper we intend to add a new physical approach viewpoint to the cusp-core problem of dwarf galaxies using the properties of the polytropic spheres in modelling DM haloes, namely the influence of the polytropic index on the energy density radial profiles demonstrated in Figure~\ref{fig:density}

In the lines of these warm or self-interacting possibilities, we consider the possible range of the values of the polytropic index $n$. The most extended study of the dwarf galaxies and their observed velocity profiles based on the standard cold DM models of haloes is presented in LITTLE THINGS \citep{Oh-etal:2015:AJ:}, and we test our polytrope model for the selected 20 members of the LITTLE THINGS, using a coherent approach of treating the observational data in the well-defined extended study of \citet{Oh-etal:2015:AJ:}. In the LITTLE THINGS study of \citet{Oh-etal:2015:AJ:}, 26 (including 3 from THINGS) dwarf galaxies were considered in great detail. We chose 20 galaxies, with regular rotation pattern in their HI velocity fields. We studied their data using the polytrope models, after treating the original observational data kindly submitted by the authors of \citet{Oh-etal:2015:AJ:} to make it possible to have a clear comparison of the results of our polytrope models to the results of the isothermal models of DM haloes of dwarf galaxies. The polytrope models facilitate a systematic study due to the estimated values of the polytropic index $n$ and the relativistic parameter (central density).

As shown in the simulations by \citet{Gov-etal:2012}, the degree of baryonic contributions is significantly dependent on the amount of stars. In other words, galaxies with higher past star formation activity have more significant outflows in the same total mass budget (DM + baryons) than lower past star formation activity galaxies. Therefore, to investigate the effect of baryons on the central cusp, a reliable measurement of the stellar mass in galaxy is essential. \citet{Pra-Bur:2002} showed that for the case of late-type dwarf galaxies the total kinematics is dominated by the DM, nonetheless it is still important to separate the contribution by baryons from the total rotation curve. In our case we used the velocity curves corresponding to the role of the DM haloes. As mentioned in \citet{Oh-etal:2015:AJ:}, these were achieved by using multiwavelength data sets, respectively, by using a Spitzer IRAC 3.6~$\mu\mathrm{m}$ images and ancillary optical colour information \citep{Hun-Elm:2006}. We plan to extend our polytrope DM model to the contribution of the baryonic matter in the future.

\begin{table*}
\caption{\label{tab:LittleThings_sum}Results of the LITTLE THINGS sample dwarf galaxies for the pseudo-isothermal fit by~\citet{Oh-etal:2015:AJ:}. The results are presented for the selected 20 galaxies taken from the whole LITTLE THINGS sample to compare the isothermal and polytrope models in our study.}
\begin{center}
\begin{tabular}{llccr}
\toprule
\# & Name & {$R_\mathrm{max}$} (kpc) & {$R_\mathrm{C}$} (kpc) & {$\rho_0$} $\big(10^{-3}M_\odot\,\mathrm{pc}^{-3}\big)$\\ \midrule
(1)  & CVnIdwA  & 2.59 &      2.01 $\pm$ 0.52  & 8.19    $\pm$ 1.62  \\
(2)  & DDO 50   & 9.81 &      0.15 $\pm$ 0.08  & 379.64  $\pm$ 381.14\\
(3)  & DDO 52   & 5.43 &      1.33 $\pm$ 0.07  & 48.81   $\pm$ 3.63  \\
(4)  & DDO 53   & 1.45 &      2.22 $\pm$ 1.95  & 25.10   $\pm$ 5.63  \\
(5)  & DDO 87   & 7.39 &      2.46 $\pm$ 0.11  & 13.91   $\pm$ 0.77  \\
(6)  & DDO 101  & 1.95 &      0.32 $\pm$ 0.01  & 849.14  $\pm$ 77.23 \\
(7)  & DDO 126  & 3.99 &      1.33 $\pm$ 0.10  & 21.59   $\pm$ 2.00  \\
(8)  & DDO 133  & 3.48 &      0.83 $\pm$ 0.06  & 73.69   $\pm$ 7.61  \\
(9)  & DDO 154  & 7.32 &      0.95 $\pm$ 0.03  & 53.21   $\pm$ 3.19  \\
(10) & DDO 168  & 3.14 &      2.81 $\pm$ 0.83  & 39.81   $\pm$ 6.37  \\
(11) & DDO 210  & 0.31 &      0.20 $\pm$ 0.05  & 116.43  $\pm$ 23.05 \\
(12) & DDO 216  & 1.12 &      0.15 $\pm$ 0.03  & 127.02  $\pm$ 43.35 \\
(13) & IC 10    & 0.54 &      0.27 $\pm$ 0.12  & 190.4   $\pm$ 76.4  \\
(14) & IC 1613  & 2.71 &      0.20 $\pm$ 0.04  & 19.25   $\pm$ 3.45  \\
(15) & NGC 1569 & 3.05 &      2.71 $\pm$ 0.81  & 15.23   $\pm$ 2.45  \\
(16) & NGC 2366 & 8.08 &      1.21 $\pm$ 0.04  & 43.89   $\pm$ 2.51  \\
(17) & NGC 3738 & 1.75 &      0.45 $\pm$ 0.04  & 2132.36 $\pm$ 277.75\\
(18) & UGC 8508 & 1.86 &      1.95 $\pm$ 0.21  & 45.29   $\pm$ 2.38  \\
(19) & WLM      & 3.04 &      0.74 $\pm$ 0.01  & 57.46   $\pm$ 1.57  \\
(20) & Haro 36  & 3.16 &      8.40 $\pm$ 12.17 & 16.99   $\pm$ 2.78\\%
\bottomrule
\end{tabular}
\end{center}
\end{table*}

To compare the fitting of the velocity curves observed in the LITTLE THINGS by the polytrope models, we used the results obtained for the standard core-like, spherical pseudo-isothermal model with constant-density core, with the radial profile given by the relation~\citep{Oh-etal:2015:AJ:}
\begin{equation}\label{eq:psedoiso}
    \rho_\mathrm{iso}(R) = \frac{\rho_0}{1 + \left(R/R_\mathrm{C}\right)^2}\, ,
\end{equation}
where $\rho_0$ and $R_\mathrm{C}$ are the constant core density and core radius of the halo. The corresponding rotational velocity is given by
\begin{equation}\label{eq:vel_psedoiso}
    v_\mathrm{iso}(R) = \sqrt{4\pi G\rho_0 R_\mathrm{C}^2\left[1 - \frac{R_\mathrm{C}}{R} \tan^{-1}\left(\frac{R}{R_\mathrm{C}}\right)\right]}\, .
\end{equation}
The results of the fitting procedure are also used to compare the polytrope and ISO models. The results of the fitting procedure of the ISO model on all the members of the LITTLE THINGS are presented in Table~\ref{tab:LittleThings_sum}.
\section{Matching the velocity curves observed in dwarf galaxies from LITTLE THINGS}
To match the observational data reflecting measured velocity curves in the dwarf galaxies belonging to the LITTLE THINGS, we use the standard $\chi$-squares method, that is we are looking for the lowest value of the test representing differences between the observed data and velocity curves predicted by the polytrope models.

The so-called reduced $\chi^2$ statistics is defined as follows:
\begin{equation}
  \chi^2_k = \sum_{i=1}^{N} \left(\frac{v_\mathrm{obs}(r_i)-v_\mathrm{pol}(r_i)}{e_i}\right)^2\Big/k\, ,
\end{equation}
where $k$ is the degree of the freedom, N is the number of the data, $v_\mathrm{obs}(r_{i})$ ($v_\mathrm{pol}(r_{i})$) is the observed (polytropy-predicted) velocity at the radius $r_{i}$, and $e_{i}$ is the error of the measurement of the velocity $v_\mathrm{obs}(r_{i})$. The velocity $v_\mathrm{pol}(r_{i})$ is governed by the radial profile of the circular geodesic motion in the polytrope model given by equation~(\ref{eq:pol_radial_velocity}) and the $v_\mathrm{iso}(R)$ by equations~(\ref{eq:psedoiso},\ref{eq:vel_psedoiso}).

\subsection{Fitting by polytropes}
In the matching procedure we are considering the values of the polytropic index $0 < n < 4$, and whole the range of the relativistic parameter $\sigma$ restricted by the causality limit; however, only very small values of $\sigma$ were found to be relevant, corresponding to the non-relativistic limit for the matter content. For each considered dwarf galaxy, we present a map of the $\xi$-parameter in dependence on the parameters $n$--$\sigma$, the values of the parameters $n$, $\sigma$, and the value of $\chi$ of the best match in the figure, representing the corresponding theoretical velocity profile giving the best match to the observational data related to the velocity curve of the considered dwarf galaxy.

\begin{table*}
\caption{\label{tab:vel_dw_result} Results of the polytropic spheres matching the observation data of LITTLE THINGS. The second and third column correspond to the best-fitting polytropic index $n$ and relativistic parameter $\sigma$. The computed values of the $\chi^2$ for our model and the pseudo-isothermal are given in the fourth and fifth columns. The values of mass and maximal velocities are also presented.}
        \begin{tabular*}{\textwidth}{@{\extracolsep{\fill}}lcccccccccc@{}}
\toprule
\multirow{2}{*}{Name} & \multirow{2}{*}{$n$} & {$\sigma$} & {$\rho_\mathrm{c}$}  & \multirow{2}{*}{$\chi_\mathrm{pol}$} & \multirow{2}{*}{$\chi_\mathrm{iso}$} & {$M$} & {$M_\mathrm{iso}$} & {$v^\mathrm{max}_\mathrm{pol}$} & {$v^\mathrm{max}_\mathrm{iso}$} & {$v^\mathrm{max}_\mathrm{obs}$} \\[-0.5ex]
 & & {$\left(10^{-9}\right)$} & {$\left(10^{-3}M_\odot\,\mathrm{pc}^{-3}\right)$} & & & {$\left(10^8\,M_\odot\right)$} & {$\left(10^8\,M_\odot\right)$} & {$\left(\mathrm{km\,s}^{-1}\right)$} & {$\left(\mathrm{km\,s}^{-1}\right)$} & {$\left(\mathrm{km\,s}^{-1}\right)$} \\
 \midrule
\multicolumn{11}{l}{\emph{The first class of models}} \\
CVnIdwA  & 0.1   &   3.36  &      4.949 & 2.73 &   3.79  &   3.68  &   3.16 &   24.70  &    24.66  &   24.91  \\
Haro 36  & 0.3   &  23.30  &      17.730 & 0.27 &   0.27  &  44.39  &  20.72 &   65.73  &    53.24  &   55.87  \\
IC 1613  & 0.5   &   0.94  &      2.423 & 2.45 &   4.28  &   1.10  &   0.23 &   13.26  &     8.60  &   16.01  \\
DDO 53   & 0.7   &   2.98  &      29.091 & 1.07 &   1.86  &   2.00  &   2.57 &   23.74  &    27.71  &   27.31  \\
UGC 8508 & 0.7   &  12.20  &      42.728 & 0.26 &   0.21  &  13.63  &   8.11 &   48.03  &    43.45  &   45.56  \\
DDO 168  & 0.9   &  18.10  &      43.319 & 0.99 &   5.43  &  26.81  &  30.71 &   58.69  &    65.02  &   58.03  \\
DDO 210  & 0.95  &   0.57  &      93.449 & 0.82 &   0.64  &   0.10  &   0.06 &   10.39  &     9.68  &   10.64  \\
\multicolumn{11}{l}{\emph{The second class of models \hrulefill}} \\
NGC 1569 & 1.25 &   1.56  &      7.180 & 0.18 &   1.66  &   1.93  &  10.71 &   17.36  &    39.00  &   25.00  \\
DDO 126  & 1.3  &   5.83  &      15.306 & 0.20 &   0.46  &   9.70  &  11.18 &   33.57  &    34.84  &   34.13  \\
IC 10    & 1.7  &   1.21  &     16.599 & 0.16 &   0.34  &   0.26  &   0.42 &   15.40  &    18.49  &   18.60  \\
DDO 87   & 1.95 &  12.50  &      10.460 & 0.62 &   0.27  &  45.49  &  45.66 &   49.63  &    51.71  &   53.54  \\
\multicolumn{11}{l}{\emph{The third class of models \hrulefill}} \\
DDO 52   & 2.15 &  15.40  &   29.697 & 0.27 &   0.16  &  39.01  &  39.71 &   55.21  &  56.46  &   57.47  \\
NGC 3738 & 2.25 &  74.10  & 1539.748 & 2.70 &   1.69  &  58.75  &  62.75 &  121.26  & 126.09  &  129.69  \\
WLM      & 2.3  &   5.57  &   40.630 & 0.21 &   0.07  &   7.56  &   8.12 &   33.27  &  34.97  &   36.27  \\
DDO 133  & 2.3  &   8.64  &   48.091 & 1.37 &   0.97  &  13.42  &  15.12 &   41.44  &  43.50  &   42.87  \\
DDO 101  & 2.65 &  19.80  &  376.840 & 5.38 &  12.69  &  18.24  &  16.38 &   63.06  &  62.37  &   63.83  \\
DDO 216  & 2.7  &   0.80  &   69.366 & 0.65 &   0.31  &   0.35  &   0.32 &   12.67  &  13.05  &   17.07  \\
NGC 2366 & 2.75 &  13.10  &   23.240 & 0.44 &   0.37  &  40.32  &  51.35 &   51.27  &  52.56  &   52.44  \\
DDO 154  & 2.95 &  10.20  &   28.574 & 0.50 &   0.18  &  26.04  &  35.91 &   45.24  &  46.22  &   47.55  \\
DDO 50   & 3.95 &   3.43  &   59.010 & 1.66 &   3.86  &   4.43  &  10.28 &   26.50  &  21.90  &   30.97  \\
\bottomrule
\end{tabular*}
\end{table*}

\begin{figure*}[hp]
\begin{minipage}[t]{0.49\linewidth}
\begin{minipage}[t]{\linewidth}\raggedright
   \begin{minipage}[t]{0.35\linewidth}\scriptsize
      {\large (1) CVnIdwA}
   \end{minipage}
   \begin{minipage}[t]{0.60\linewidth}\raggedleft\scriptsize
      Right Ascension: \ascension{12}{38}{39}{2}\\
          Declination: \declination{+32}{45}{41}{0}\\
          Distance: 3.6~Mpc\\
          Absolute magnitude: $-$12.4~mag\\[-5mm]\mbox{}
   \end{minipage}
\end{minipage}
        \begin{center}
        \begin{minipage}{\linewidth}
                        \includegraphics[width=\linewidth]{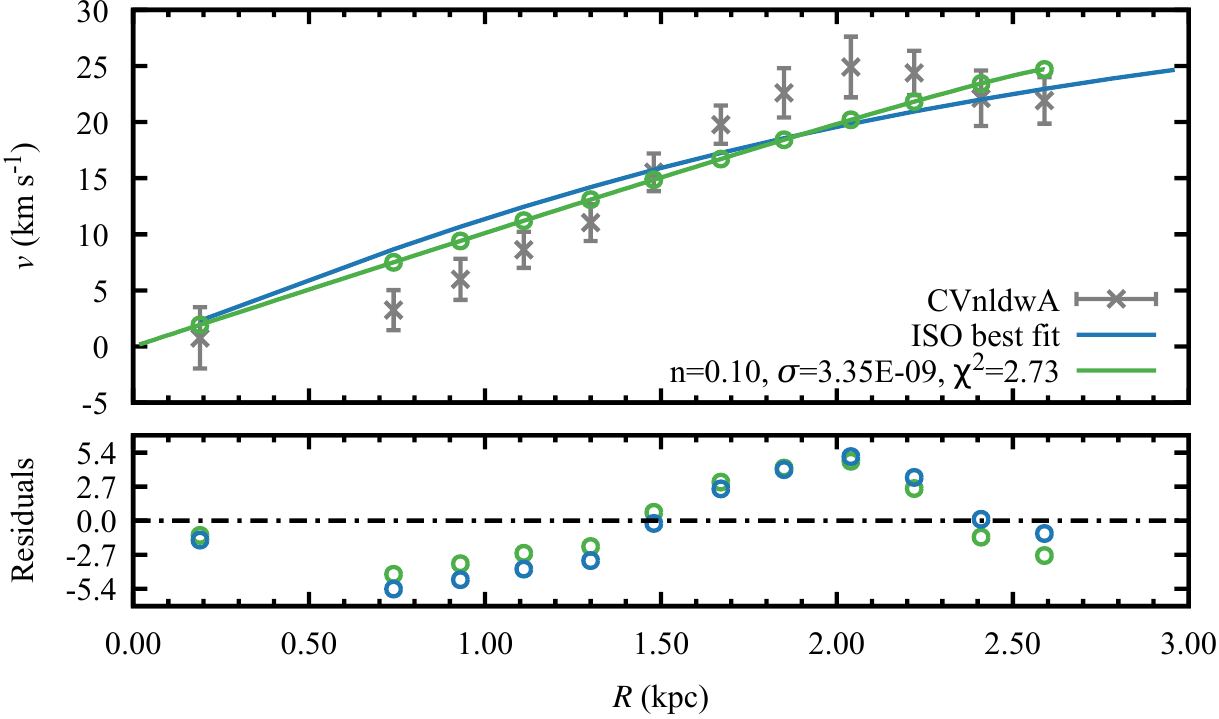}\\
                        \includegraphics[width=.49\linewidth]{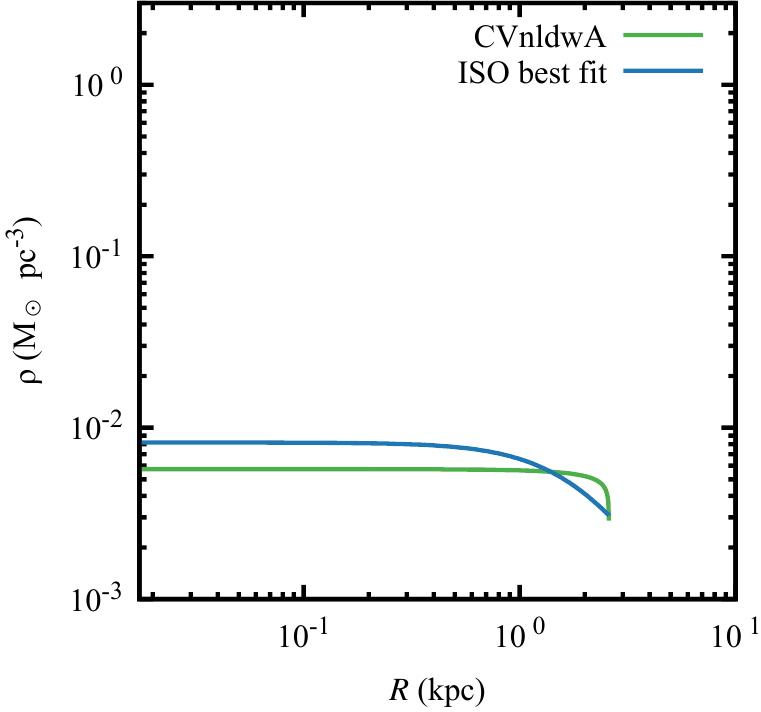}
                        \includegraphics[width=.49\linewidth]{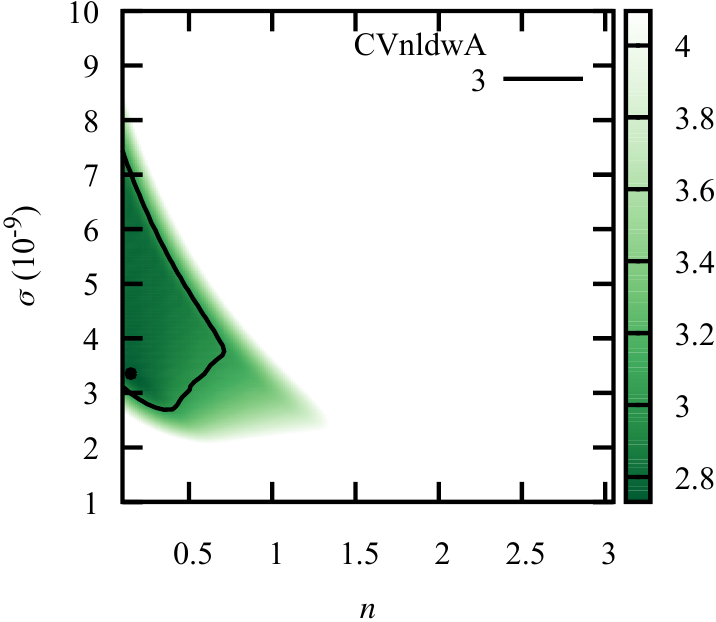}\\[-6.2mm]\mbox{}
        \end{minipage}
        \end{center}    
\end{minipage}
\begin{minipage}[t]{.49\linewidth}
\begin{minipage}[t]{\linewidth}\raggedright
   \begin{minipage}[t]{0.35\linewidth}
    {\large (2) DDO 50}\\
   \end{minipage}
        \begin{minipage}[t]{0.60\linewidth}\raggedleft\scriptsize
    Right Ascension: \ascension{08}{19}{03}{7}\\
        Declination: \declination{+70}{43}{24}{6}\\
        Distance: 3.4~Mpc\\
        Absolute magnitude: $-$16.6~mag\\[-5mm]\mbox{}
        \end{minipage}
\end{minipage}
        \begin{center}
        \begin{minipage}{\linewidth}
                        \includegraphics[width=\linewidth]{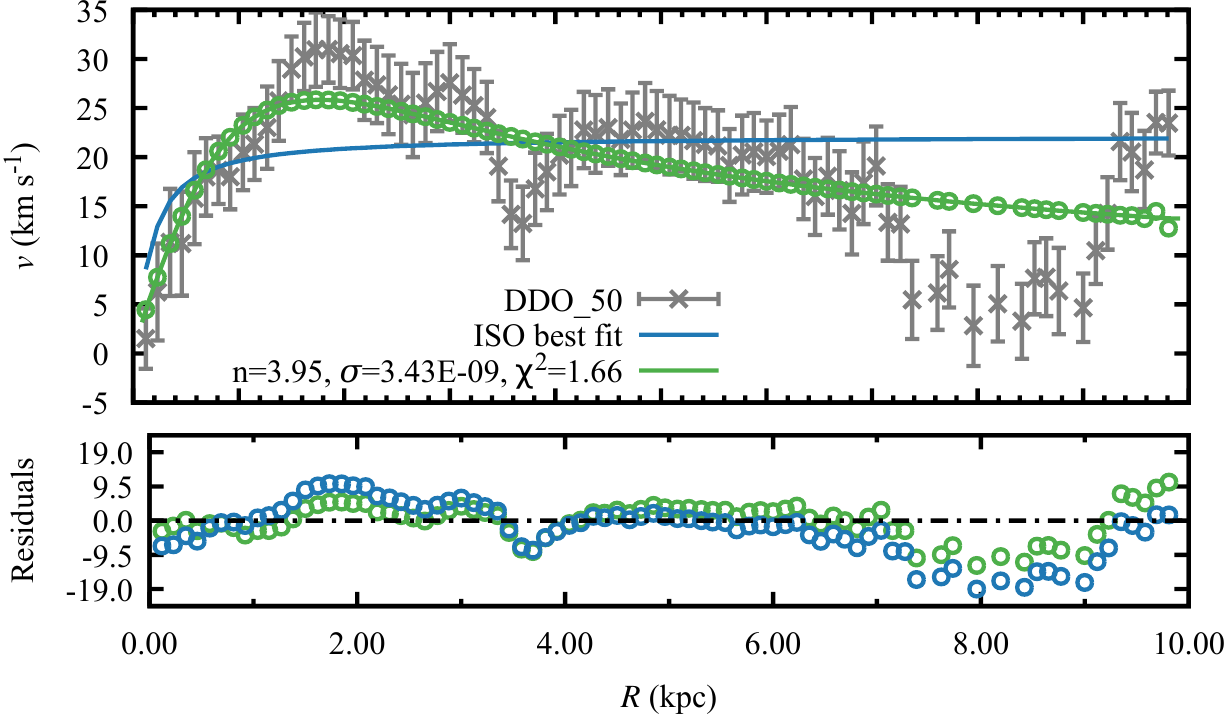}\\
                        \includegraphics[width=.49\linewidth]{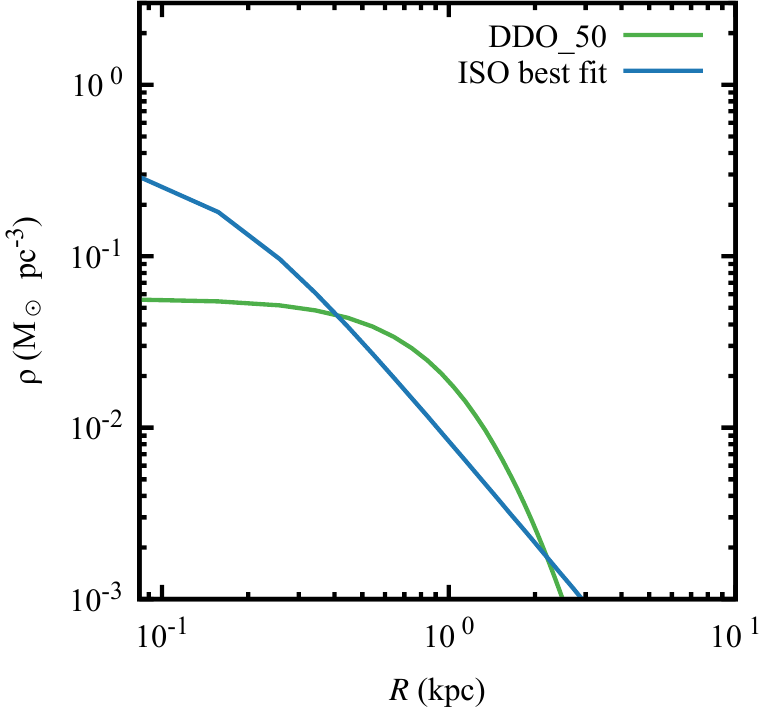}
                        \includegraphics[width=.49\linewidth]{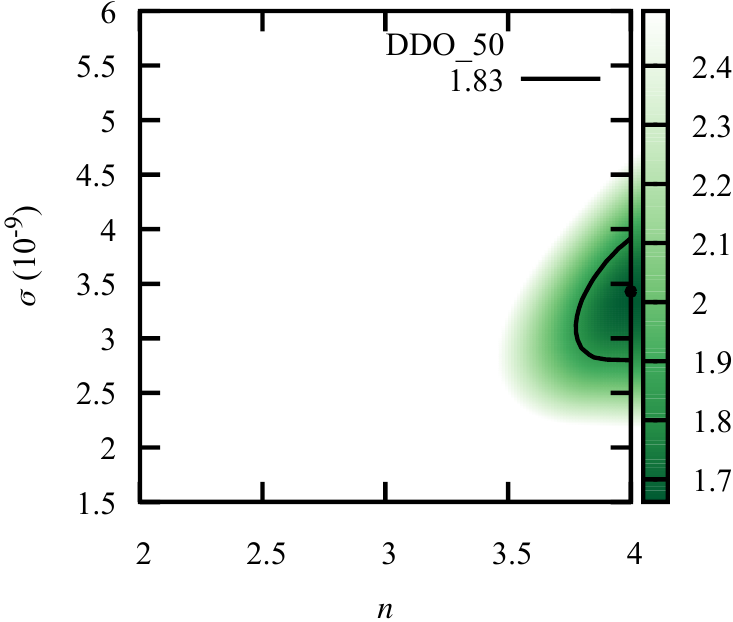}\\[-6.2mm]\mbox{}
        \end{minipage}
        \end{center}    
\end{minipage}
\\[3ex]
\begin{minipage}[t]{.49\linewidth}
\begin{minipage}[t]{\linewidth}\raggedright
   \begin{minipage}[t]{0.35\linewidth}
        {\large (3) DDO 52}
   \end{minipage}
        \begin{minipage}[t]{0.60\linewidth}\raggedleft\scriptsize
        Right Ascension: \ascension{08}{28}{28}{4}\\
        Declination: \declination{+41}{51}{26}{5}\\
        Distance: 10.3~Mpc\\
        Absolute magnitude: $-$15.4~mag\\[-5mm]\mbox{}
        \end{minipage}
\end{minipage}
        \begin{center}          
        \begin{minipage}{\linewidth}
                        \includegraphics[width=\linewidth]{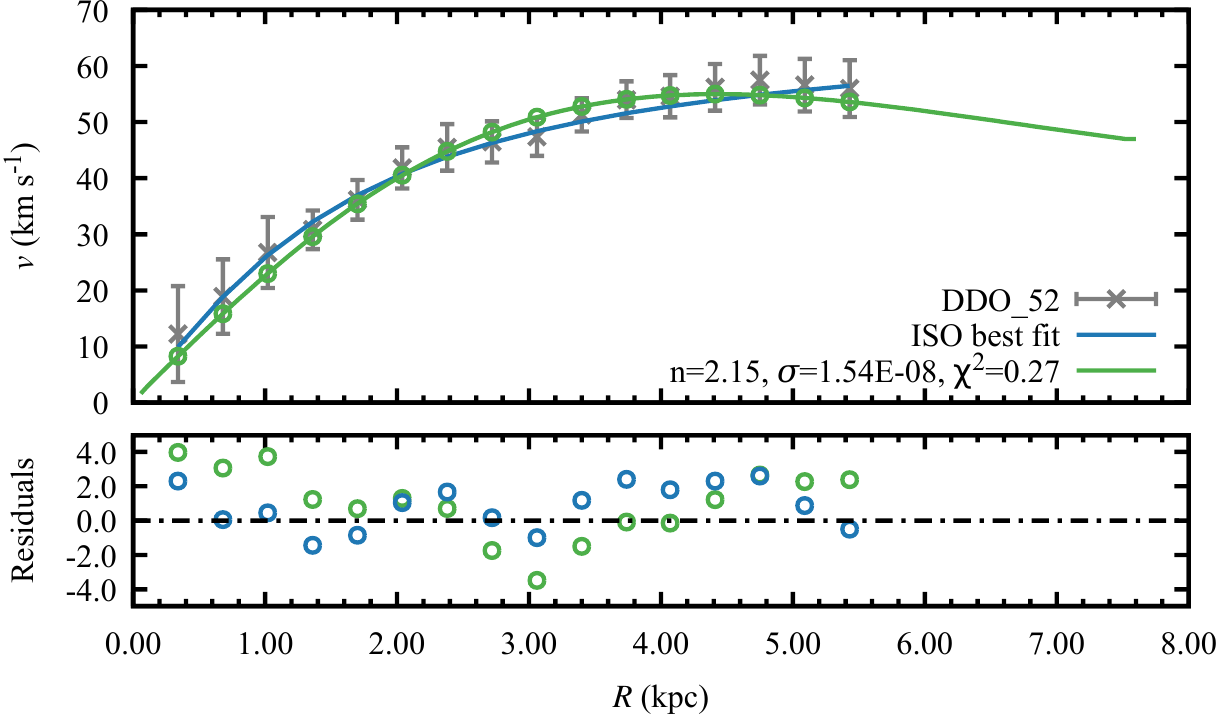}\\
                        \includegraphics[width=.49\linewidth]{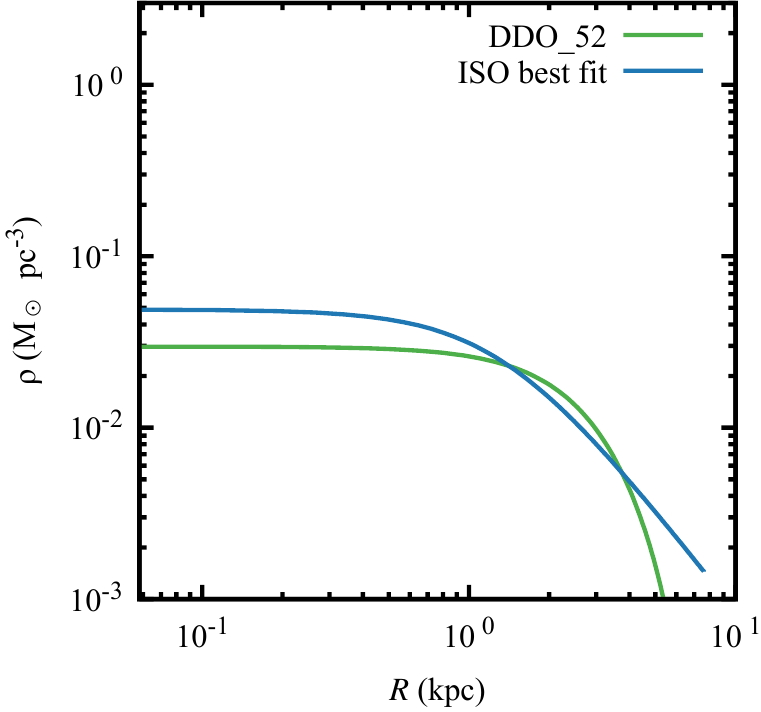}
                        \includegraphics[width=.49\linewidth]{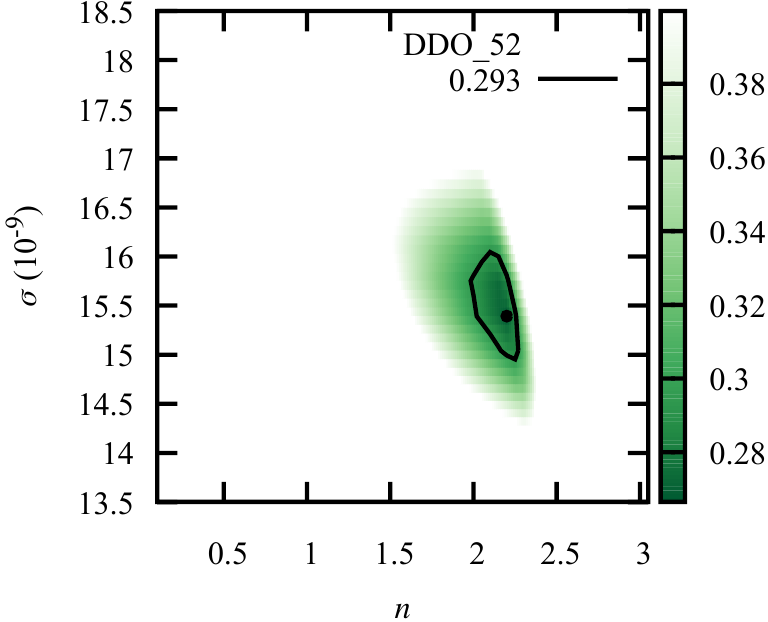}\\[-6.2mm]\mbox{}
        \end{minipage}
        \end{center}
\end{minipage}
\begin{minipage}[t]{.49\linewidth}
\begin{minipage}[t]{\linewidth}\raggedright
   \begin{minipage}[t]{0.35\linewidth}
    {\large (4) DDO 53}
   \end{minipage}
        \begin{minipage}[t]{0.60\linewidth}\raggedleft\scriptsize
    Right Ascension: \ascension{08}{34}{06}{4}\\
        Declination: \declination{+66}{10}{47}{9}\\
        Distance: 3.6~Mpc\\
        Absolute magnitude: $-$13.8~mag\\[-5mm]\mbox{}
        \end{minipage}
\end{minipage}
        \begin{center}
        \begin{minipage}{\linewidth}
                        \includegraphics[width=\linewidth]{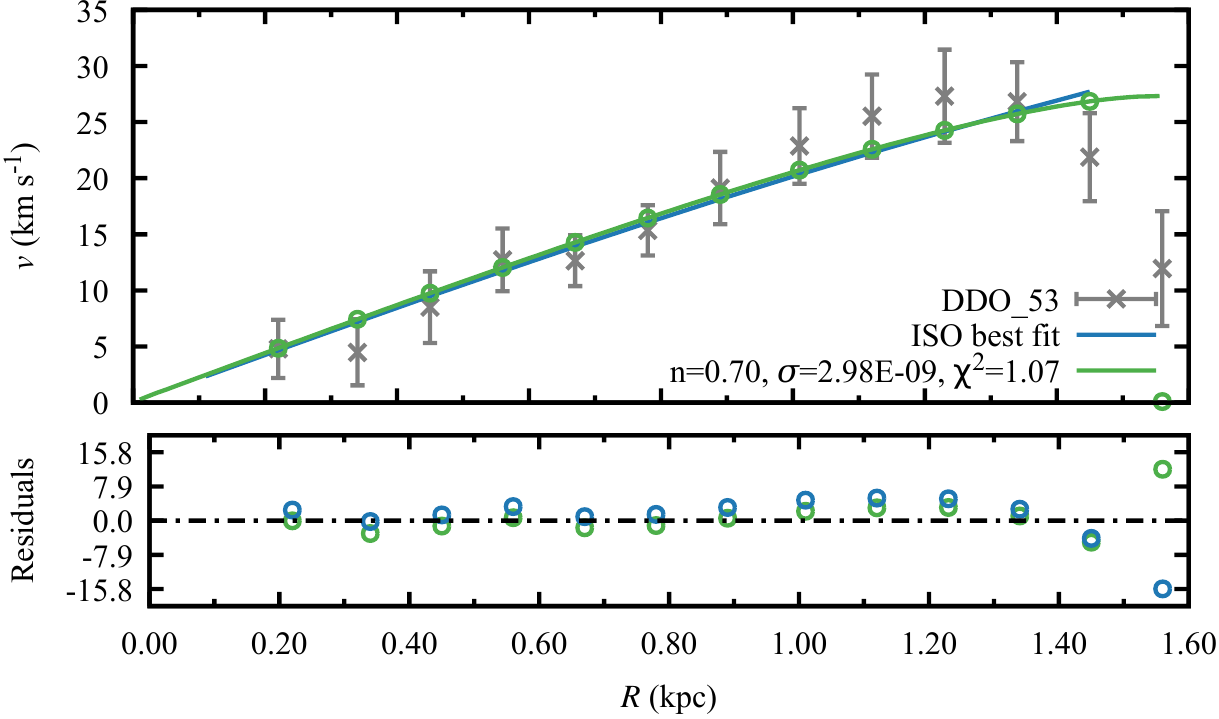}\\
                        \includegraphics[width=.49\linewidth]{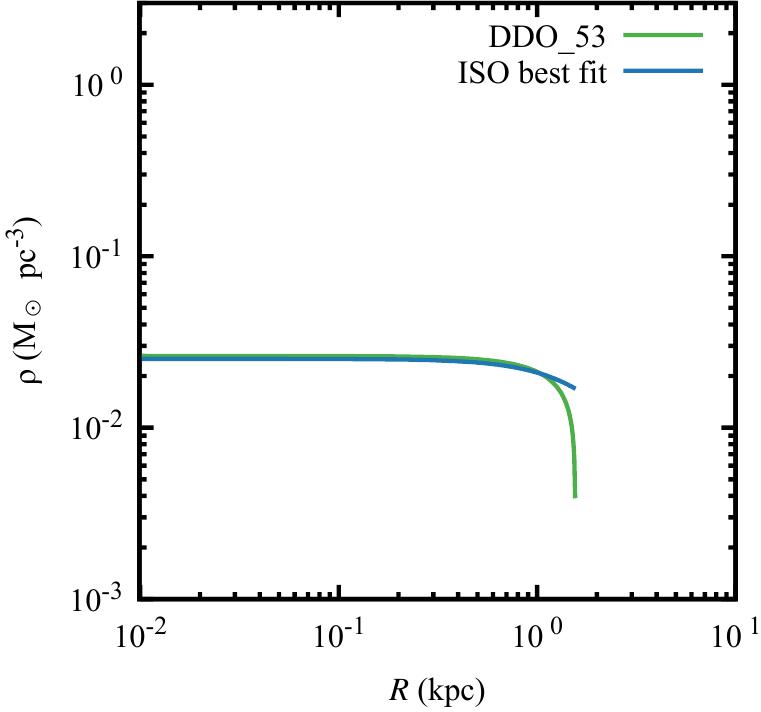}
                        \includegraphics[width=.49\linewidth]{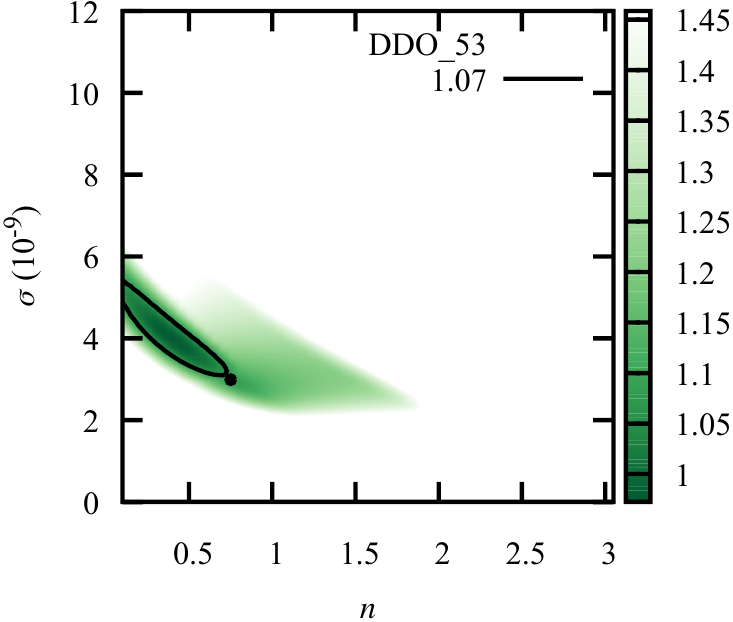}
        \end{minipage}
        \end{center}    
\end{minipage}
\caption{\label{fig2}Velocity curve, mass density profile, and contour map of the best free parameters $\sigma$ and $n$; for CVnIdwA (top left), DDO\,50 (top right), DDO\,52 (bottom left) and DDO\,53 (bottom right).}
\end{figure*}

\begin{figure*}[htb!]
\begin{minipage}[t]{0.49\linewidth}
\begin{minipage}[t]{\linewidth}\raggedright
   \begin{minipage}[t]{0.35\linewidth}
    {\large (5) DDO 87}
   \end{minipage}
        \begin{minipage}[t]{0.60\linewidth}\raggedleft\scriptsize
    Right Ascension: \ascension{10}{49}{34}{9}\\
        Declination: \declination{+65}{31}{47}{9}\\
        Distance: 7.7~Mpc\\
        Absolute magnitude: $-$15.0~mag\\[-5mm]\mbox{}
        \end{minipage}
\end{minipage}
        \begin{center}
        \begin{minipage}{\linewidth}
                        \includegraphics[width=\linewidth]{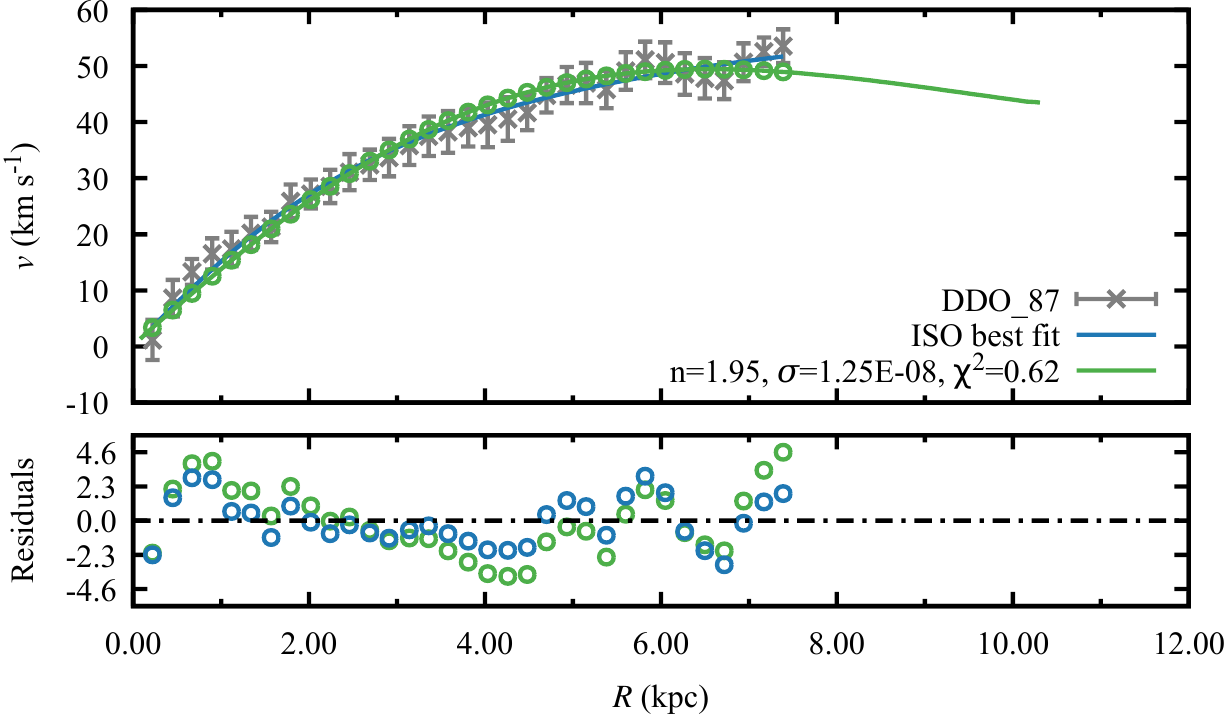}\\
                        \includegraphics[width=.49\linewidth]{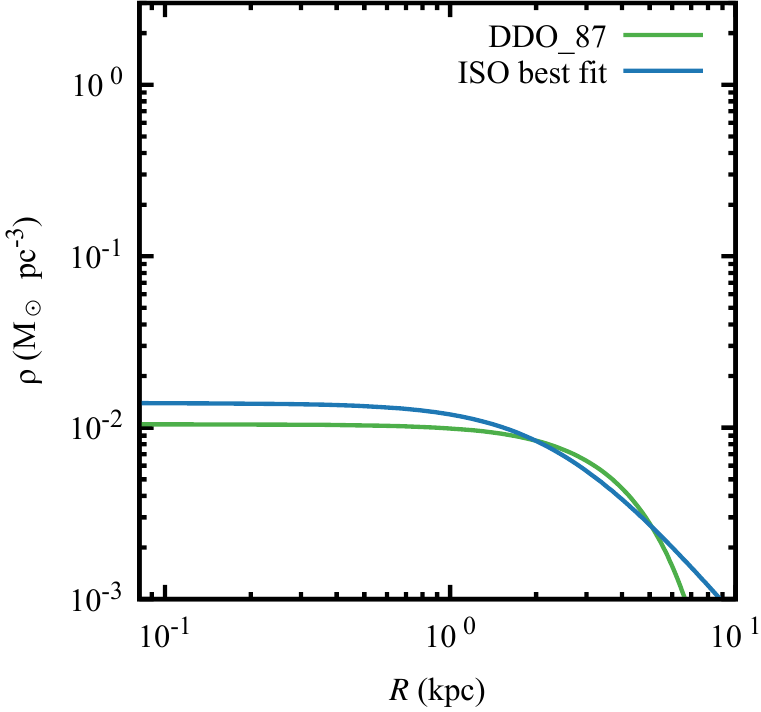}
                        \includegraphics[width=.49\linewidth]{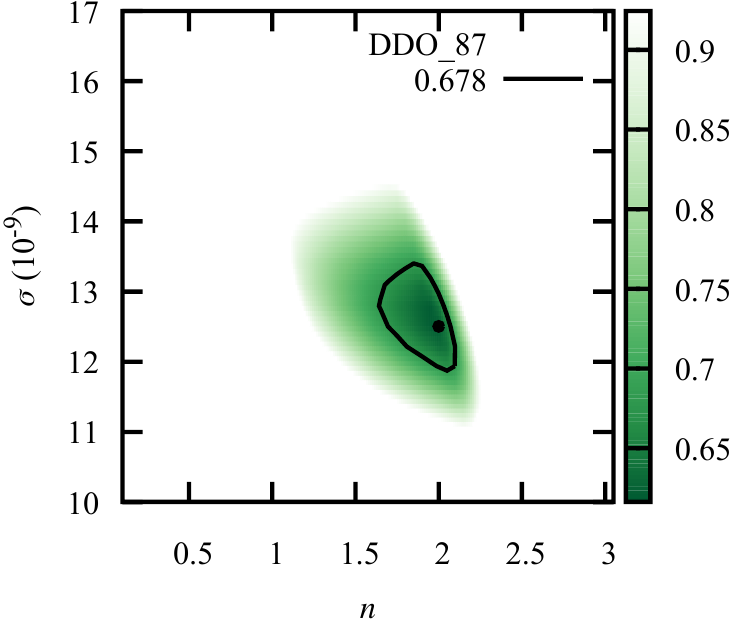}\\[-6.2mm]\mbox{}
        \end{minipage}
        \end{center}    
\end{minipage}
\begin{minipage}[t]{.49\linewidth}
\begin{minipage}[t]{\linewidth}\raggedright
   \begin{minipage}[t]{0.35\linewidth}
    {\large (6) DDO 101}
   \end{minipage}
        \begin{minipage}[t]{0.60\linewidth}\raggedleft\scriptsize
    Right Ascension: \ascension{11}{55}{39}{1}\\
        Declination: \declination{+31}{31}{09}{9}\\
        Distance: 6.4~Mpc\\
        Absolute magnitude: $-$15.0~mag\\[-5mm]\mbox{}
        \end{minipage}
\end{minipage}
        \begin{center}  
        \begin{minipage}{\linewidth}
                        \includegraphics[width=\linewidth]{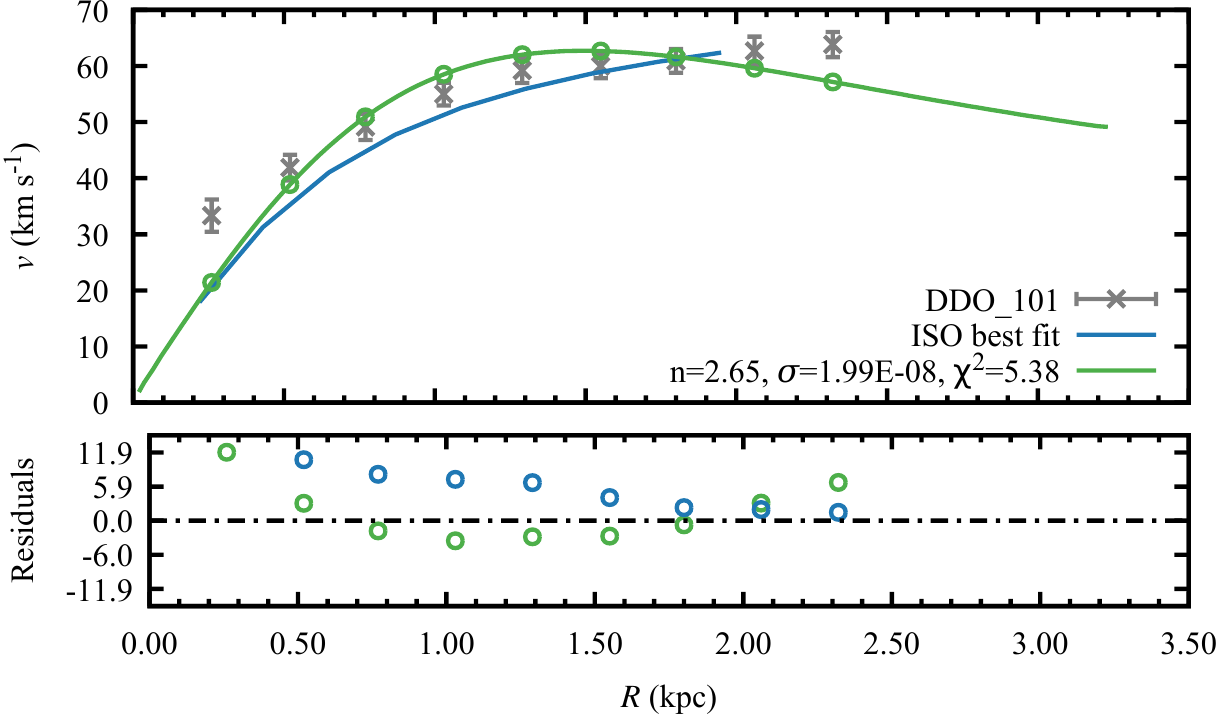}\\
                        \includegraphics[width=.49\linewidth]{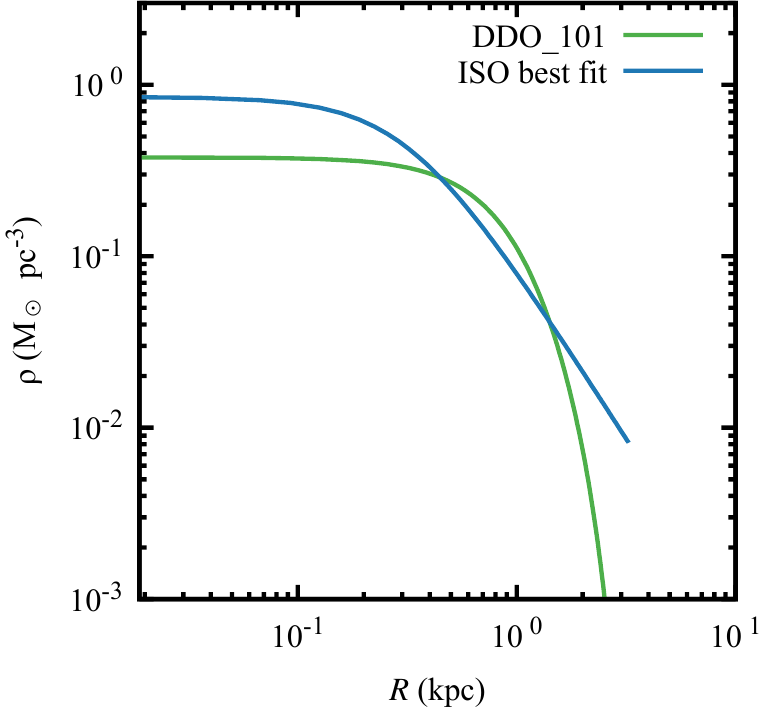}
                        \includegraphics[width=.49\linewidth]{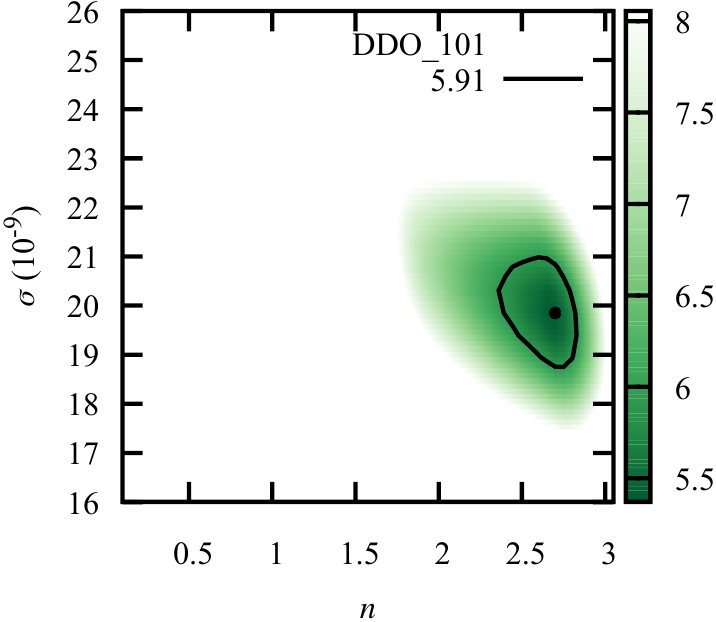}\\[-6.2mm]\mbox{}
        \end{minipage}
        \end{center}
\end{minipage}
\\[3ex]
\begin{minipage}[t]{.49\linewidth}
\begin{minipage}[t]{\linewidth}\raggedright
   \begin{minipage}[t]{0.35\linewidth}
{\large (7) DDO 126}
   \end{minipage}
        \begin{minipage}[t]{0.60\linewidth}\raggedleft\scriptsize
    Right Ascension: \ascension{12}{27}{06}{6}\\
        Declination: \declination{+37}{08}{15}{9}\\
        Distance: 4.9~Mpc\\
        Absolute magnitude: $-$14.9~mag\\[-5mm]\mbox{}
        \end{minipage}
\end{minipage}
        \begin{center}
        \begin{minipage}{\linewidth}
                        \includegraphics[width=\linewidth]{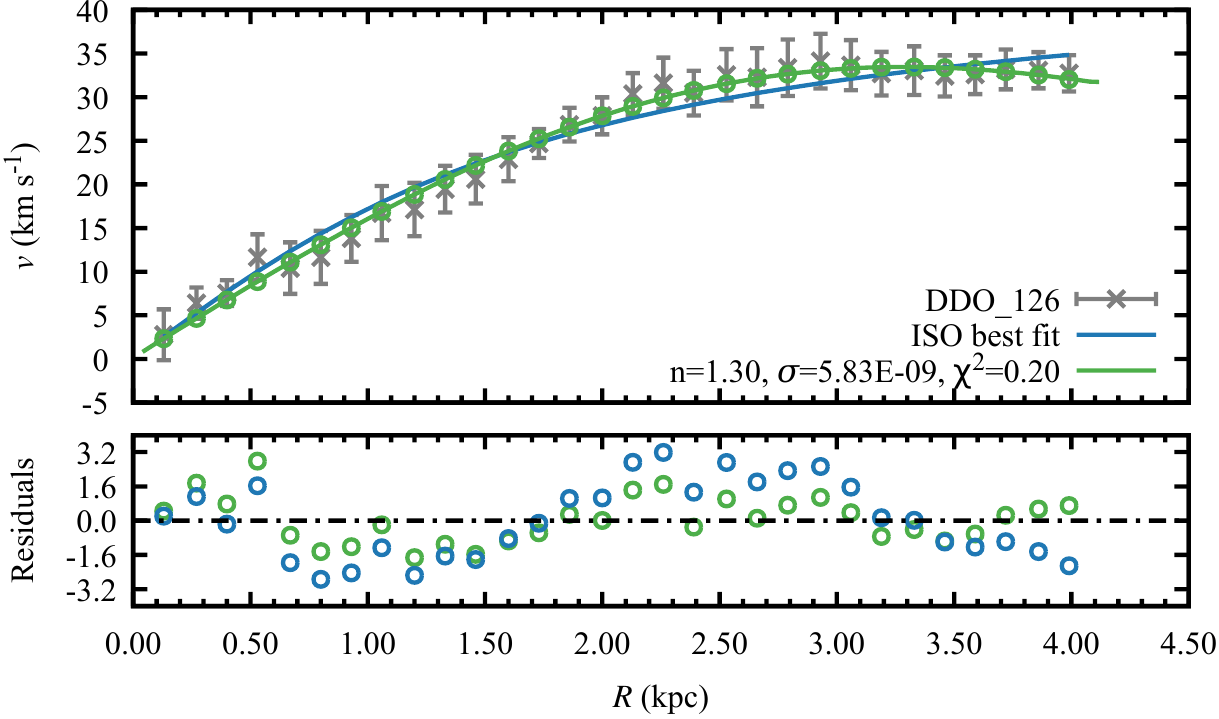}\\
                        \includegraphics[width=.49\linewidth]{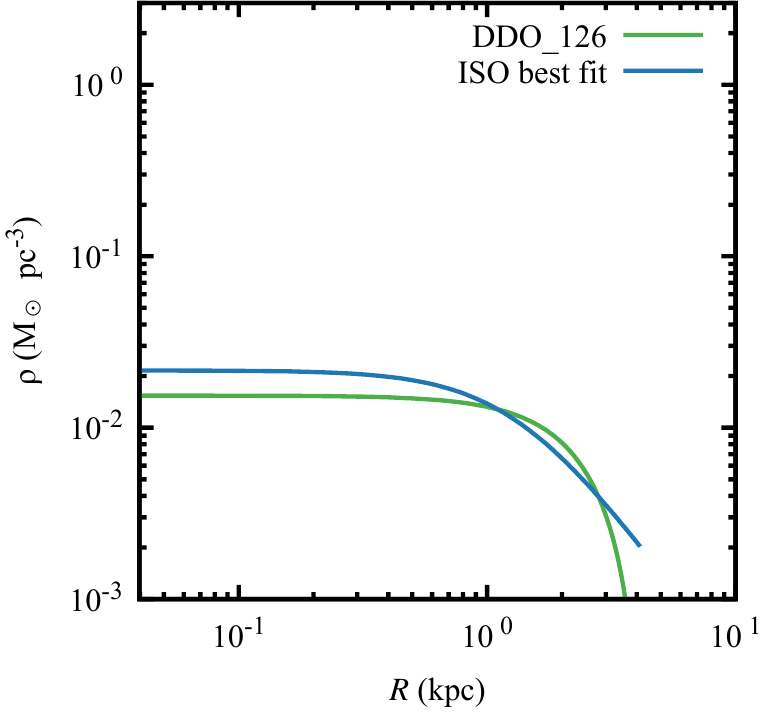}
                        \includegraphics[width=.49\linewidth]{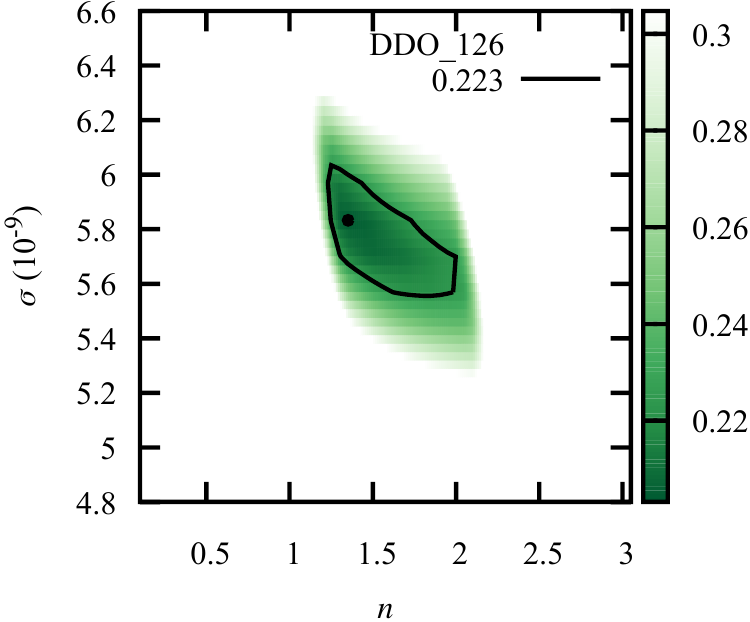}\\[-6.2mm]\mbox{}
        \end{minipage}
        \end{center}    
\end{minipage}
\begin{minipage}[t]{.49\linewidth}
\begin{minipage}[t]{\linewidth}\raggedright
   \begin{minipage}[t]{0.35\linewidth}
    {\large (8) DDO 133}
   \end{minipage}
        \begin{minipage}[t]{0.60\linewidth}\raggedleft\scriptsize
    Right Ascension: \ascension{12}{32}{55}{2}\\
        Declination: \declination{+31}{32}{19}{1}\\
        Distance: 3.5~Mpc\\
        Absolute magnitude: $-$14.8~mag\\[-5mm]\mbox{}
        \end{minipage}
\end{minipage}
        \begin{center}          
        \begin{minipage}{\linewidth}
                        \includegraphics[width=\linewidth]{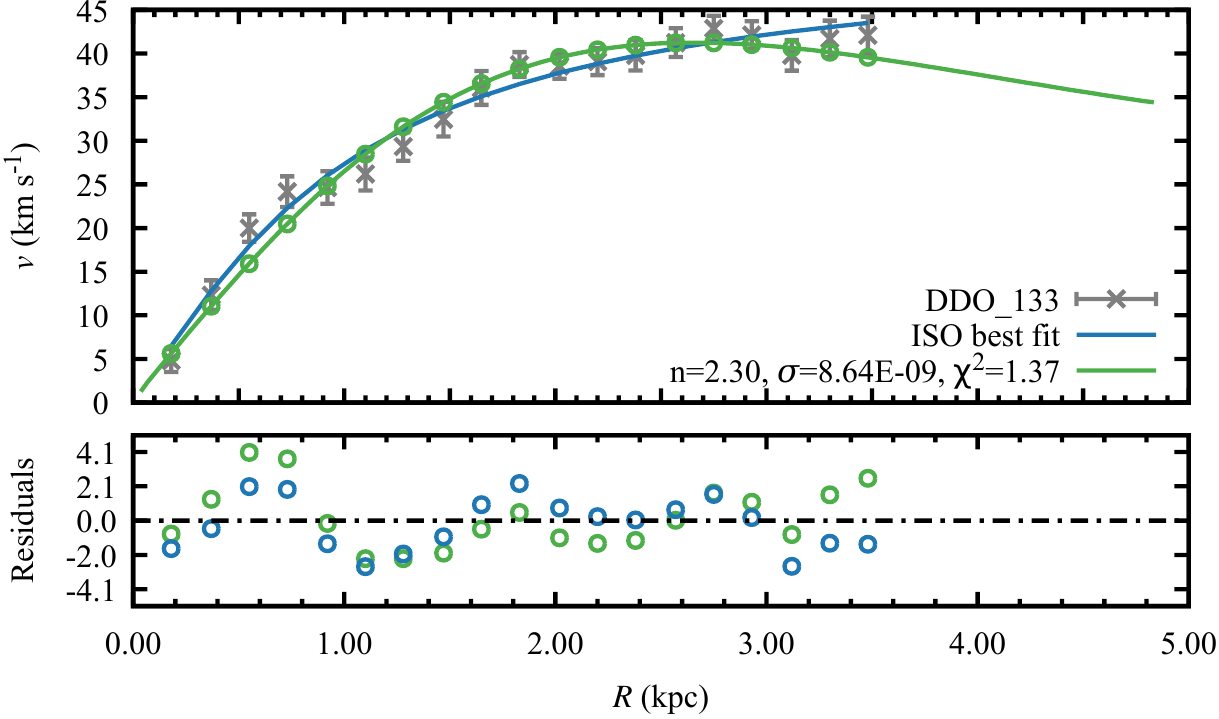}\\
                        \includegraphics[width=.49\linewidth]{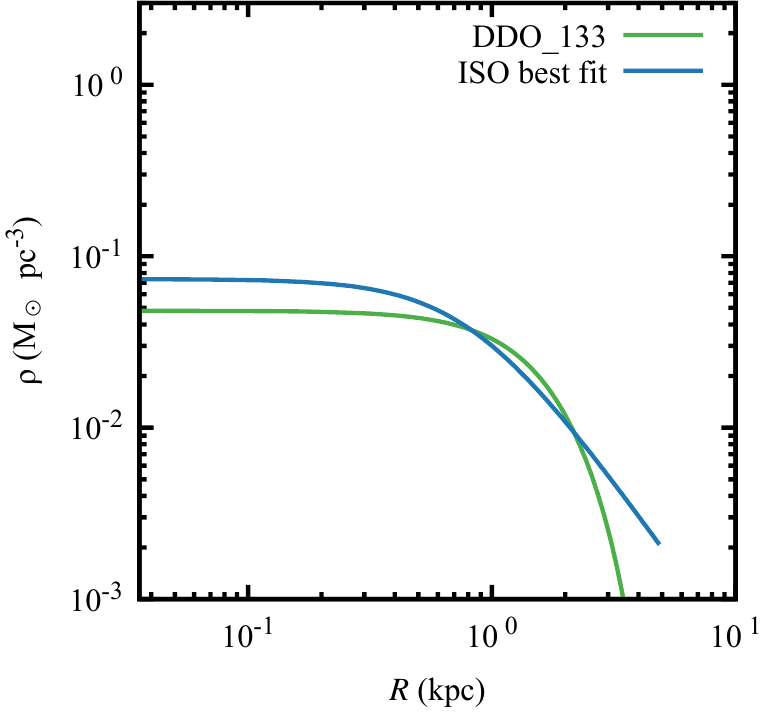}
                        \includegraphics[width=.49\linewidth]{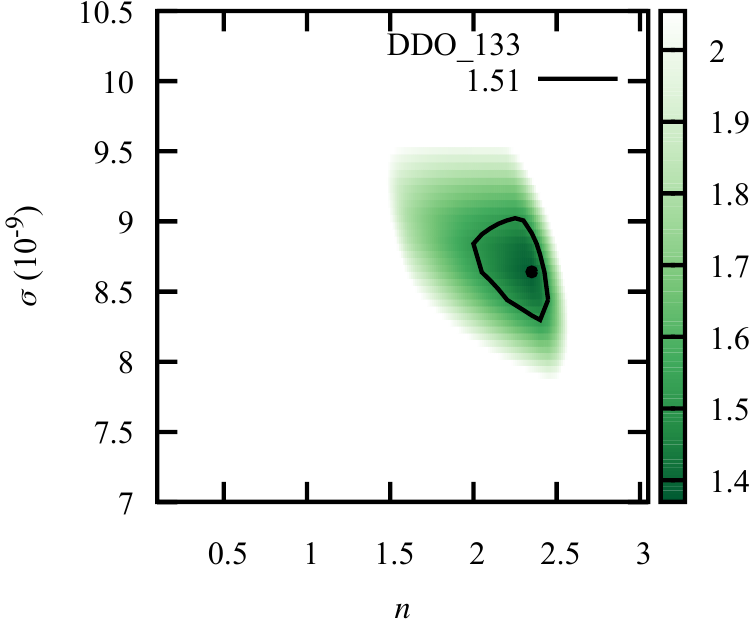}
        \end{minipage}
        \end{center}
\end{minipage}
\caption{\label{fig3}Velocity curve, mass density profile, and contour map of the best free parameters $\sigma$ and $n$; for DDO\,87 (top left), DDO\,101 (top right), DDO\,126 (bottom left) and DDO\,133 (bottom right).}
\end{figure*}

\begin{figure*}[p]
\begin{minipage}[t]{0.49\linewidth}
\begin{minipage}[t]{\linewidth}\raggedright
   \begin{minipage}[t]{0.35\linewidth}
    {\large (9) DDO 154}
   \end{minipage}
        \begin{minipage}[t]{0.60\linewidth}\raggedleft\scriptsize
    Right Ascension: \ascension{12}{54}{05}{7}\\
        Declination: \declination{+27}{09}{09}{9}\\
        Distance: 3.7~Mpc\\
        Absolute magnitude: $-$14.2~mag\\[-5mm]\mbox{}
        \end{minipage}
\end{minipage}
        \begin{center}
        \begin{minipage}{\linewidth}
                        \includegraphics[width=\linewidth]{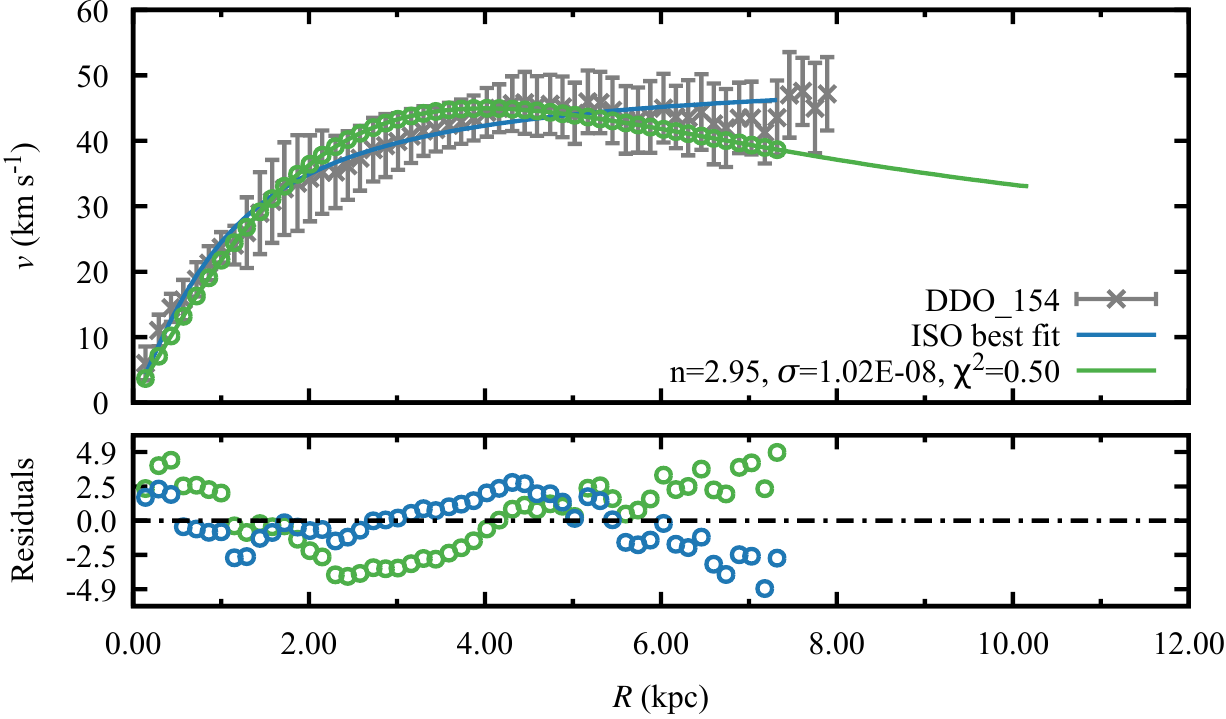}\\
                        \includegraphics[width=.49\linewidth]{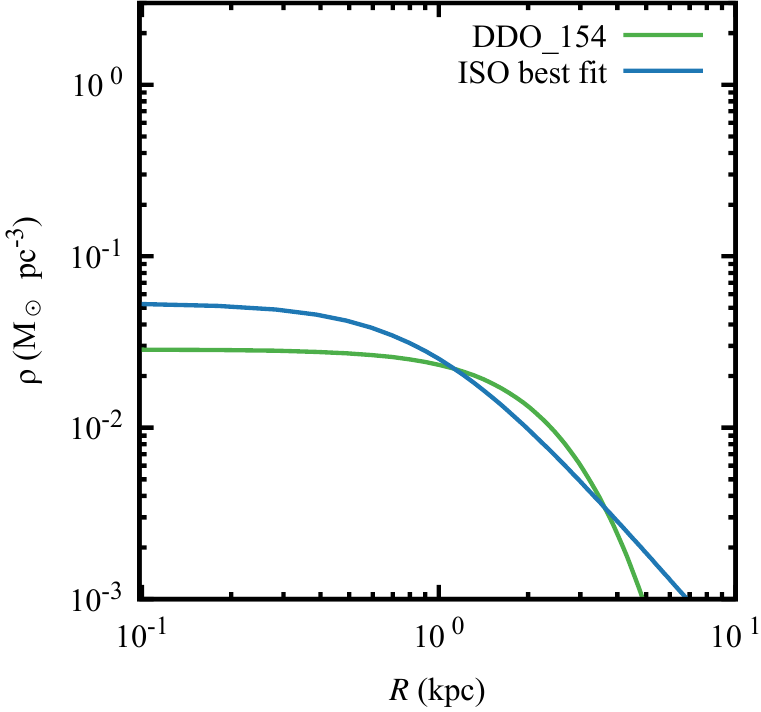}
                        \includegraphics[width=.49\linewidth]{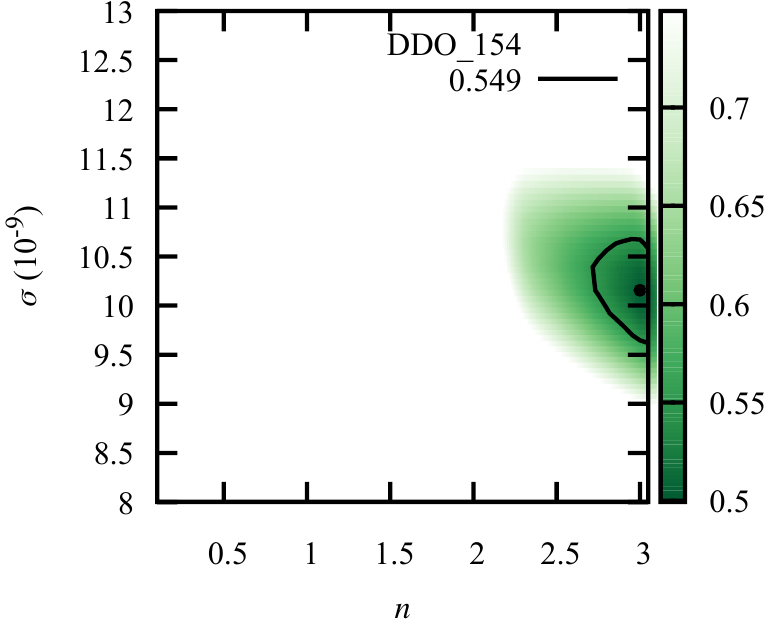}\\[-6.2mm]\mbox{}
        \end{minipage}
        \end{center}    
\end{minipage}
\begin{minipage}[t]{.49\linewidth}
\begin{minipage}[t]{\linewidth}\raggedright
   \begin{minipage}[t]{0.35\linewidth}
    {\large (10) DDO 168}
   \end{minipage}
        \begin{minipage}[t]{0.60\linewidth}\raggedleft\scriptsize
    Right Ascension: \ascension{13}{14}{27}{3}\\
        Declination: \declination{+45}{55}{37}{3}\\
        Distance: 4.3~Mpc\\
        Absolute magnitude: $-$15.7~mag\\[-5mm]\mbox{}
        \end{minipage}
\end{minipage}
        \begin{center}                  
        \begin{minipage}{\linewidth}
                        \includegraphics[width=\linewidth]{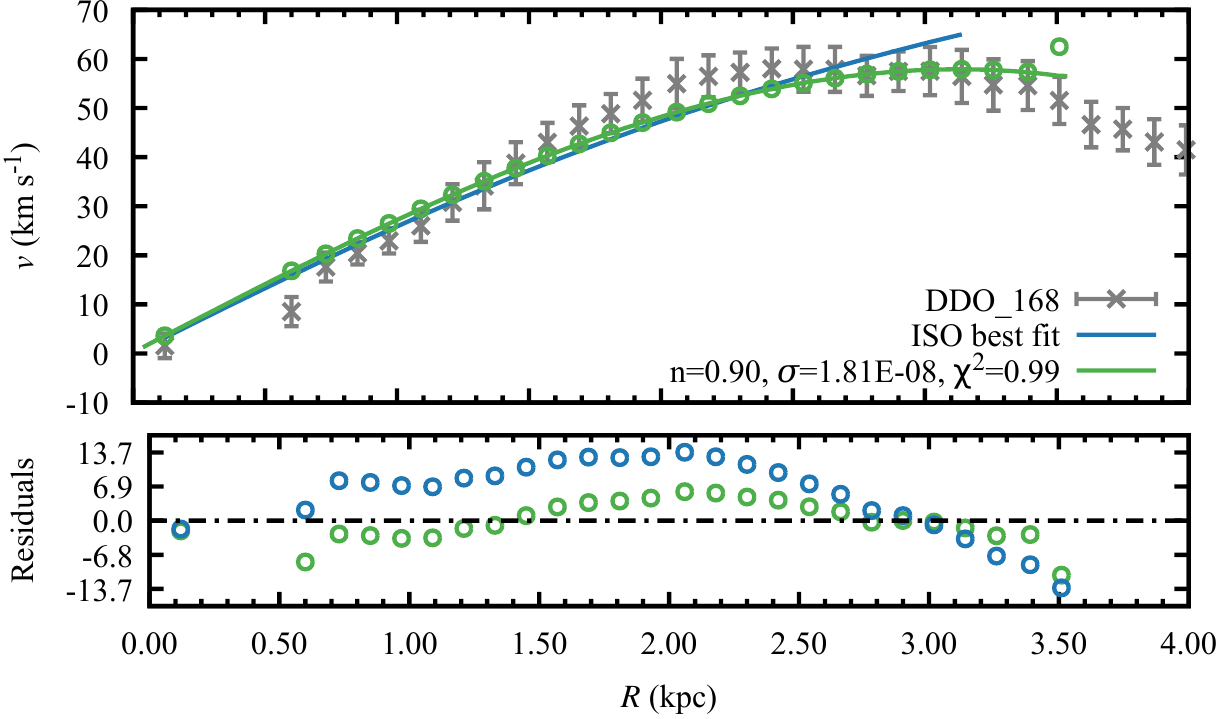}\\
                        \includegraphics[width=.49\linewidth]{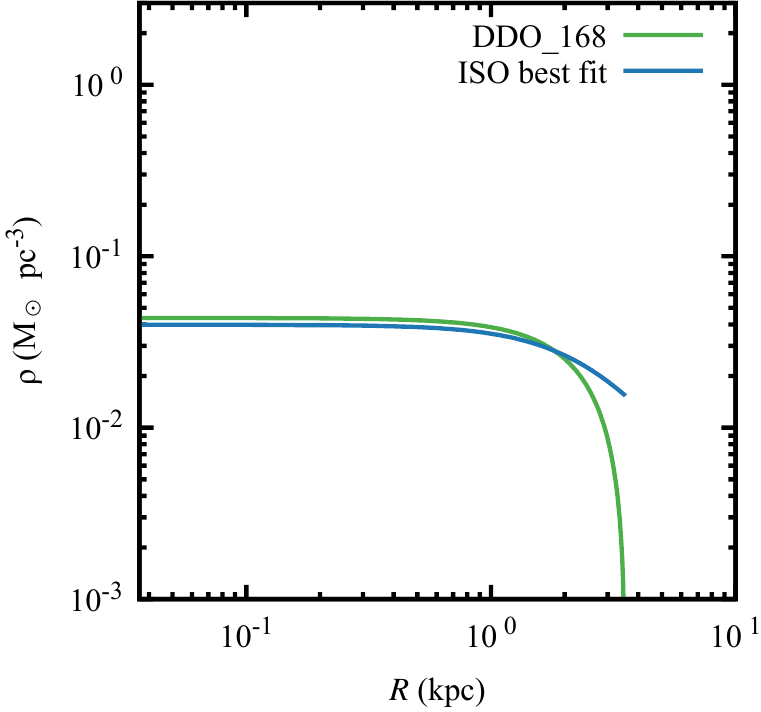}
                        \includegraphics[width=.49\linewidth]{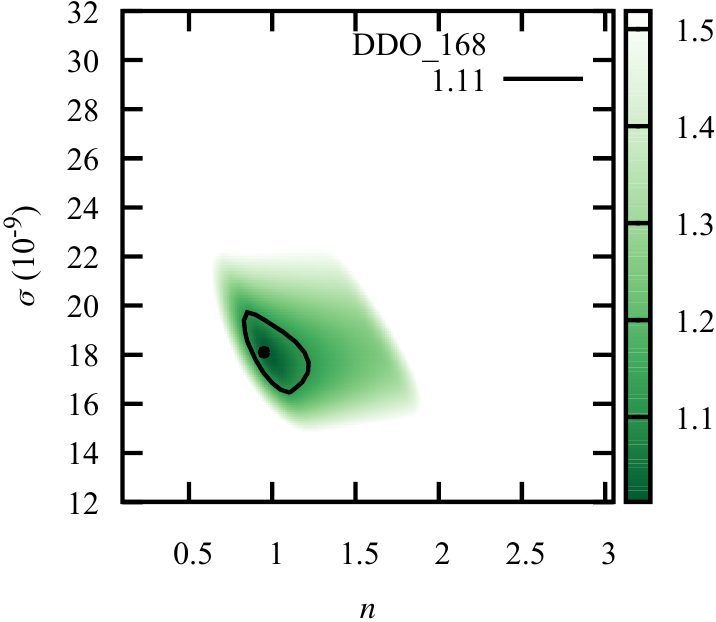}\\[-6.2mm]\mbox{}
        \end{minipage}
        \end{center}
\end{minipage}
\\[3ex]
\begin{minipage}[t]{.49\linewidth}
\begin{minipage}[t]{\linewidth}\raggedright
   \begin{minipage}[t]{0.35\linewidth}
    {\large (11) DDO 210}
   \end{minipage}
        \begin{minipage}[t]{0.60\linewidth}\raggedleft\scriptsize
    Right Ascension: \ascension{20}{46}{51}{6}\\
        Declination: \declination{$-$12}{50}{57}{7}\\
        Distance: 0.9~Mpc\\
        Absolute magnitude: $-$10.9~mag\\[-5mm]\mbox{}
        \end{minipage}
\end{minipage}
        \begin{center}
        \begin{minipage}{\linewidth}
                        \includegraphics[width=\linewidth]{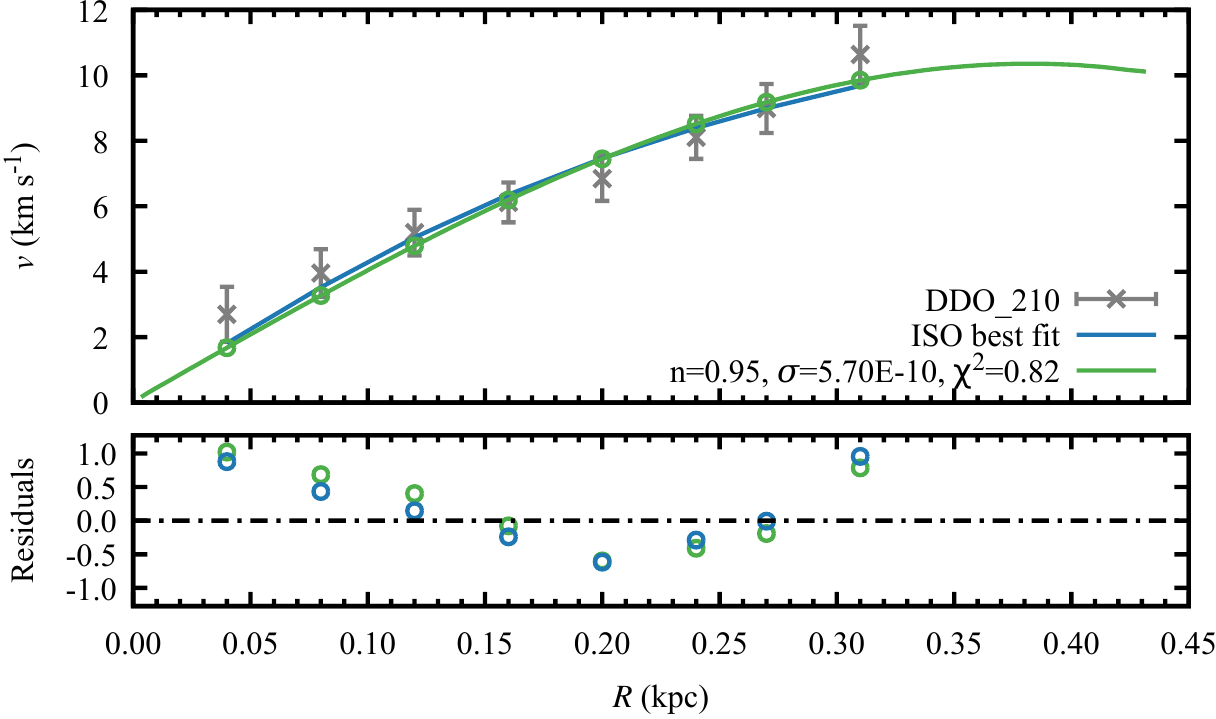}\\
                        \includegraphics[width=.49\linewidth]{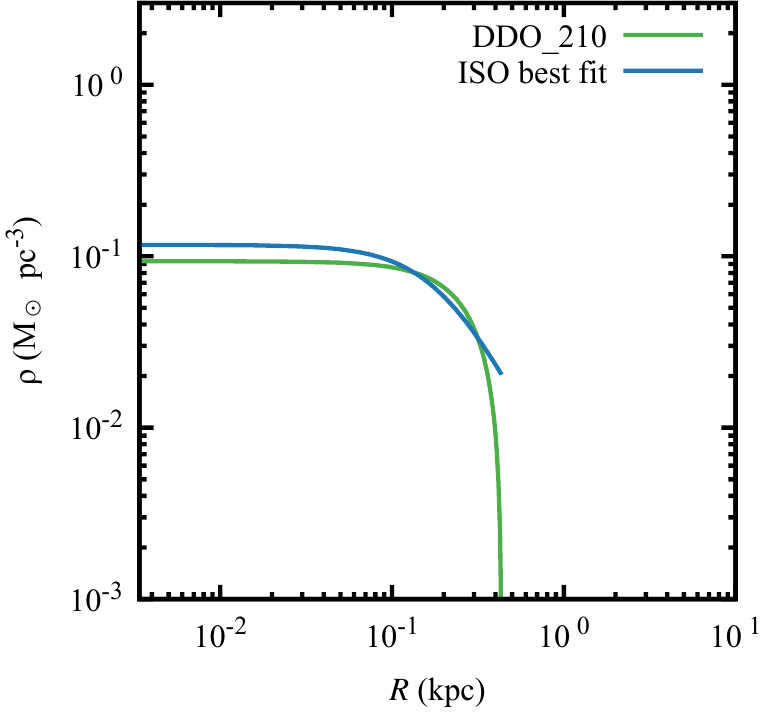}
                        \includegraphics[width=.49\linewidth]{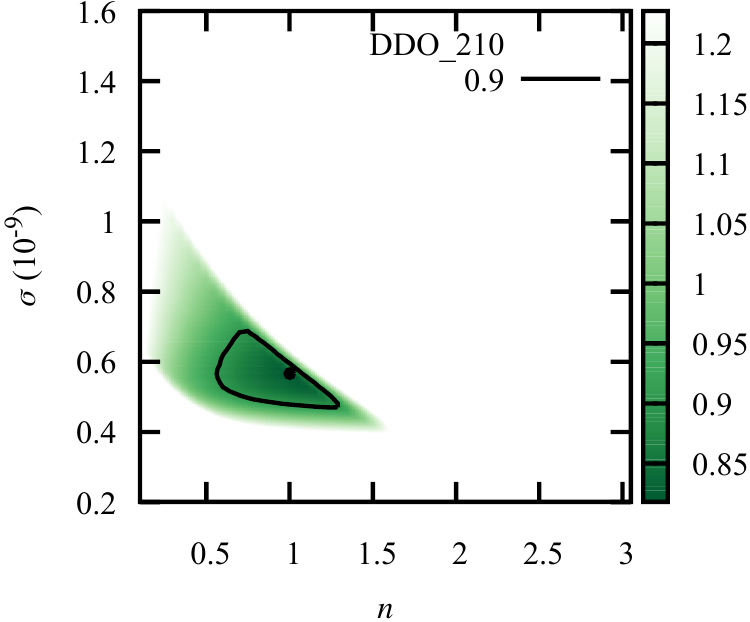}\\[-6.2mm]\mbox{}
        \end{minipage}
        \end{center}    
\end{minipage}
\begin{minipage}[t]{.49\linewidth}
\begin{minipage}[t]{\linewidth}\raggedright
   \begin{minipage}[t]{0.35\linewidth}
    {\large (12) DDO 216}
   \end{minipage}
        \begin{minipage}[t]{0.60\linewidth}\raggedleft\scriptsize
    Right Ascension: \ascension{23}{28}{34}{7}\\
        Declination: \declination{+14}{44}{56}{2}\\
        Distance: 1.1~Mpc\\
        Absolute magnitude: $-$13.7~mag\\[-5mm]\mbox{}
        \end{minipage}
\end{minipage}
        \begin{center}          
        \begin{minipage}{\linewidth}
                        \includegraphics[width=\linewidth]{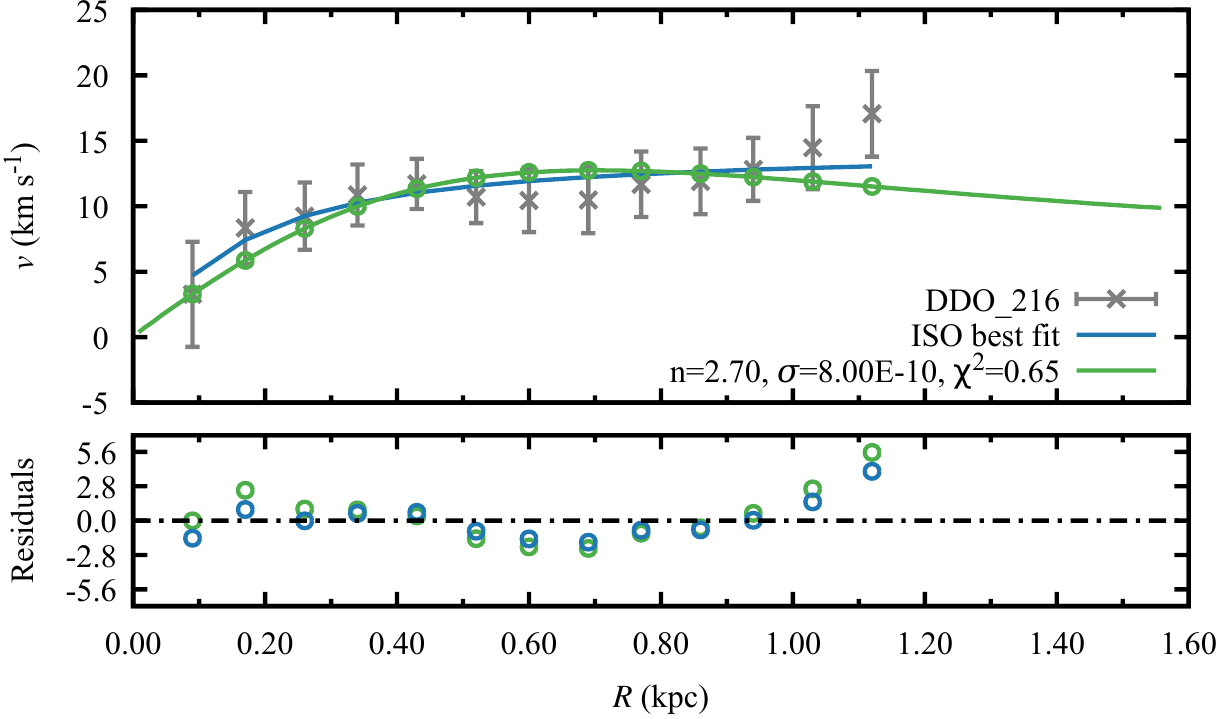}\\
                        \includegraphics[width=.49\linewidth]{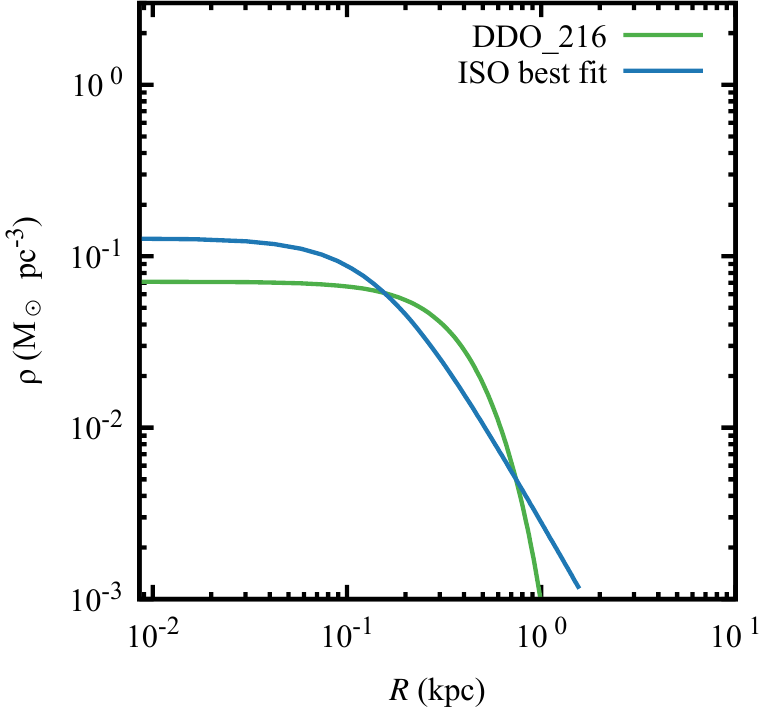}
                        \includegraphics[width=.49\linewidth]{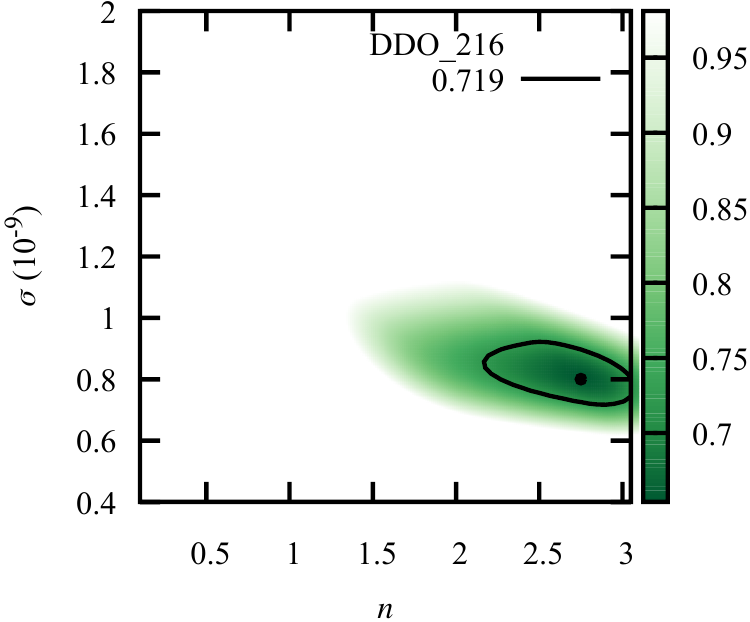}
        \end{minipage}
        \end{center}
\end{minipage}
\caption{\label{fig4}Velocity curve, mass density profile, and contour map of the best free parameters $\sigma$ and $n$; for DDO\,154 (top left), DDO\,168 (top right), DDO\,210 (bottom left) and DDO\,216 (bottom right).}
\end{figure*}

\begin{figure*}[p]
\begin{minipage}[t]{0.49\linewidth}
\begin{minipage}[t]{\linewidth}\raggedright
   \begin{minipage}[t]{0.35\linewidth}
    {\large (13) IC 10}
   \end{minipage}
        \begin{minipage}[t]{0.60\linewidth}\raggedleft\scriptsize
    Right Ascension: \ascension{00}{20}{18}{9}\\
        Declination: \declination{+59}{17}{49}{9}\\
        Distance: 0.7~Mpc\\
        Absolute magnitude: $-$16.3~mag\\[-5mm]\mbox{}
        \end{minipage}  
\end{minipage}
        \begin{center}
        \begin{minipage}{\linewidth}
                        \includegraphics[width=\linewidth]{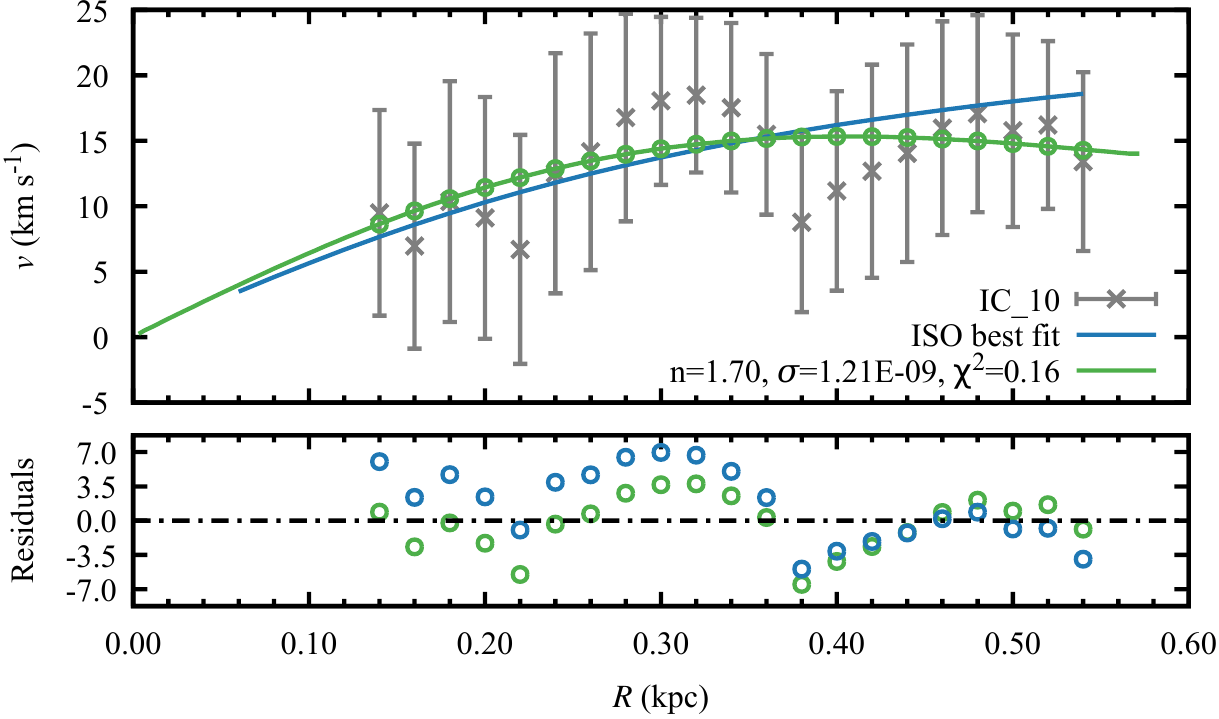}\\
                        \includegraphics[width=.49\linewidth]{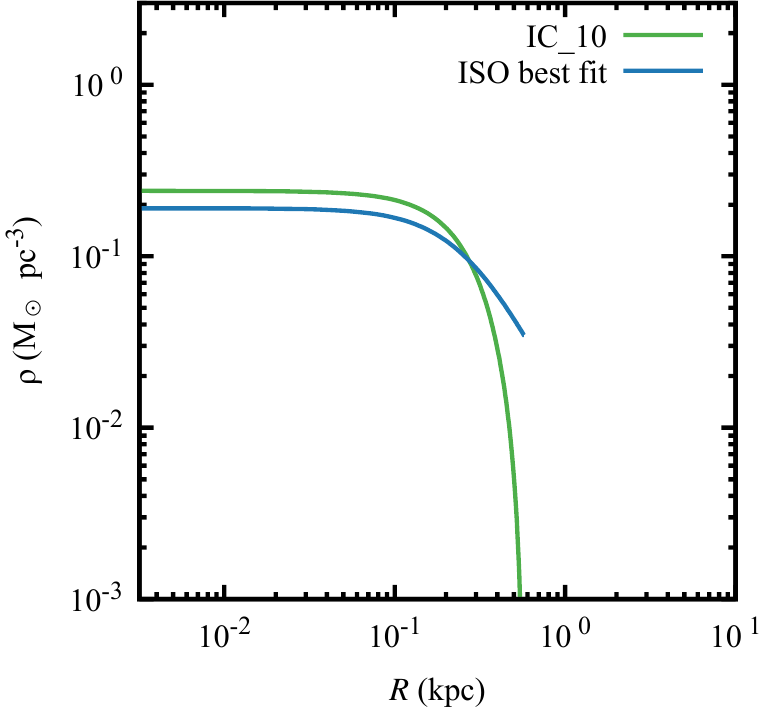}
                        \includegraphics[width=.49\linewidth]{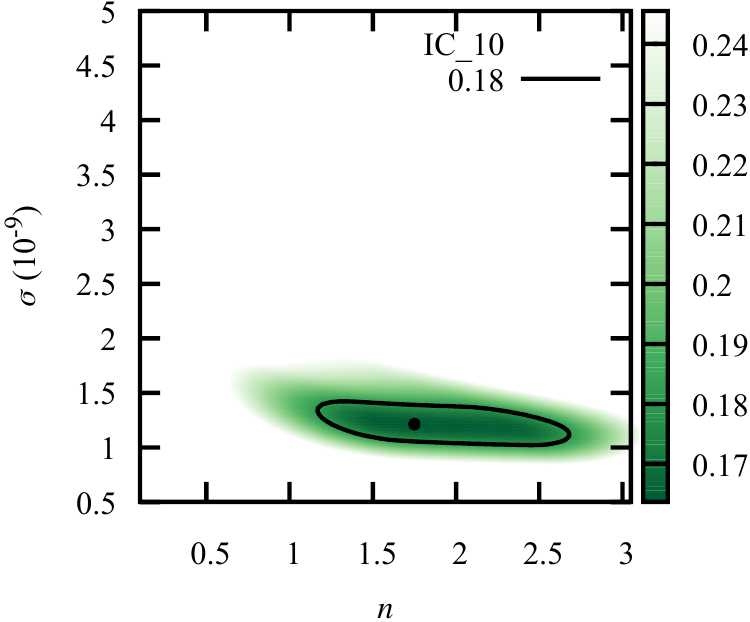}\\[-6.2mm]\mbox{}
        \end{minipage}
        \end{center}
\end{minipage}
\begin{minipage}[t]{.49\linewidth}
\begin{minipage}[t]{\linewidth}\raggedright
   \begin{minipage}[t]{0.35\linewidth}
    {\large (14) IC 1613}
   \end{minipage}
        \begin{minipage}[t]{0.60\linewidth}\raggedleft\scriptsize
    Right Ascension: \ascension{01}{04}{49}{6}\\
        Declination: \declination{+02}{08}{14}{1}\\
        Distance: 0.7~Mpc\\
        Absolute magnitude: $-$14.6~mag\\[-5mm]\mbox{}
        \end{minipage}  
\end{minipage}
        \begin{center}  
        \begin{minipage}{\linewidth}
                        \includegraphics[width=\linewidth]{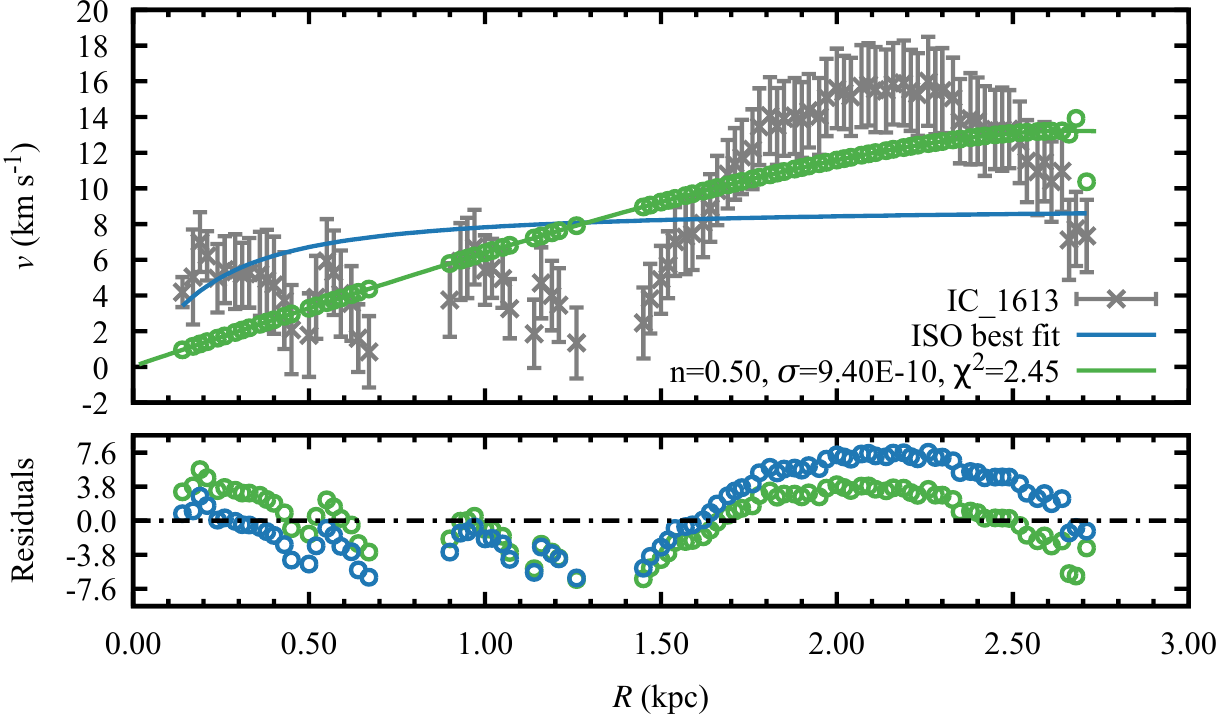}\\
                        \includegraphics[width=.49\linewidth]{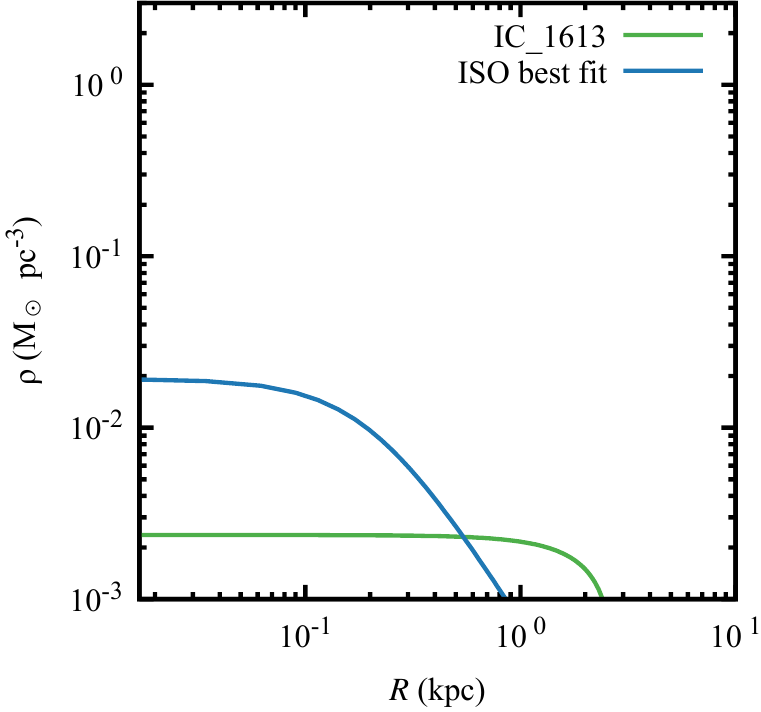}
                        \includegraphics[width=.49\linewidth]{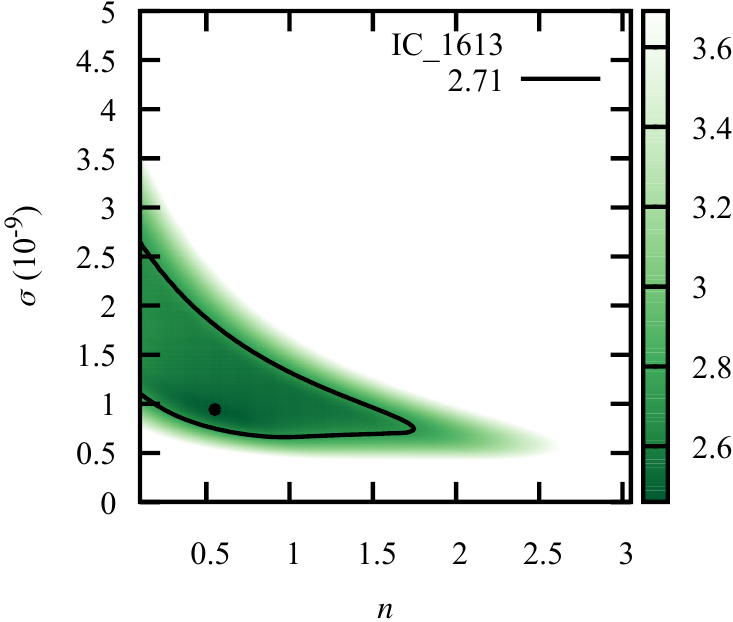}\\[-6.2mm]\mbox{}
        \end{minipage}
        \end{center}
\end{minipage}
\\[3ex]
\begin{minipage}[t]{.49\linewidth}
\begin{minipage}[t]{\linewidth}\raggedright
   \begin{minipage}[t]{0.35\linewidth}
    {\large (15) NGC 1569}
   \end{minipage}
        \begin{minipage}[t]{0.60\linewidth}\raggedleft\scriptsize
    Right Ascension: \ascension{04}{30}{46}{2}\\
        Declination: \declination{+64}{51}{10}{3}\\
        Distance: 3.4~Mpc\\
        Absolute magnitude: $-$18.2~mag\\[-5mm]\mbox{}
        \end{minipage}  
\end{minipage}
        \begin{center}
        \begin{minipage}{\linewidth}
                        \includegraphics[width=\linewidth]{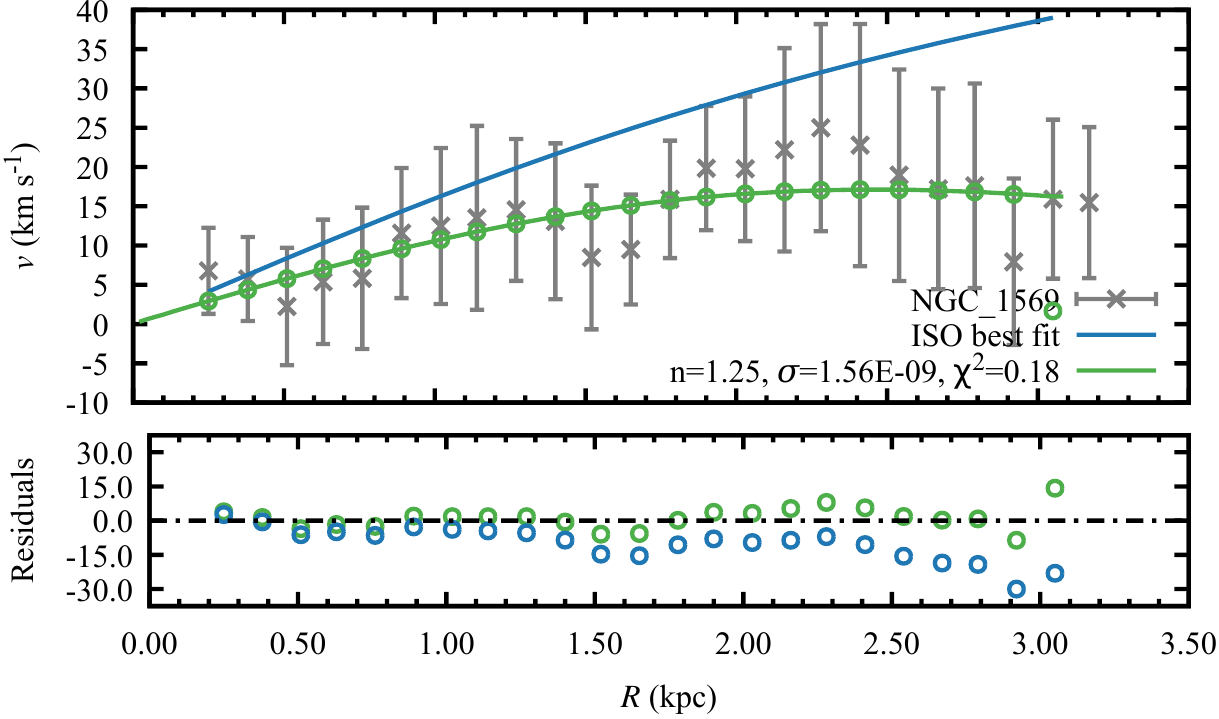}\\
                        \includegraphics[width=.49\linewidth]{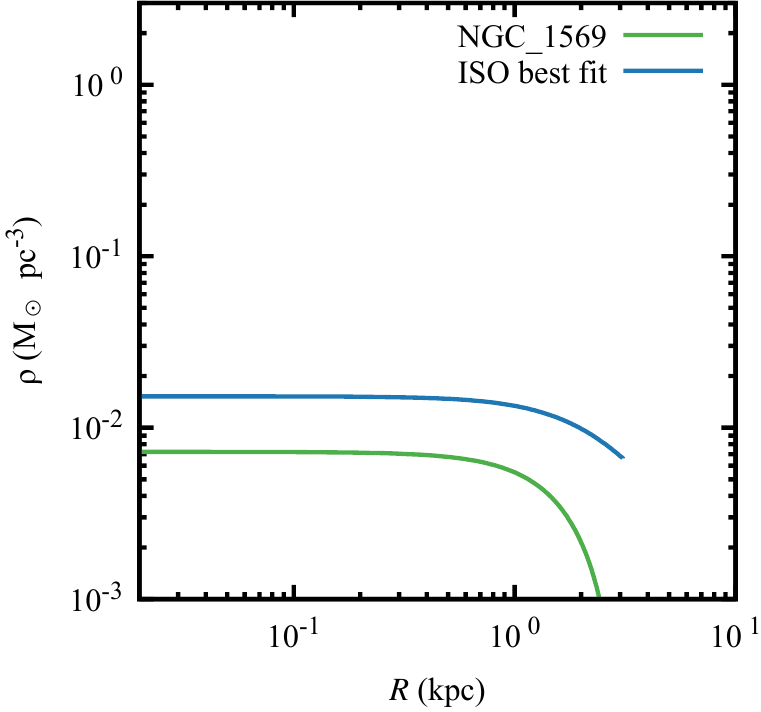}
                        \includegraphics[width=.49\linewidth]{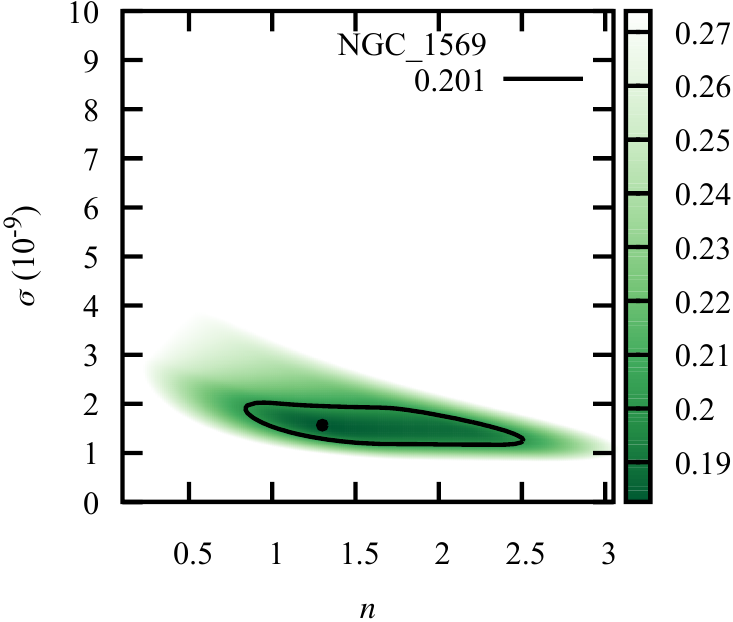}\\[-6.2mm]\mbox{}
        \end{minipage}
        \end{center}
\end{minipage}
\begin{minipage}[t]{.49\linewidth}
\begin{minipage}[t]{\linewidth}\raggedright
   \begin{minipage}[t]{0.35\linewidth}
    {\large (16) NGC 2366}
   \end{minipage}
        \begin{minipage}[t]{0.60\linewidth}\raggedleft\scriptsize
    Right Ascension: \ascension{07}{28}{53}{4}\\
        Declination: \declination{+69}{12}{49}{6}\\
        Distance: 3.4~Mpc\\
        Absolute magnitude: $-$16.8~mag\\[-5mm]\mbox{}
        \end{minipage}  
\end{minipage}
        \begin{center}          
        \begin{minipage}{\linewidth}
                        \includegraphics[width=\linewidth]{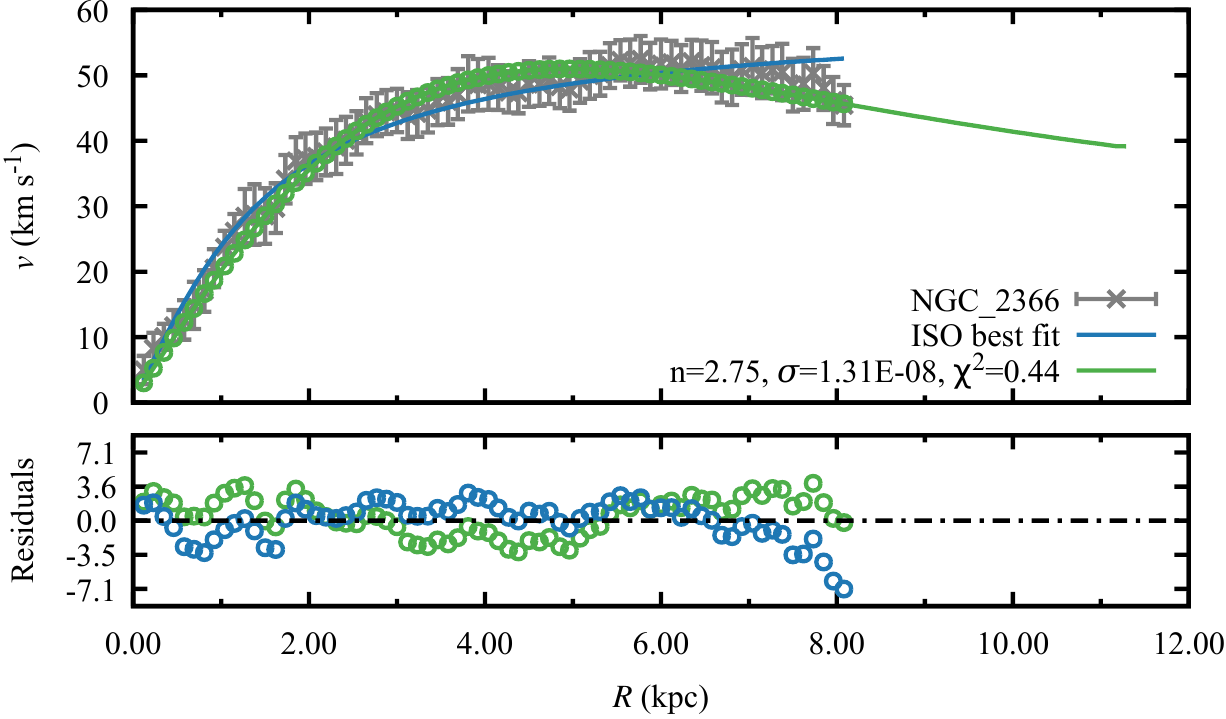}\\
                        \includegraphics[width=.49\linewidth]{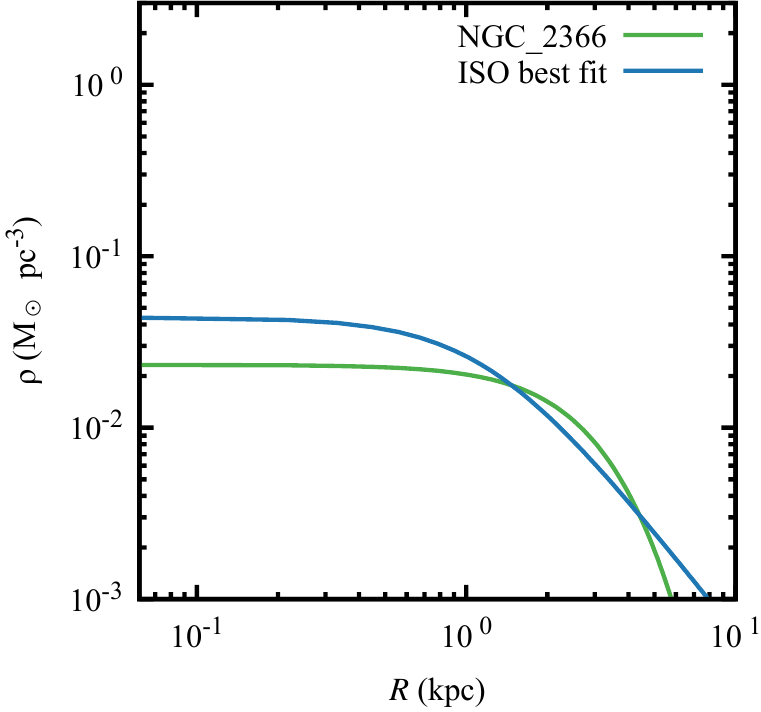}
                        \includegraphics[width=.49\linewidth]{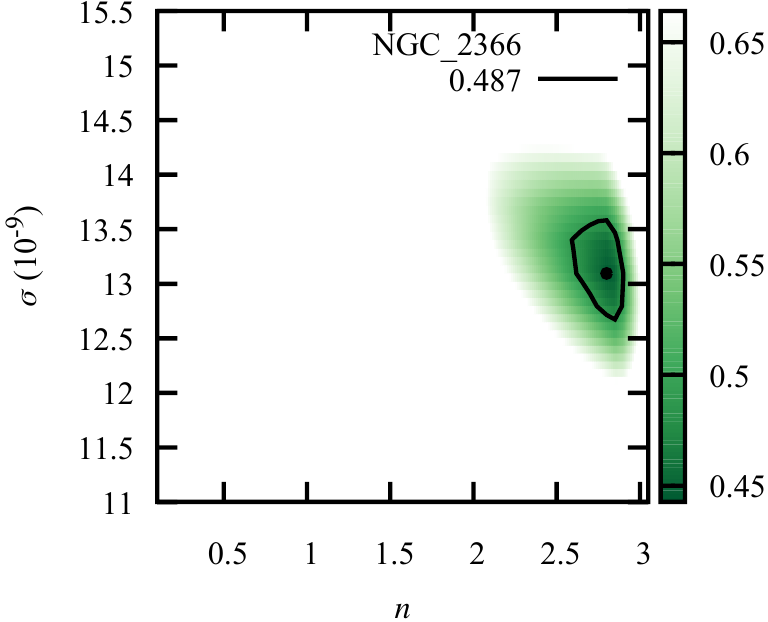}
        \end{minipage}
        \end{center}
\end{minipage}
\caption{\label{fig5}Velocity curve, mass density profile, and contour map of the best free parameters $\sigma$ and $n$; for IC\,10 (top left), IC\,1613 (top right), NGC\,1569 (bottom left) and NGC\,2366 (bottom right).}
\end{figure*}

\begin{figure*}[p]
\begin{minipage}[t]{0.49\linewidth}
\begin{minipage}[t]{\linewidth}\raggedright
   \begin{minipage}[t]{0.35\linewidth}
    {\large (17) NGC 3738}
   \end{minipage}
        \begin{minipage}[t]{0.60\linewidth}\raggedleft\scriptsize
    Right Ascension: \ascension{11}{35}{46}{9}\\
        Declination: \declination{+54}{31}{44}{8}\\
        Distance: 4.9~Mpc\\
        Absolute magnitude: $-$17.1~mag\\[-5mm]\mbox{}
        \end{minipage}  
\end{minipage}
        \begin{center}
        \begin{minipage}{\linewidth}
                        \includegraphics[width=\linewidth]{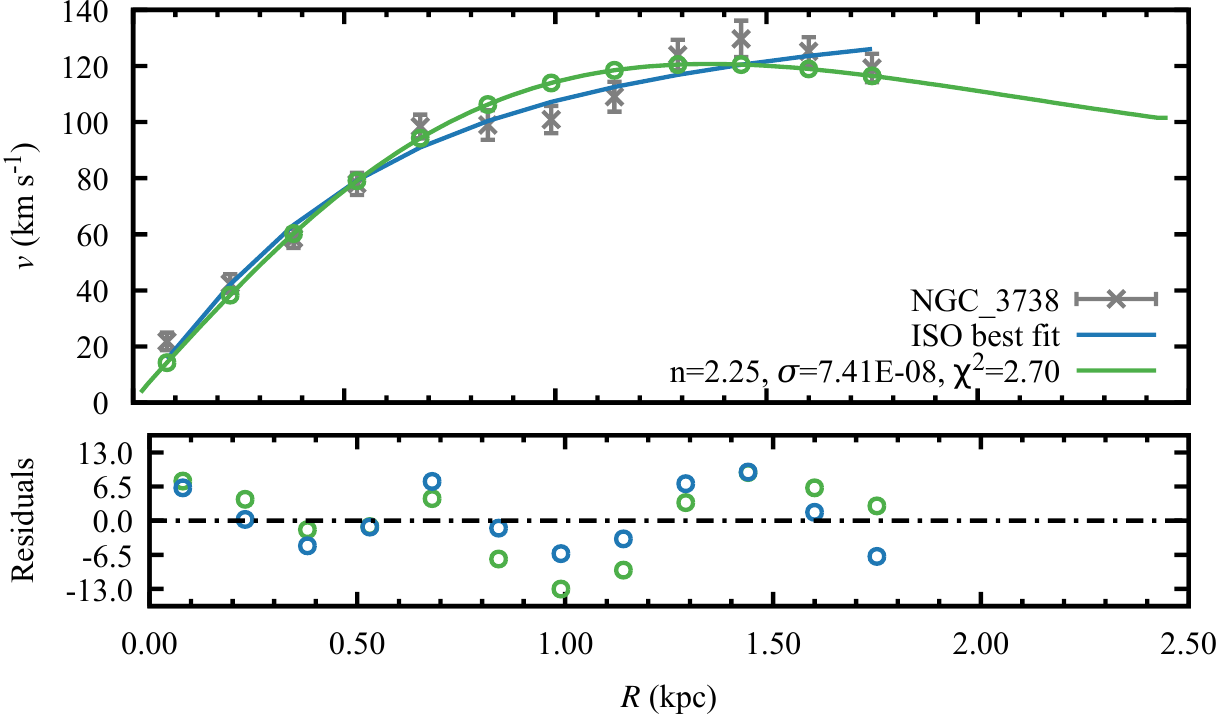}\\
                        \includegraphics[width=.49\linewidth]{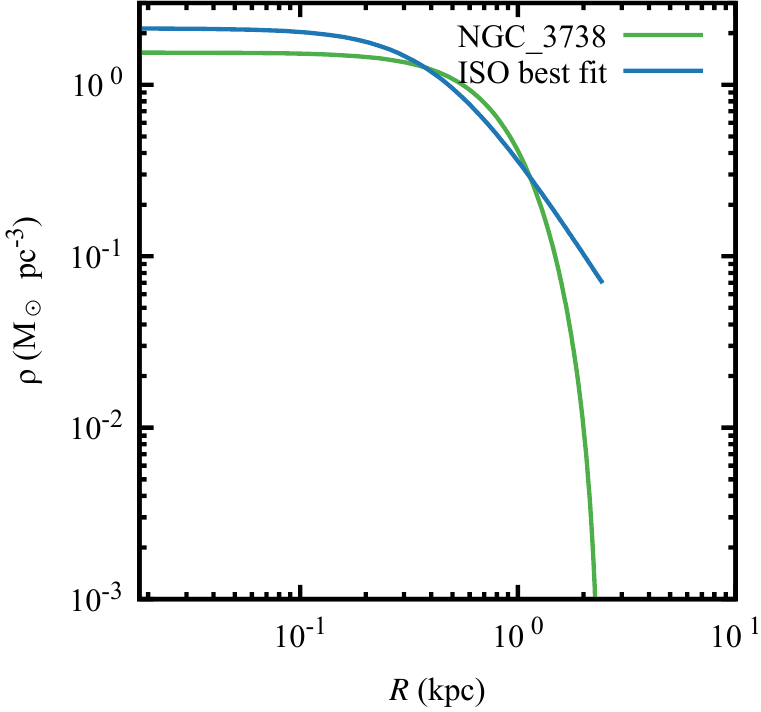}
                        \includegraphics[width=.49\linewidth]{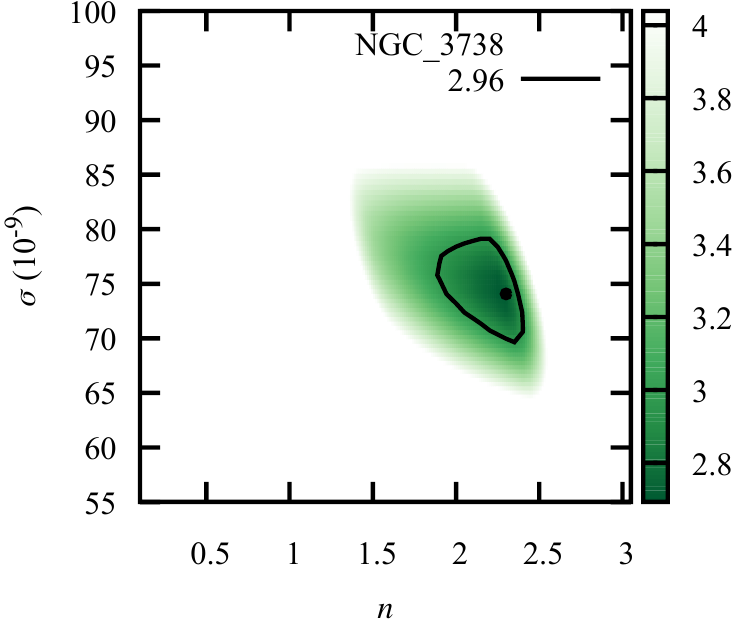}\\[-6.2mm]\mbox{}
        \end{minipage}
        \end{center}
\end{minipage}
\begin{minipage}[t]{.49\linewidth}
\begin{minipage}[t]{\linewidth}\raggedright
   \begin{minipage}[t]{0.35\linewidth}
    {\large (18) UGC 8508}
   \end{minipage}
        \begin{minipage}[t]{0.60\linewidth}\raggedleft\scriptsize
    Right Ascension: \ascension{13}{30}{44}{9}\\
        Declination: \declination{+54}{54}{32}{4}\\
        Distance: 2.6~Mpc\\
        Absolute magnitude: $-$13.6~mag\\[-5mm]\mbox{}
        \end{minipage}  
\end{minipage}
        \begin{center}          
        \begin{minipage}{\linewidth}
                        \includegraphics[width=\linewidth]{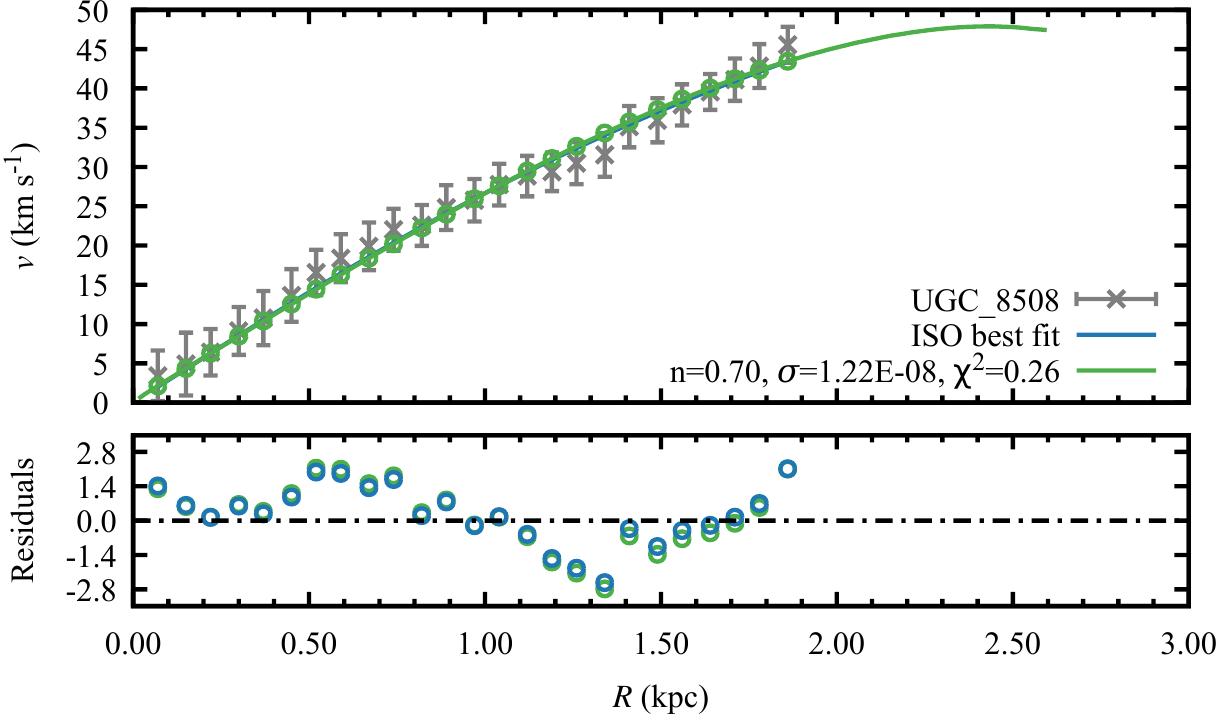}\\
                        \includegraphics[width=.49\linewidth]{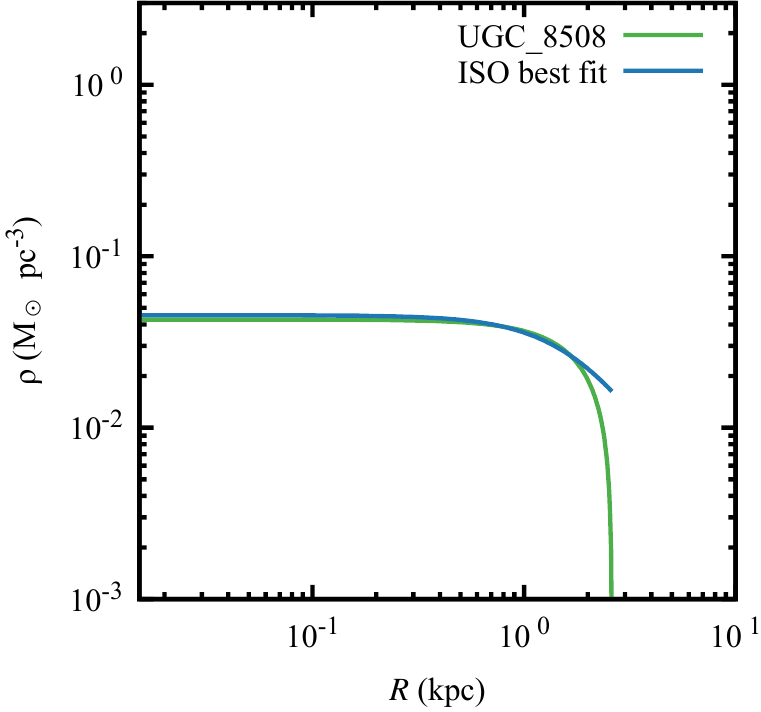}
                        \includegraphics[width=.49\linewidth]{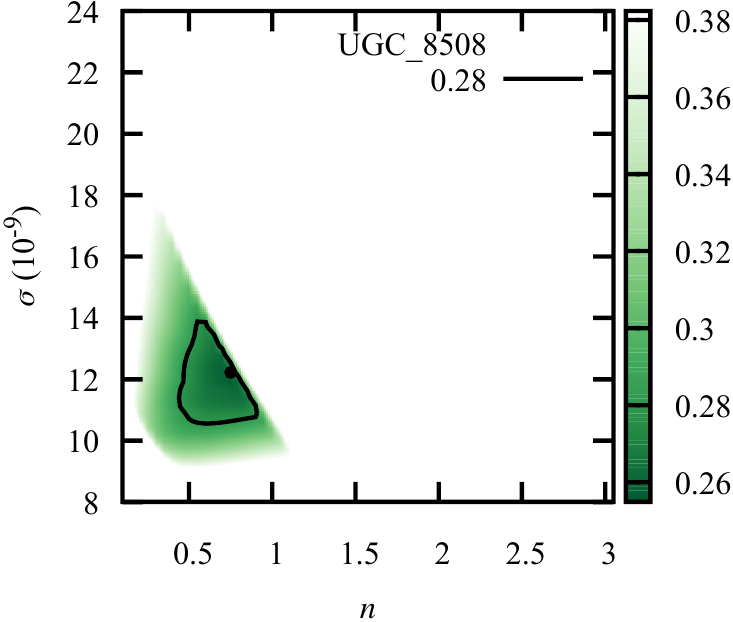}\\[-6.2mm]\mbox{}
        \end{minipage}
        \end{center}
\end{minipage}
\\[3ex]
\begin{minipage}[t]{.49\linewidth}
\begin{minipage}[t]{\linewidth}\raggedright
   \begin{minipage}[t]{0.35\linewidth}
    {\large (19) WLM}
   \end{minipage}
        \begin{minipage}[t]{0.60\linewidth}\raggedleft\scriptsize
    Right Ascension: \ascension{00}{01}{59}{9}\\
        Declination: \declination{$-$15}{27}{57}{2}\\
        Distance: 1.0~Mpc\\
        Absolute magnitude: $-$14.4~mag\\[-5mm]\mbox{}
        \end{minipage}  
\end{minipage}
        \begin{center}
        \begin{minipage}{\linewidth}
                        \includegraphics[width=\linewidth]{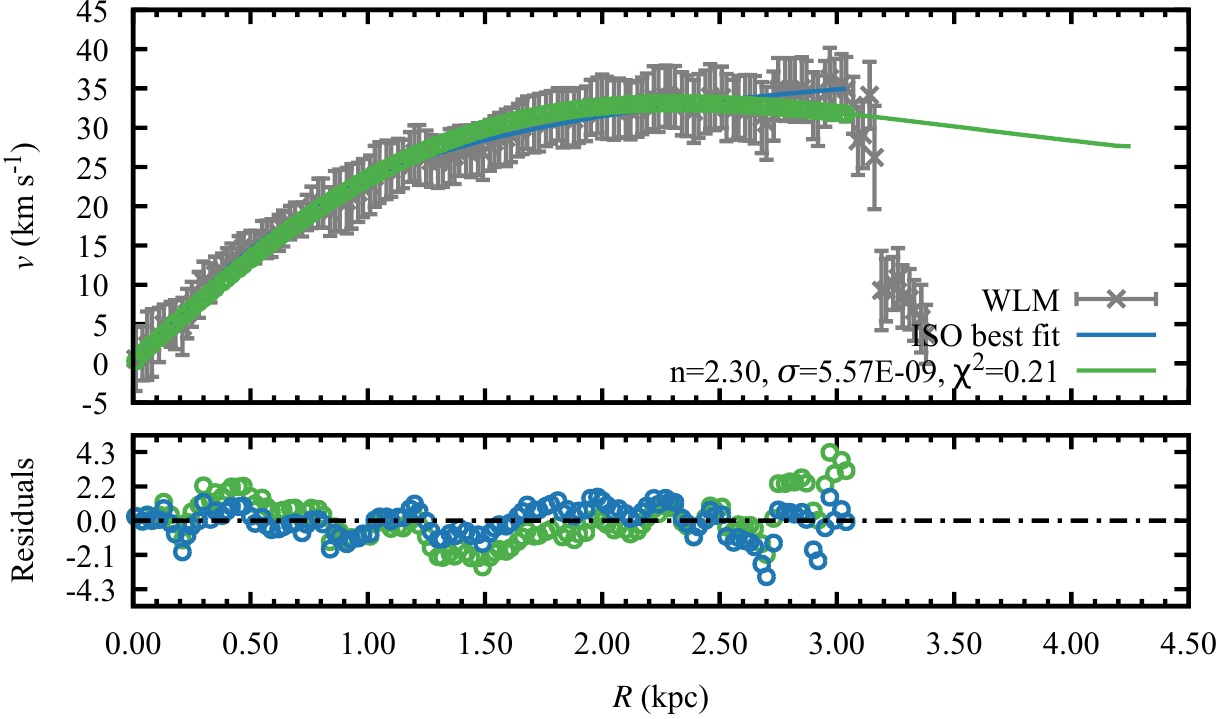}\\
                        \includegraphics[width=.49\linewidth]{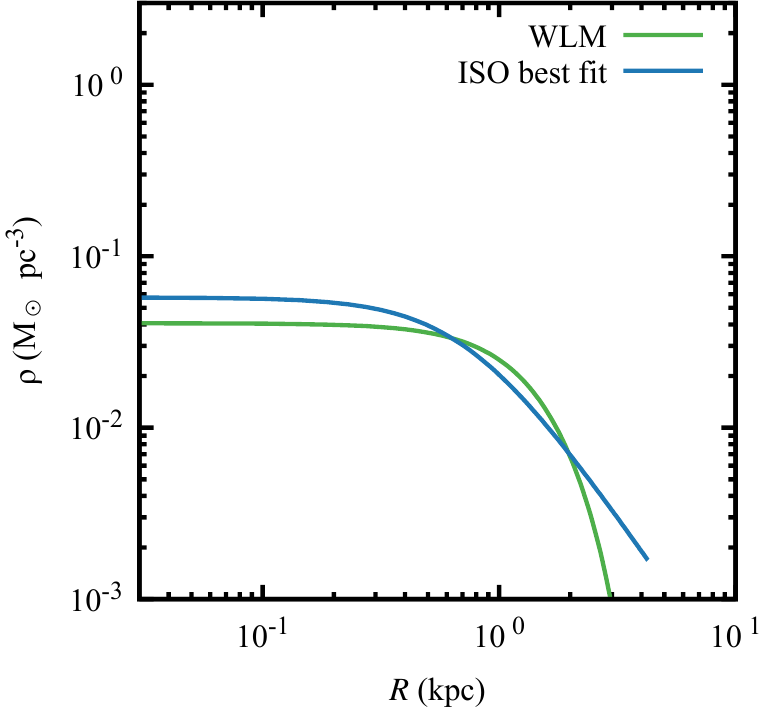}
                        \includegraphics[width=.49\linewidth]{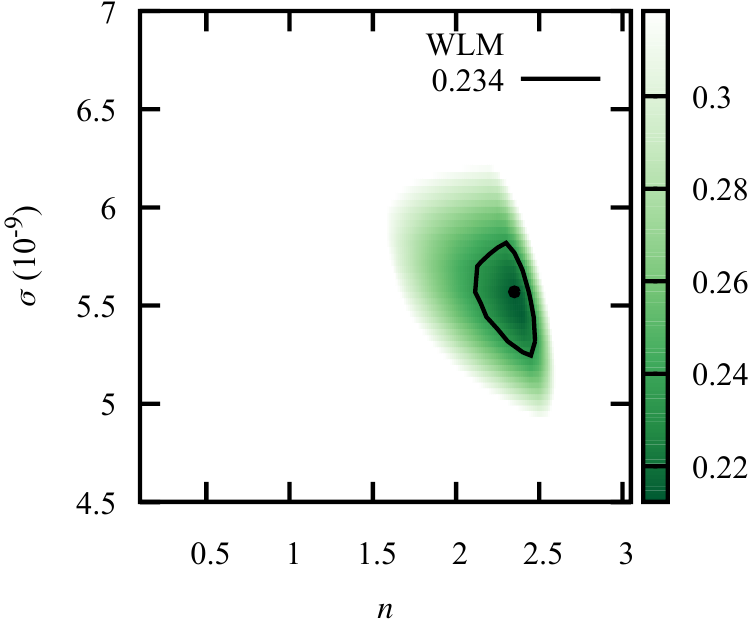}\\[-6.2mm]\mbox{}
        \end{minipage}
        \end{center}
\end{minipage}
\begin{minipage}[t]{.49\linewidth}
\begin{minipage}[t]{\linewidth}\raggedright
   \begin{minipage}[t]{0.35\linewidth}
    {\large (20) Haro 36}
   \end{minipage}
        \begin{minipage}[t]{0.60\linewidth}\raggedleft\scriptsize
    Right Ascension: \ascension{12}{46}{56}{6}\\
        Declination: \declination{+51}{36}{47}{3}\\
        Distance: 9.3~Mpc\\
        Absolute magnitude: $-$15.9~mag\\[-5mm]\mbox{}
        \end{minipage}  
\end{minipage}\mbox{ }
        \begin{center}  
        \begin{minipage}{\linewidth}
                        \includegraphics[width=\linewidth]{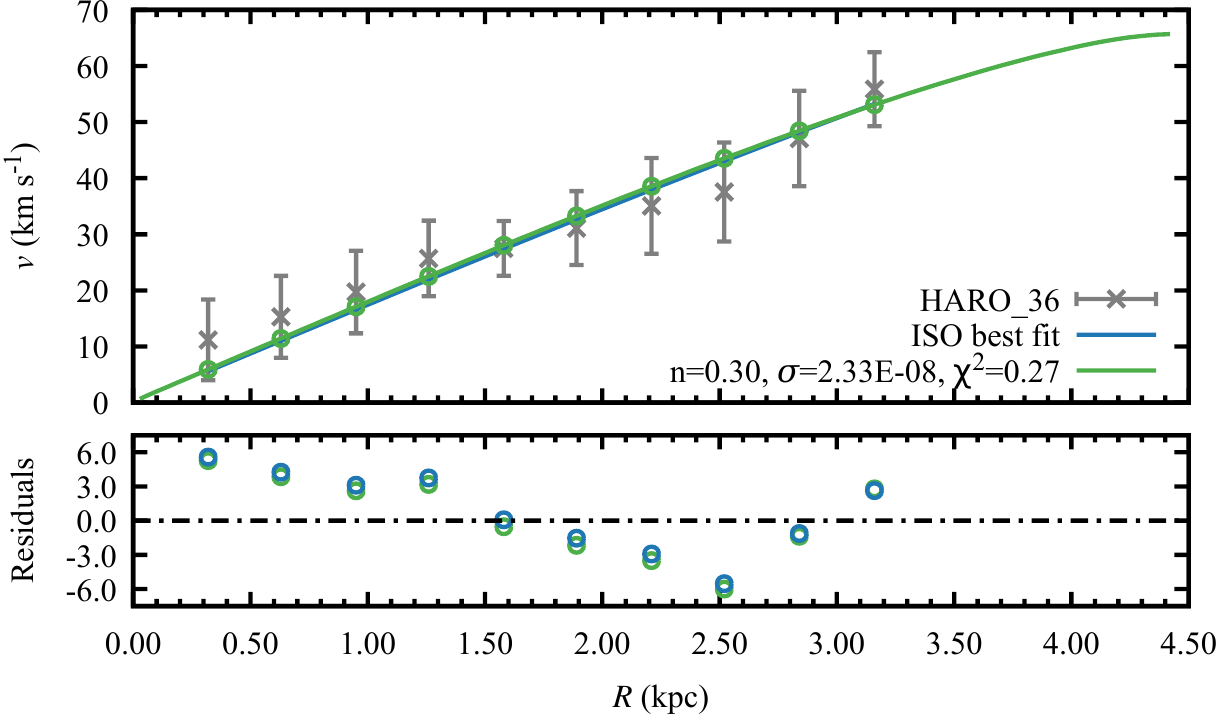}\\
                        \includegraphics[width=.49\linewidth]{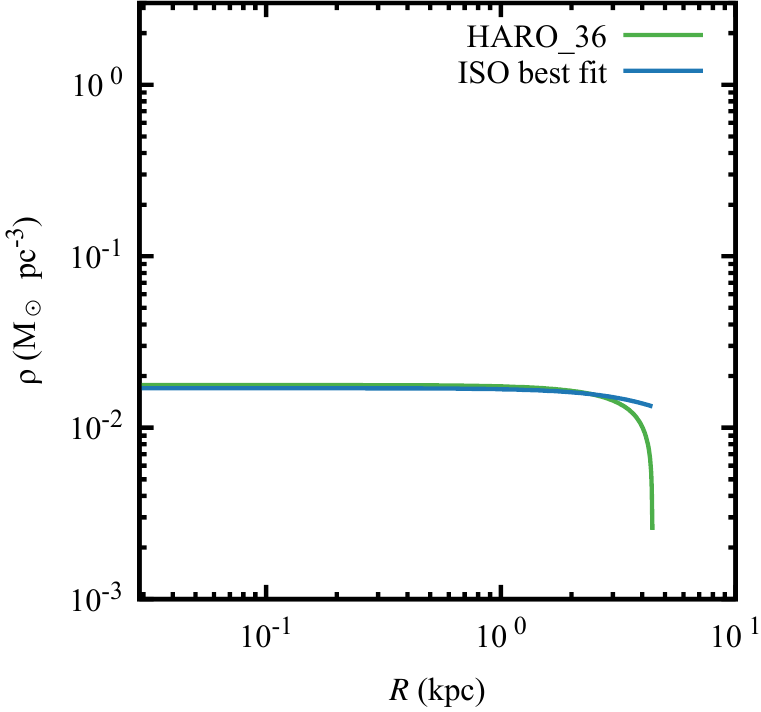}
                        \includegraphics[width=.49\linewidth]{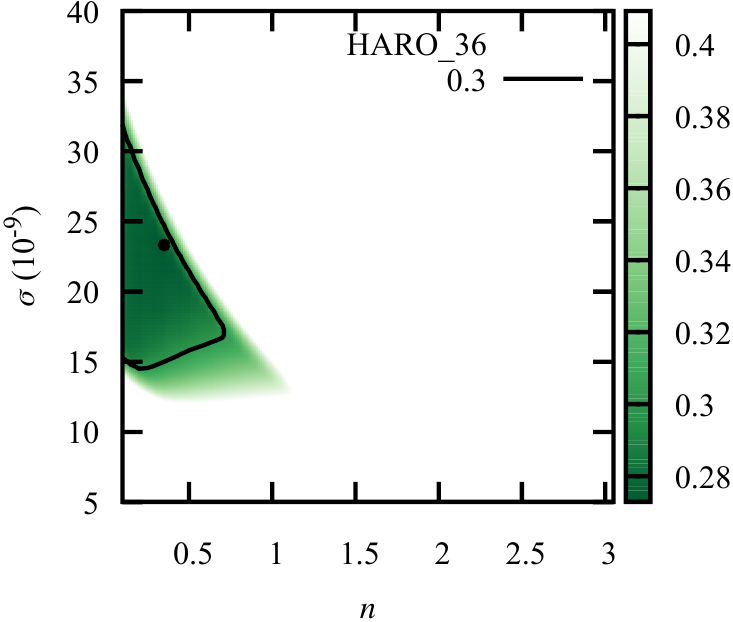}
        \end{minipage}
        \end{center}
\end{minipage}
\caption{\label{fig6}Velocity curve, mass density profile, and contour map of the best free parameters $\sigma$ and $n$; for NGC\,3738 (top left), UGC\,8508 (top right), WLM (bottom left) and Haro\,36 (bottom right).}
\end{figure*}

The $\chi^2$ tests are only applied for the data presented in \citet{Oh-etal:2015:AJ:}. For each of the dwarf galaxies of the LITTLE THINGS, we separately give the results of the fitting procedure and comparison of our polytrope fitting velocity curves with those predicted by the pseudo-isothermal model. The results are presented in Figs.~\ref{fig2}--\ref{fig6}. For more involved comparison of the methods, we add a different comparison of each point from the observational data. For each of the considered dwarf galaxies, the mass density profile of the polytrope obtained by the fitting procedure is given, and the area of the $97\%$ confidence (the green contour plot in the bottom right corner in Figs.~\ref{fig2}--\ref{fig6}) of the free polytrope parameters $n$ and $\sigma$ is added. All the results are summarised in Table~\ref{tab:vel_dw_result}.

We note that usually the region of 97\% confidence is restricted to a relative small range of the parameter $n$, which is around $\Delta n\leq 0.5$, thereby giving relevant restrictions. However, there are a few cases, (see Figs.~\ref{fig4}--\ref{fig5}) in which $\Delta n \sim 1.5$ , giving signals of significant effects of factors other than DM, perhaps of the strong effect of baryonic matter.

We can see that the velocity curves predicted by the polytrope models fit in all the considered galaxies the observed velocity curves whose quality of the fit is comparable with the quality of the fit by the pseudo-isothermal models. For some galaxies the $\chi$-square is little bit better owing to the polytrope model, for the other galaxies the pseudo-isothermal model fit is better; the relation is 10:10. Clearly, the quality of the fit to observational data in polytrope models are on the same level as the pseudo-isothermal model. Considering the interval of confidence, %
the variability of the polytropic index $n$ and the relativistic parameter $\sigma$ is around $25$\% of the best value.

Our results of the fitting procedure give different values of $n$ and $\sigma$ for each of the considered dwarf galaxies, with a large extension of the polytropic index that covers whole the range of the considered values, $n \in (0,4)$. This result is very contradictory to our expectation that the values of $n$ of the selected polytrope models are coupled around one or two characteristic values in accordance with the assumption of a unique physical mechanism hidden behind the DM halo model. However, this enables an interesting conclusion that the dwarf galaxies modelled by the polytrope spheres can be separated into three classes according to the value of the polytropic index $n$, governing the behaviour of the velocity profile of the circular geodesics of the polytrope interior.

In the first class of models, where $n \in (0,1)$, the velocity profile increases nearly linearly from the centre to the halo edge. In the second class of models, where $n \in (1,2)$, the velocity profile reaches an outer region where it has almost flat dependence on the radius. In the third class of models, where $n > 2$, the velocity profile reaches a maximum with increasing radius and then follows a slow decline. Inside all of the three polytrope classes, the parameter $\sigma$ related to individual members (dwarf galaxy models) varies significantly, but remains  very
small ($\sigma < 10^{-8}$) in all the considered cases; the variations of $\sigma$ has only minor effect on the qualitative character of the velocity profiles predicted by the model.

Of course, the observational data contain four galaxies demonstrating an apparently irregular (perturbative) behaviour of the observed velocity curves. It is very interesting that the polytrope models implied by the fitting procedure predict, in two of these cases, the values of the relevant polytropic index on the edges of the allowed range. A value of $n = 3.95$ is  predicted for the DDO~50 galaxy and $n = 0.5$ is predicted for the IC~1613 galaxy. In both of these cases the perturbations of the smooth theoretical velocity profiles by the observational data are strongly irregular. The two other galaxies of this kind have $n = 1.70$ (IC~10) and $n = 1.25$ (NGC~1569) and demonstrate in observational data perturbations from the smooth theoretical velocity profiles having almost periodic character. There is also an interesting phenomenon that occurs in the behaviour of the 97\% confidence range of the polytropic index, $\Delta n_\mathrm{conf}$, for these four irregular galaxies. While the confidence range is very high, $\Delta n_\mathrm{conf} \sim 1.5,$ for the galaxies modelled by the polytropes with low polytropic indexes $n = 0.5, 1.25, 1.70$, in the case of the galaxy corresponding to the polytrope with high index $n = 3.95$, the confidence range is only  $\Delta n_\mathrm{conf} \sim 0.5$, similar to the rest of the selected 20 dwarf galaxies.

If we assume that the irregularities are caused by the presence of a significant amount of baryonic matter, then these limiting cases indicate that the existence of these irregularities is probably caused by local conditions, not by fundamental physics related to the behaviour of the polytropes.


The distribution of the considered dwarf galaxies into the introduced classes is rather irregular: the first class contains seven galaxies, the second class only four galaxies, while the third class contains  nine galaxies. The polytropic index $n$ is distributed across whole the range of $n$ corresponding to the classes; the distribution is almost regular, with an exception of one very high value of $n=3.95$ (for the dwarf galaxy with a strongly irregular observed velocity curve) in the third class where the rest of the polytrope models have $n \in (2,3)$. The distribution of the parameter $\sigma$ in the first class covers more than one order around $\sigma = 10^{-9}$, in the second class the distribution of $\sigma$ covers one order, while in the third class it covers more than two orders. The central density of the selected polytrope models is distributed around the value of $\rho_\mathrm{c} = 10^{-24}~\mathrm{g\, cm}^3$, with similar characteristics of the distribution as in the case of the parameter $\sigma$.

We observe no correlation between the values of the free parameters $n, \sigma, \rho_\mathrm{c}$; we thus assume they are governed mainly by the local conditions. The masses, velocity profiles, and the maximal velocities of the profiles of the polytrope models are, of course, fully related to the values of the polytrope parameters through structure equations, and their distribution reflects properties of the polytropes. We note that the second class could be considered an intermediate state between the first and third classes; the second class is possibly closer to the first class because there is no maximum velocity profile for both the first and second classes. We thus could speculate that at least two physical mechanisms (or two types of DM) could be relevant for binding dwarf galaxies. Of course, such an idea has to be confronted with the particle physics models if realistic candidates of the DM can be found experimentally. On the other hand, we have to recall that such a dichotomy in explanation of DM haloes is also considered in the case of large galaxies because the DM haloes related to both cold DM and hot DM have recently been in play. In the case of large galaxies, the polytrope models also predict high values of the relativistic parameter $\sigma > 0.1$ \citep{Stu-Hle-Nov:2016:PHYSR4:,Stu-Nov-Hla-Hle:2021:preparation:}; in the case of the dwarf galaxies, the DM must be in a strongly non-relativistic state with $\sigma < 10^{-8}$. However, the study of relativistic polytropic model with high levels of $\sigma$ are important \citep{Stu-Nov-Hla-Hle:2021:preparation:}, as the recent studies of the second data release (DR2) of the Gaia mission \citep{GAIA_RD2} shows that the Galactic halo is composed of disrupted dwarf galaxies \citep{Hay-etal:2018,Hel-etal:2018:Nature}. This is especially the case if we take into account that the scaling laws of dwarf galaxies return a DM density for the Galactic halo that is consistent with measurements from the vertical Jeans equations \citep{Cas:2020:APJ}

\begin{figure}[tb!]
        \begin{center}
                \includegraphics[width=\linewidth]{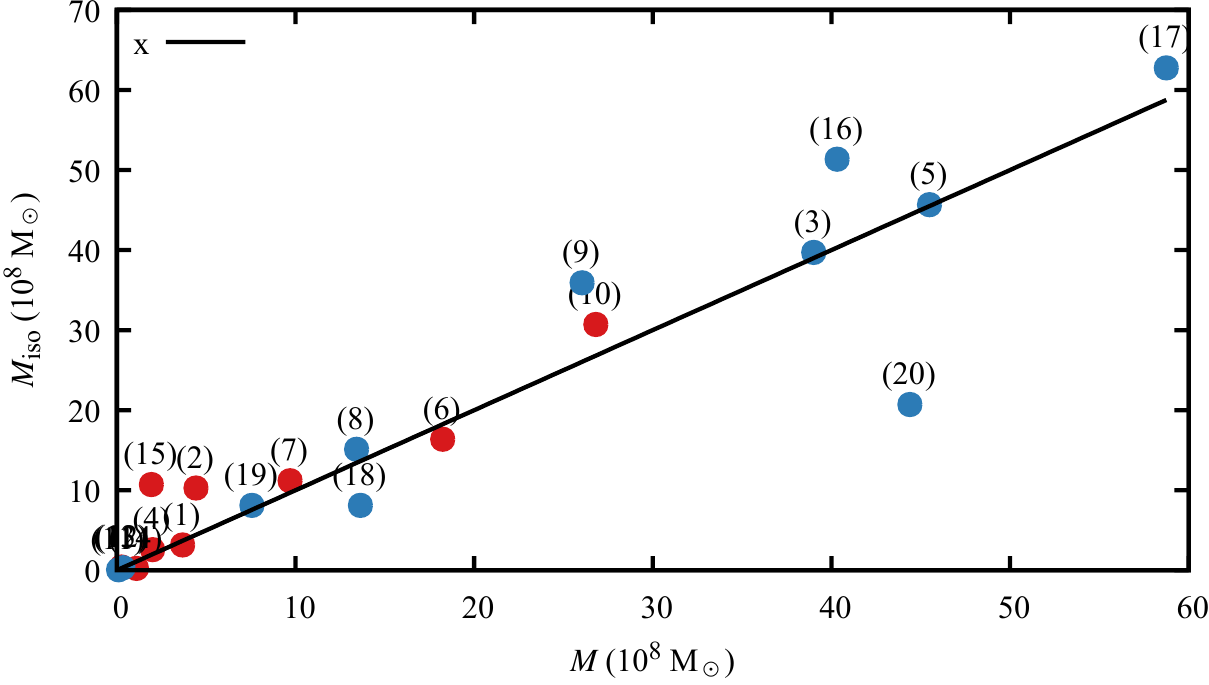}\\
                \mbox{}\vfill%
                \includegraphics[width=\linewidth]{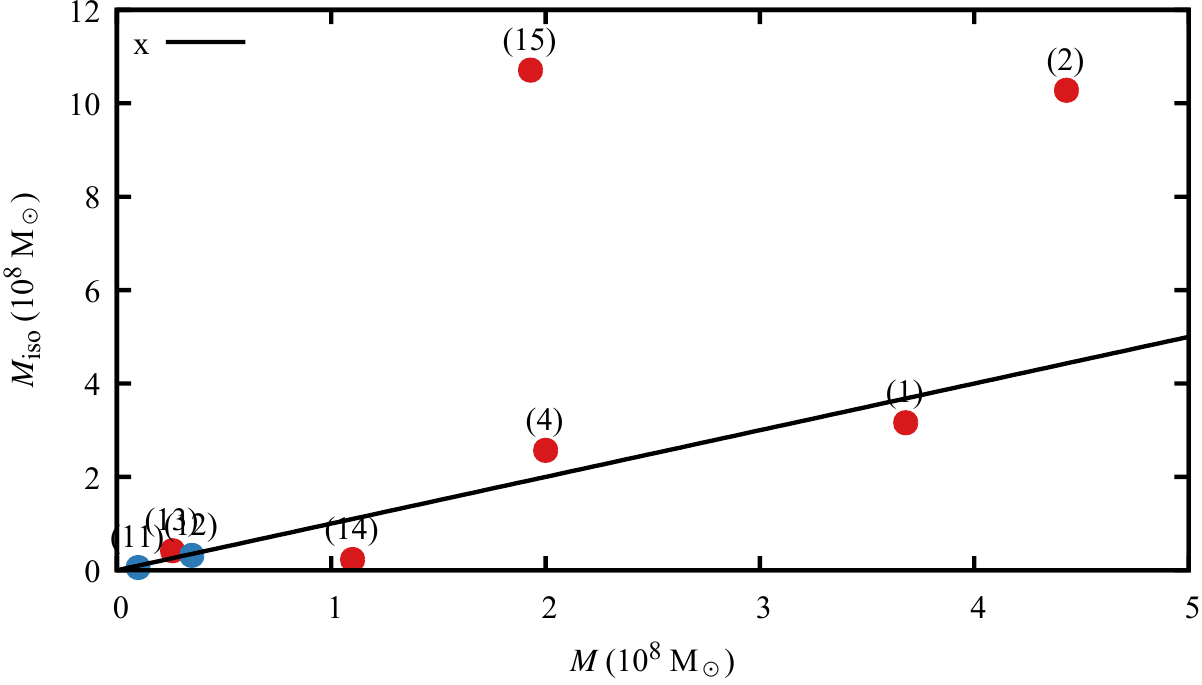}
        \end{center}    
                \caption{\label{fig22}Comparison of the mass of the polytropic sphere with the mass of pseudo-isothermal model for each tested dwarf galaxy. Each point is accompanied by a number in brackets representing the dwarf galaxy (see Table~\ref{tab:LittleThings_sum} for the name). The cases where the polytrope model has a lower value of the $\chi^2$ test compared to the pseudo-isothermal model are depicted as red points, while the blue points indicate a better fit of the pseudo-isothermal model. The black line corresponds to the line of equality. The bottom figure is just a zoomed view of the top figure.}
\end{figure}

\begin{figure}[tb!]
        \begin{center}
                \includegraphics[width=\linewidth]{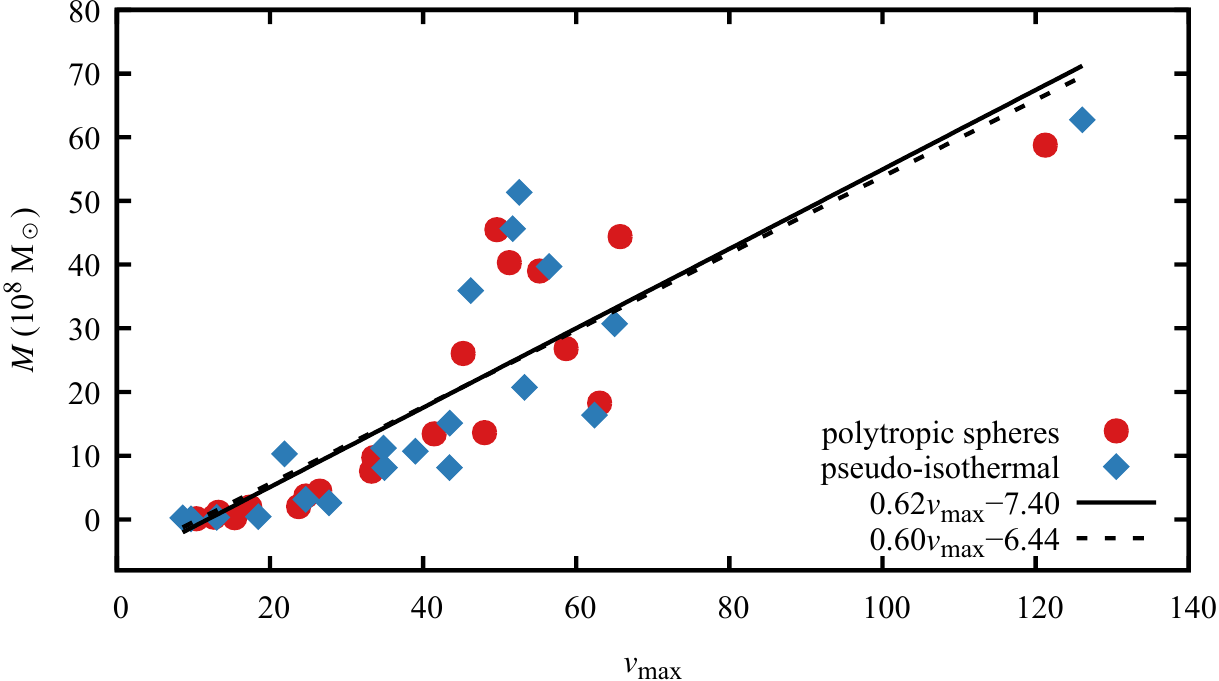}\\
                \mbox{}\vfill%
                \includegraphics[width=\linewidth]{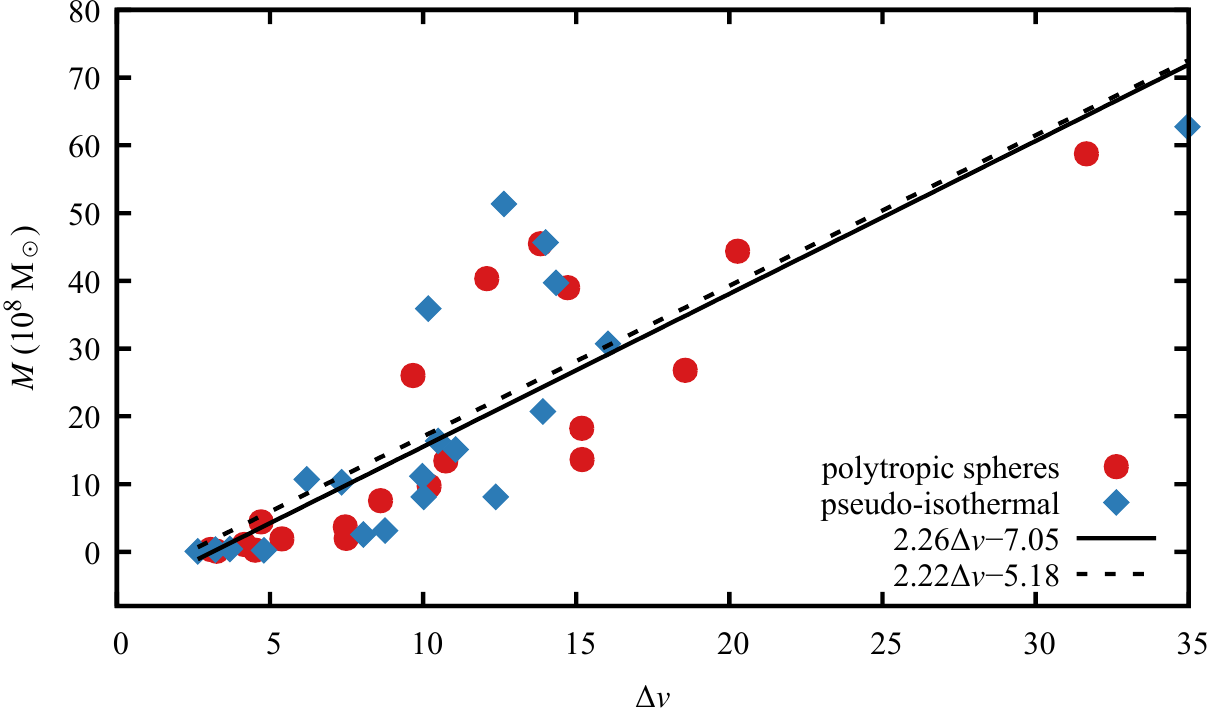}
        \end{center}    
                \caption{\label{fig23}Comparison of the velocities and the mass of the polytropic sphere. \textit{Top}: Relation of the maximal velocity $v_\mathrm{max}$ with the mass $M$ for both models. The red points correspond to the polytrope model and the blue points to the pseudo-isothermal model. \textit{Bottom}: Correlation between the velocity dispersion $\Delta v$ and the mass $M$. The same schematic colour as in the top figure is used.}
\end{figure}

\subsection{Comparison with pseudo-isothermal model}
Interesting results can also be obtained by comparing the fitting procedures by the pseudo-isotropic and polytrope models. We thus study the inner slope of the DM density $\alpha$ for our polytrope models, and compare these to those given by the pseudo-isothermal models.

During the numerical fitting procedure, the central density $\rho_\mathrm{c}$ was, for the best fit, selected to approximately match the outer radius of the dwarf galaxy. For a concrete galaxy, such an approach enables us to compare the mass of the concrete polytrope model with the mass of the corresponding pseudo-isothermal model that is defined by the relation
\begin{equation}
    M_\mathrm{iso} = 4\pi \int_0^R \rho_\mathrm{iso}(r)r^2\,\mathrm{d}r\, .
\end{equation}

The comparison of the masses $M_\mathrm{pol}$ and $M_\mathrm{iso}$ is presented in Fig.~\ref{fig22}; each point represents a dwarf galaxy, above the equality line there is $M_\mathrm{pol} < M_\mathrm{iso}$. We can see that the polytrope model mass is mostly equal or slightly lower than the isothermal model mass. For completeness, we give in Fig.~\ref{fig23} the relation of the maximal velocity $v_\mathrm{max}$, and the velocity dispersion $\Delta v$, to the mass parameter M, for both the polytrope (red points) and pseudo-isothermal (blue points) models. The relationship of the velocity dispersion $\Delta v$ and matter is well known; for example around the $\mathrm{SgrA}^\star$ super-massive black hole in the Galaxy centre $\Delta v {\sim} 75~\mbox{km}\,\mbox{s}^{-1}$, while for the Andromeda super-massive central black hole $\Delta v {\sim} 160~\mbox{km}\,\mbox{s}^{-1}$. For the group of dwarf galaxies studied in our paper,  $\Delta v < 35~\mbox{km}\,\mbox{s}^{-1}$, in accordance with the typical elliptical dwarf galaxies in the Coma cluster where $\Delta v < 80~\mbox{km}\,\mbox{s}^{-1}$ \citep{kou-etal:2012:MNRAS:velDwarfs}.

\begin{figure*}[tbp]
        \begin{center}
                        \hfill\includegraphics[width=.4\linewidth]{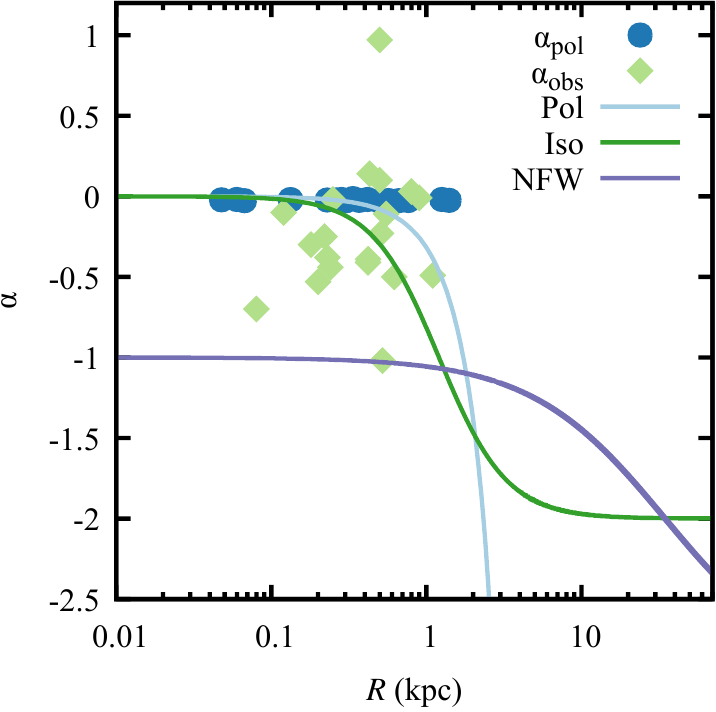}\hfill%
                        \includegraphics[width=.4\linewidth]{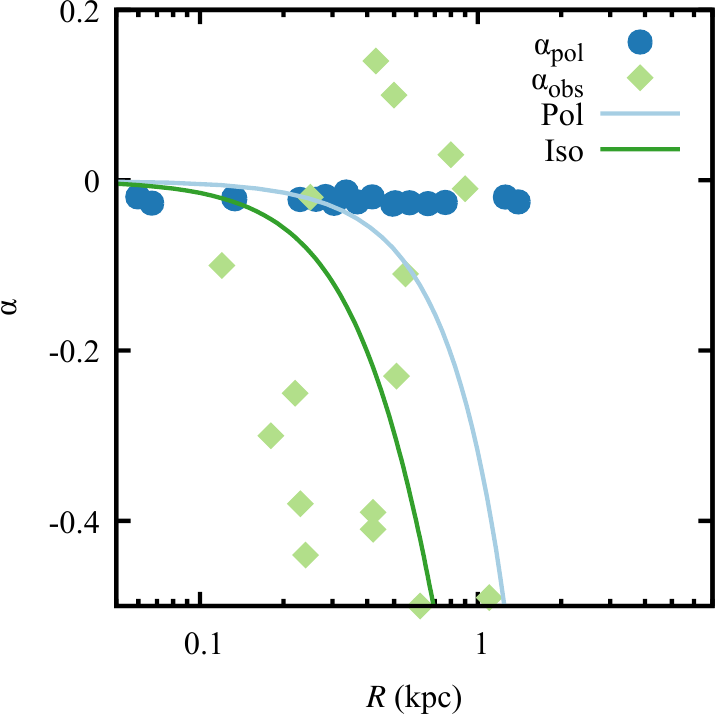}\hfill\mbox{}\\
                        \caption{\label{fig24}Inner slope of the DM density profiles $\alpha$ against the inner radius for polytropic sphere and pseudo-isothermal model. The blue points correspond to the mean value of inner slope $\alpha_\mathrm{pol}$ within the break radius $R^\mathrm{pol}_\mathrm{inner}$ for each dwarf galaxy. Similarly, the rhombus represents the inner slope $\alpha_\mathrm{obs}$ within the radius $R^\mathrm{obs}_\mathrm{inner}$ derived from the observational data. The typical behaviour of the slope $\alpha$ against radius for the polytropical sphere, pseudo-isothermal, and NFW models are also presented. On the left side  the profile of whole maximal radius of a specific galaxy is illustrated, while on the right side a closer view is presented.}
        \end{center}
\end{figure*}

The mass density profile of the polytrope models demonstrates a slower decrease in comparison with the ``core'' pseudo-isothermal density profile defined by Eq.~(\ref{eq:psedoiso}). We also determine the inner slope $\alpha$ of the matter density profile within the ``break radius'' of the DM density profiles for both models used in the fitting procedure. The break radius is defined as the location at which the slope of the matter density radial profile changes most rapidly \citep{Blok-Bos:2002:AA:}. Therefore, we define  the radius $R{}_\mathrm{inner}^\mathrm{pol}$
for the polytrope models density profile and compute the mean inner slope $\alpha$ of the polytrope model for each dwarf galaxy. For the inner slope $\alpha_\mathrm{iso}$ and the inner radius $R{}_\mathrm{inner}^\mathrm{obs}$, we are using those given in \citet{Oh-etal:2015:AJ:}. In Fig.~\ref{fig24} we compare the inner slope $\alpha$ against the observed radius $R$ for both models and all considered dwarf galaxies.

As clear from Fig.~\ref{fig24}, most of the LITTLE THINGS galaxies demonstrate significant deviation from the NFW model of the $\Lambda$-CDM haloes --- the slope $\alpha$ related to the density profiles of the polytrope and pseudo-isothermal models do not agree with the cusp predicted by the NFW model corresponding to the simulations based on the $\Lambda$-CDM model. The mean value of the polytrope slope $\alpha_\mathrm{pol}{\sim}-0.06$ and the observed value $\alpha_\mathrm{obs} {\sim}-0.26$, both differ significantly from those related to the NFW model, $\alpha_\mathrm{NFW} {\sim} 1$. The polytrope models thus predict, similar to the pseudo-isothermal models, core-like haloes. In the polytrope model, the slope is concentrated near the zero value for a longer range of radius in comparison with the other models, and the break radius is located at larger distance from the centre. All the results discussed above are summarised in Tables~\ref{tab:vel_dw_result} and \ref{tab:alpha_slope}. These results clearly demonstrate the relevance of the polytrope models (separated into three classes) to explain the properties of the dwarf galaxies from the LITTLE THINGS.

\begin{table}
\caption{\label{tab:alpha_slope}Results of the inner slope of the matter density profiles $\alpha$ and the break radii $R_\mathrm{inner}$ for both DM haloes models.}
\begin{center}
\begin{tabular}{lcccc}
\toprule
Name & $\alpha_\mathrm{pol}$ & $\alpha_\mathrm{obs}$ & $R_\mathrm{inner}^\mathrm{pol}$ (kpc)& $R_\mathrm{inner}^\mathrm{obs}$ (kpc) \\
\midrule
CVnIdwA  & $-$0.036 & \phantom{$-$}0.03 & 1.60 & 0.80 \\
DDO 50   & $-$0.025 & \phantom{$-$}0.10 & 0.09 & 0.50 \\
DDO 52   & $-$0.064 &       $-$0.49 & 0.80 & 1.10 \\
DDO 53   & $-$0.067 & \phantom{$-$}0.14 & 0.48 & 0.14 \\
DDO 87   & $-$0.068 &       $-$0.01 & 1.28 & 0.90 \\
DDO 101  & $-$0.056 &       $-$1.02 & 0.23 & 0.52 \\
DDO 126  & $-$0.068 &       $-$0.39 & 0.79 & 0.42 \\
DDO 133  & $-$0.062 &       $-$0.11 & 0.46 & 0.55 \\
DDO 154  & $-$0.064 &       $-$0.41 & 0.59 & 0.42 \\
DDO 168  & $-$0.069 & \phantom{$-$}0.97 & 0.92 & 0.50 \\
DDO 210  & $-$0.066 &       $-$0.70 & 0.11 & 0.08 \\
DDO 216  & $-$0.053 &       $-$0.30 & 0.11 & 0.18 \\
IC 10    & $-$0.072 &       $-$0.25 & 0.09 & 0.22 \\
IC 1613  & $-$0.063 &       $-$0.10 & 0.99 & 0.12 \\
NGC 1569 & $-$0.069 &       $-$0.23 & 0.61 & 0.51 \\
NGC 2366 & $-$0.062 &       $-$0.53 & 0.79 & 0.20 \\
NGC 3738 & $-$0.072 &       $-$0.44 & 0.26 & 0.24 \\
UGC 8508 & $-$0.066 &       $-$0.38 & 0.79 & 0.23 \\
WLM      & $-$0.062 &       $-$0.02 & 0.40 & 0.25 \\
Haro 36  & $-$0.054 &       $-$0.50 & 1.98 & 0.62 \\
\bottomrule
\end{tabular}
\end{center}
\end{table}

\section{Conclusions}
We briefly presented the polytropic models of DM haloes and related construction of the velocity profiles of the circular test particle motion that can be applied to match the observed velocity curves in galaxies. We realized matching the theoretical velocity profiles constructed for the DM halo polytropic models with the observational data related to the LITTLE THINGS of dwarf galaxies \citep{Oh-etal:2015:AJ:}, indicating three types of the polytropic models of DM in dwarf galaxies, and consequently possibility of existence of three types of the dwarf galaxies, may be governed by slightly different physics lying behind the simple polytrope models.

We discovered that the polytrope model gives satisfactory fits of the observational data related to the rotational velocity curves for the dwarf galaxies from the LITTLE THINGS, which are fully comparable to those presented from the standard pseudo-isothermal model. We demonstrated that for all the considered dwarf galaxies, the inner slope of the DM density of the polytrope model is generally close to the slope of the pseudo-isothermal (core-like) model, having $\alpha \sim 0$. In fact, the mean value of the polytrope model $\alpha_\mathrm{pol} \sim 0.06$ and the mean slope related to the observational data $\alpha_\mathrm{obs} \sim 0.23$ are both much lower than those corresponding to the NFW model, where $\alpha_\mathrm{NFW} \sim 1$. The break radius, where the density profile changes most rapidly is located at a larger distance from the centre for the polytrope models than for the pseudo-isothermal models.

The matching procedure implies that the polytropic models could be of comparable precision with the matches obtained via the standard phenomenological pseudo-isothermal model of DM haloes in dwarf galaxies, which is used in the original study of LITTLE THINGS \citep{Oh-etal:2015:AJ:}. The most important, interesting, and surprising result concerns the fact that the matching separates (in relation    to the character of the velocity profiles) the polytrope models (and dwarf galaxies) into three classes depending on the magnitude of the polytropic index $n$. Moreover, to be able to describe our tested ensemble of dwarf galaxies, we have to consider the polytropic spheres with the polytropic index $n$ starting at very low value of $n = 0.1$ and ending at an unexpected large value of $n = 3.95$. The irregular character of the velocity profiles observed in two dwarf galaxies requiring the polytropic index from the low and high value edges of the allowed range ($n = 0.5$ and $n = 3.95$) could indicate a possible significant role of local conditions for the final structure of the dwarf galaxies.

In all of the estimated polytropic models, the relativistic parameter $\sigma$ has to be very low, corresponding to strongly non-relativistic matter content, as there is always $\sigma < 10^{-8}$. The qualitatively different groups are separated into those with $n < 1$, those with $1 < n < 2$, and those with $n > 2$, giving a qualitatively different behaviour of the velocity curves. It is very interesting that the polytropes with $n > 2$ are frequently considered to be realistic models of the matter in the central parts of neutron stars \citep{Alv-Bla:2017:PRC}; of course, in the DM halo models of dwarf galaxies the relativistic parameter has to be very low, contrary to the case of the neutron stars.

The separation of the polytrope models of the dwarf galaxies demonstrated in the present paper roughly corresponds to the separation of the polytrope models of large galaxies and galaxy clusters, for which the non-relativistic models with $\sigma \ll 1$ and arbitrary $n$ can be used along with the relativistic models with $n > 3.3$ and $\sigma > 0.1$. For the tested dwarf galaxies, the relativistic parameter is always $\sigma \ll 10^{-8}$, but the polytropic index can be as high as $n = 3.95$ as in the case of the dwarf galaxy with a highly irregular velocity profile, which probably indicates a significant role of ordinary matter.

The fitting related to the selected part of the galaxies contained in the LITTLE THINGS requires a very large range of the values of the polytropic index, $n \in (0.1, 3.95)$ and demonstrates almost uniform distribution along this range. The large spread of the polytropic index $n$ in polytrope models fitting the dwarf galaxies indicates that the physical phenomenon behind the construction of DM haloes probably cannot be unique. We thus propose an idea that at least two physical phenomena (e.g. two kinds of DM) enter into  play. These are dependent on the local conditions, initialised in the very early stages of the Universe expansion, and their combined effect implies the large diversity of the polytrope models of dwarf galaxies, similar to large galaxies and galaxy clusters. Of course, such an idea demands a confirmation from a detection of at least two kinds (e.g. light and heavy) of particles constituting the DM. On the other hand, baryonic matter can be relevant, implying another effect with possibly a significant role, namely the kinematic ``heat up'' of DM at the central regions of dwarf galaxies driven by bursty star formation, which was recently discussed in \citet{Read-Wal-Ste:2019:MNRAS}.

\begin{acknowledgements}
The authors would like to express their gratitude to Dr.~Oh for kind submission of the observational data used in the work in treating the LITTLE THINGS, and the Research Centre for Theoretical Physics and Astrophysics, Institute of Physics, Silesian University in Opava for institutional support. Also, the authors would like to express their gratitude to the referee for very stimulating comments.
\end{acknowledgements}
\bibliographystyle{aa}
\bibliography{dwarfs-poly}

\end{document}